\pdfoutput=1
\documentclass[aps, prd, reprint,
superscriptaddress, amsmath, amssymb, floatfix,
showpacs, subfigure]{revtex4-2}

\usepackage[utf8]{inputenc}
\usepackage{graphicx}
\usepackage{dcolumn}

\usepackage[colorlinks, urlcolor=blue, citecolor=blue, menucolor=blue, anchorcolor=blue, linkcolor=blue, runcolor=blue]{hyperref} 
\usepackage{orcidlink}
\usepackage{bm}
\usepackage{lineno}
\usepackage{xspace}
\usepackage[inline]{enumitem}
\usepackage{placeins}
\usepackage{float}
\usepackage{array}
\usepackage{overpic}
\usepackage{xcolor}
\usepackage{placeins}
\usepackage{subfigure}
\usepackage{booktabs}
\usepackage{longtable}
\usepackage[nowatermark]{fixmetodonotes}
\usepackage{tabulary}
\usepackage{isotope}
\usepackage[T1]{fontenc}
\usepackage{multirow}
\usepackage[utf8]{inputenc}
\usepackage{graphicx}
\usepackage{dcolumn}

\usepackage{bm}
\usepackage{placeins}
\newenvironment{conditions*} 
  {\par\vspace{\abovedisplayskip}\noindent
   \tabularx{\columnwidth}{>{$}l<{$} @{${}={}$} >{\raggedright\arraybackslash}X}}
  {\endtabularx\par\vspace{\belowdisplayskip}}

\usepackage{array}
\newcolumntype{x}[1]{>{\centering\arraybackslash\hspace{0pt}}p{#1}}

\usepackage{array}
\newcolumntype{L}[1]{>{\raggedright\let\newline\\\arraybackslash\hspace{0pt}}p{#1}}
\newcolumntype{C}[1]{>{\centering\let\newline\\\arraybackslash\hspace{0pt}}p{#1}}
\newcolumntype{R}[1]{>{\raggedleft\let\newline\\\arraybackslash\hspace{0pt}}p{#1}}

\usepackage{lineno}

\newcommand{\LANL}{Los Alamos National Laboratory, Los Alamos, NM 87544, USA}
\newcommand{\Pitt}{University of Pittsburgh, Pittsburgh, PA 15260, USA}
\newcommand{\Columbia}{Columbia University in the City of New York, New York, NY 10027, USA}

\newcommand{\Chicago}{University of Chicago, Chicago, IL 60637, USA}

\renewcommand{\alt}{\text{alt}}
\newcommand{\nom}{\text{nom}}

\begin{document}
\title{Benchmarking Nucleon Production in Hadron Interactions Relevant for GeV-scale Neutrino Experiments}

\noaffiliation

\author{R.~Diurba} \affiliation{\LANL}
\author{S.~Dytman} \affiliation{\Pitt}
\author{M.\,J.~King} \affiliation{\Chicago}
\author{Y.~Liu} \affiliation{\Columbia}

\begin{abstract}
In neutrino scattering experiments, the emitted final state particles are essential for identifying interaction channels and reconstructing neutrino energies. During intranuclear propagation, neutrino-induced hadrons can interact with surrounding nucleons, altering both the number of final state particles and detectable energy depositions. As a result, neutrino experiments rely on hadron-nucleus scattering models to simulate these nuclear effects. This paper explores nucleon production predicted by hadron scattering models for common hadrons in neutrino experiments, specifically nucleons and pions. We investigate these models with simulations from the \textsc{GENIE} neutrino event generator, which includes implementations of \textsc{Geant4} Bertini Cascade and \textsc{INCL++}. For popular target nuclei, we find that the number of emitted nucleons for mesonless hadron interactions can be parameterized by simple Gaussian and exponential decay functions. We present these relationships and provide reweighting tools for addressing the hadron scattering model spread in both visible energy and final state nucleon multiplicities.
\end{abstract}

\maketitle

\section{Introduction}
\label{sec:intro}
 Neutrino-induced hadrons that re-interact inside the nucleus, known commonly as final state interactions (FSI), can alter the visible energy and final state topology of the interaction, impacting neutrino energy measurements and interaction classification for current and future neutrino experiments such as Hyper-Kamiokande (HK)~\cite{Hyper-Kamiokande:2025fci} and the Deep Underground Neutrino Experiment (DUNE)~\cite{DUNE:2020jqi,DUNE:2024wvj}. A variety of models are available, each adopting different approximations due to the non-perturbative complexity of many-body nuclear dynamics. As a result, recent studies have shown that differences in hadron scattering models used for particle transport and nuclear effects can obscure the effects of the charge-parity violating phase on the far detector neutrino spectrum, the key measurement for DUNE and HK's oscillation program~\cite{Dolan:2026dfu,Dolan:2026nlr,Yinrui}. 

FSI models in use are most often based on some combination of theory and empiricism. In this paper, we explore the four FSI models presently inside \textsc{GENIE}. A commonly used model developed for \textsc{GENIE} called hA2018 is very empirical and the default choice for many neutrino experiments. It is data-driven in that the probability for different output channels associated with hadron-nucleus interactions comes mainly from data. The choice of kinematics for the final state comes from simple models that describe prominent features in existing hadron-nucleus data. Because the entire FSI is parameterized as a single step, the model can be easily reweighted but also called into question for accuracy. Despite these problems, it provides a reasonable fit to a wide range of neutrino cross section data~\cite{Liu:2026wlw,MicroBooNE:2024yzp,MicroBooNE:2026avc}. The other FSI model internal to GENIE is hN2018. It is an intranuclear cascade (INC) model that has some theoretical corrections to the basically free hadron-nucleus cross sections. All INC models feature a succession of hadron-nucleon interactions. These have nontrivial dependence on nuclear models to account for the complicated dynamics of off-shell nucleons. \textsc{INCL++} and \textsc{Geant4} Bertini Cascade (BC) are more sophisticated INC models that have an underlying nuclear model. \textsc{Geant4} BC comes from an older model developed by Bertini~\cite{BERTINI1971670} that has been adjusted to match more modern data~\cite{Wright:2015xia}. The version of \textsc{Geant4} in \textsc{GENIE} had to be modified for use with neutrino probes; the result is roughly a 10\% decrease in hadron-nucleus cross sections. \textsc{INCL++} has been developed from a basic nuclear model~\cite{cugnon:2016ghr} that has been adjusted to describe a variety of data~\cite{mancusi:2014eia}. 

The models also differ significantly in the processes included. In particular, \textsc{Geant4} BC and \textsc{INCL++} simulate compound nucleus de-excitation, where very low-energy nucleons have a large Compton wavelength and can excite the entire nucleus with the release of a group of nucleons in isotropic directions. These cascade models also account for multinucleon pickup, where outgoing protons or neutrons merge with nucleons in the residual nucleus to form $^2\text{H}$ and higher mass nuclei. Although these processes have a significant influence on the final state, neither hA2018 nor \textsc{GENIE}'s simplified INC model (hN2018) includes either of these effects. 

 All these efforts start with free hadron-nucleon data to provide mean free paths for propagation in the residual nucleus and the kinematics of each interaction in the cascade. The nuclear medium effects are known to be important and model dependent. The ultimate criterion for any FSI model is the ability to fit a wide range of hadron-nucleus data. In this confusing situation, efforts are needed to benchmark these FSI models~\cite{Yinrui,Nikolakopoulos:2024mjj,Dytman}.
 
 As particle scattering models have different assumptions, comparing and contrasting their performance can allow experiments to better quantify model differences and develop data-driven constraints of model spread~\cite{Dytman,DUETTune}. Providing methods to alter simulations, therefore, can constrain uncertainties related to hadron scattering and ultimately improve the precision of neutrino interaction measurements.

A common way to evaluate the impact of hadronic interactions in neutrino physics is using the predicted ``visible'' hadronic energy ($E^{\mathrm{vis}}$), which sums the kinetic energy ($T$) of protons and the total energy of mesons and photons in the final state:
\begin{equation}
E^{\mathrm{vis}} = \sum_{i=p} T_i + \sum_{i=\pi, K, \gamma} E_i    
\end{equation}

\begin{figure*}
    \centering
    \includegraphics[width=0.49\linewidth]{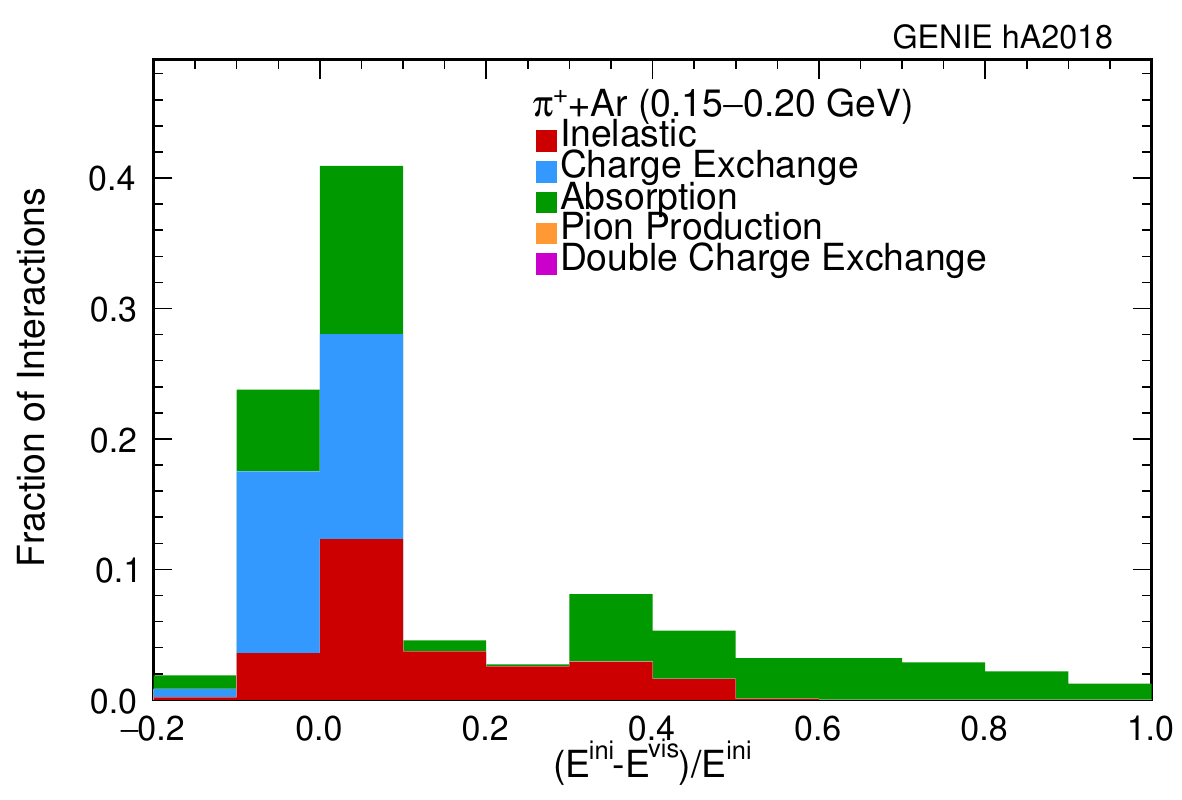}
    \includegraphics[width=0.49\linewidth]{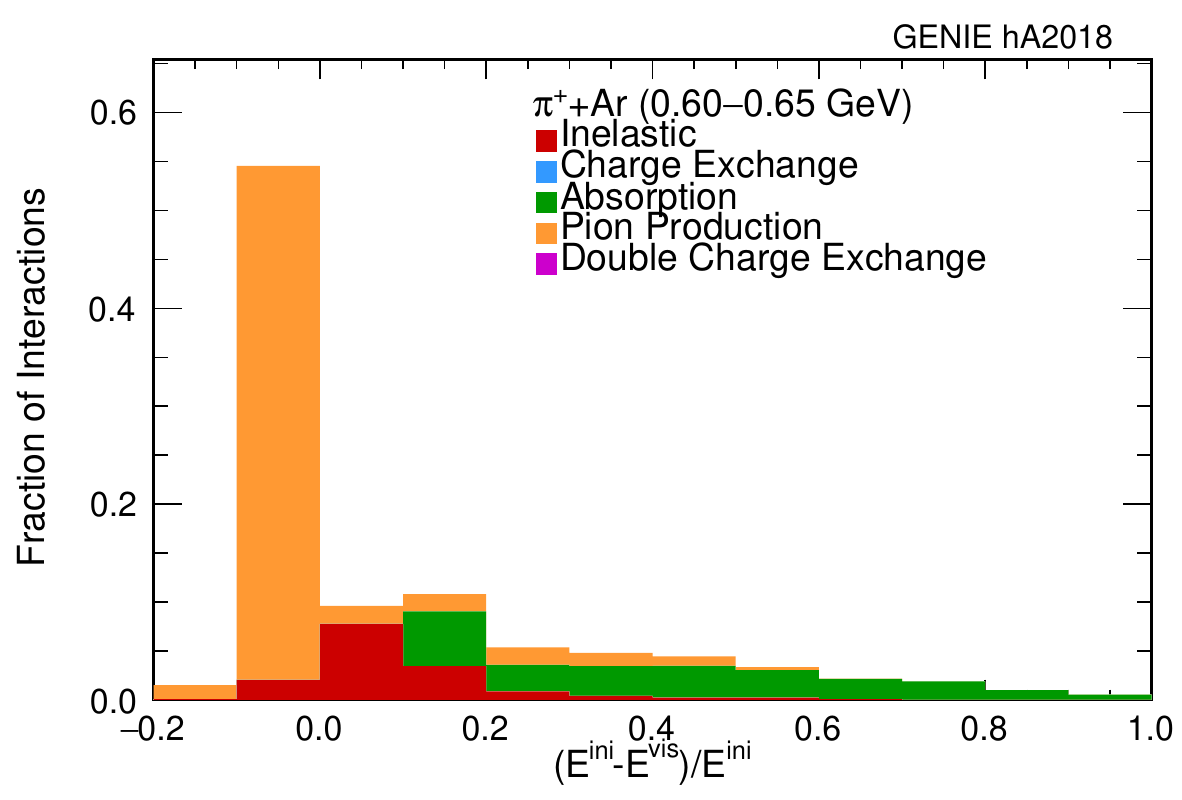}
    \caption{Fractional energy bias simulated by \textsc{GENIE} v3.4.2 hA2018 for pion energies relevant for SBN (left) and DUNE (right). A pion inelastic interaction produces one pion and an additional nucleon. Note that \textsc{GENIE} hA2018 has a feature that predicts no charge exchange interactions above kinetic energies of approximately 450 MeV.}
    \label{fig:eBias}
\end{figure*}

\begin{figure*}
    \centering
    \includegraphics[width=0.49\linewidth]{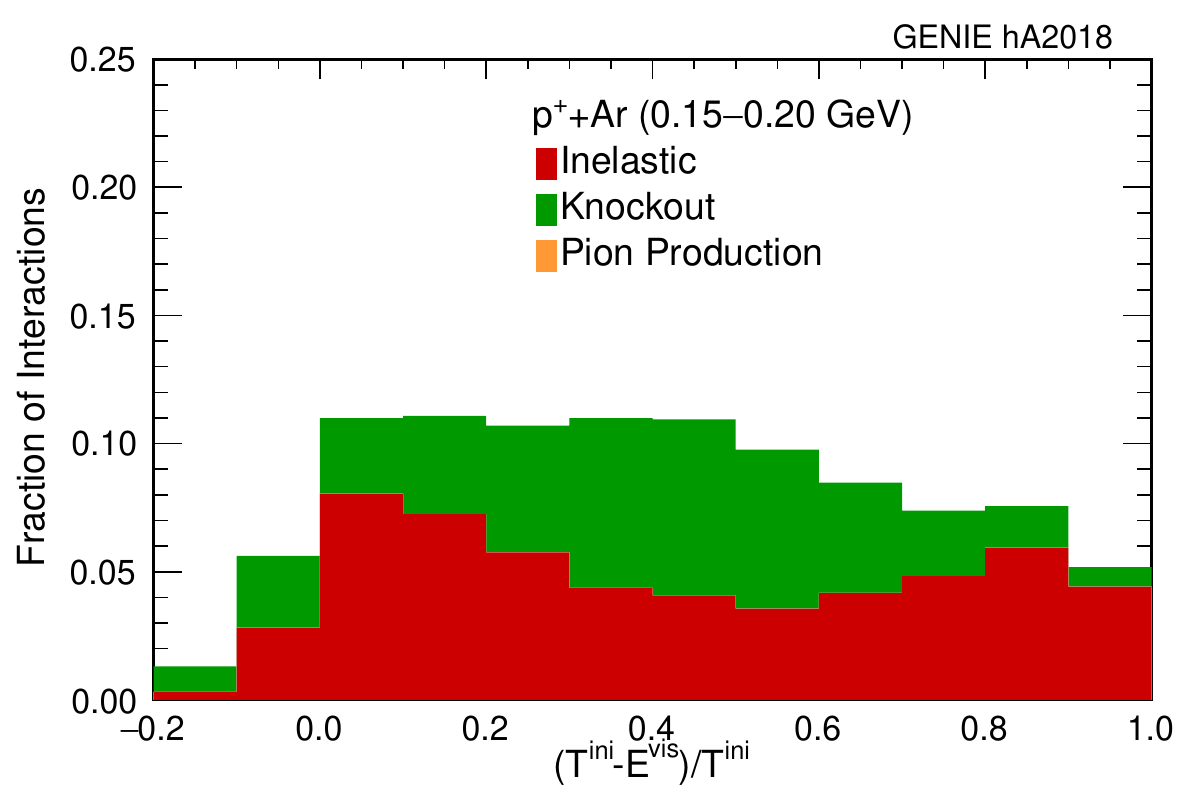}
    \includegraphics[width=0.49\linewidth]{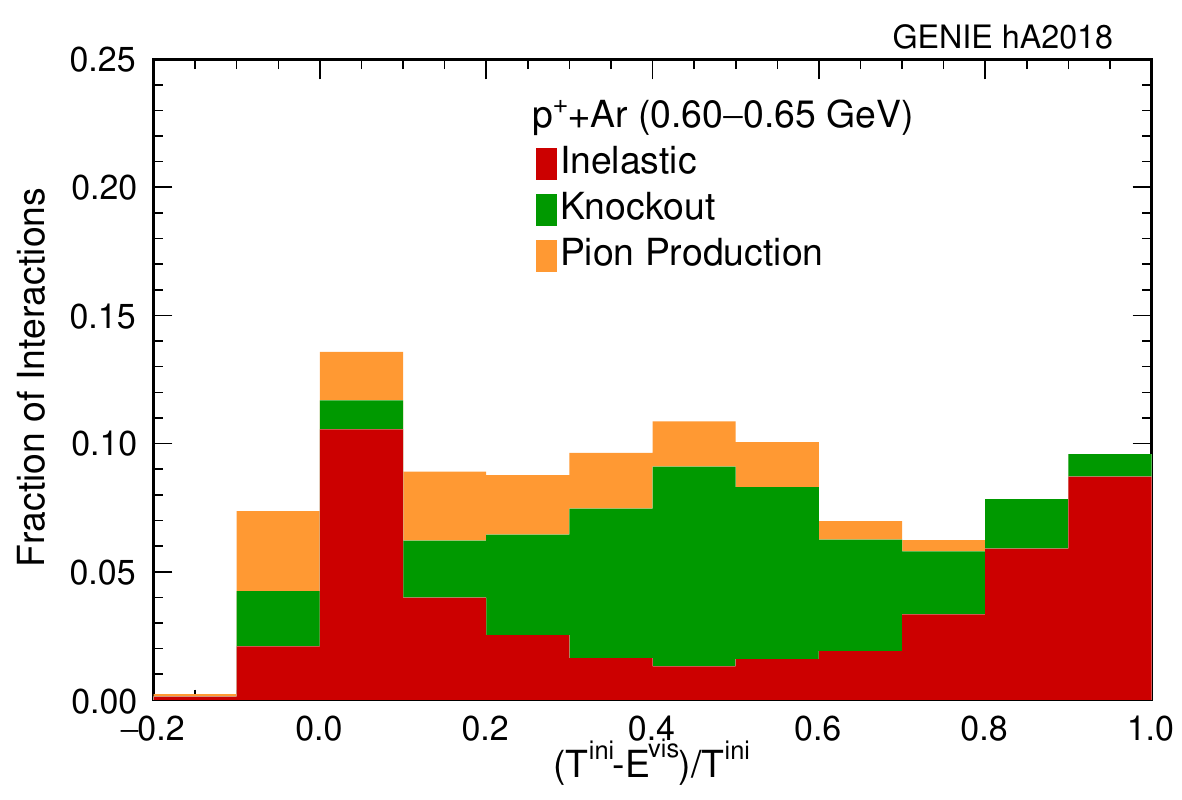}
    \caption{Fractional energy bias simulated by \textsc{GENIE} v3.4.2 hA2018 for proton energies relevant for DUNE (left) and at higher energies (right). An inelastic interaction has one proton and an additional nucleon in the final state.}
    \label{fig:eBiasP}
\end{figure*}

\noindent Figure~\ref{fig:eBias} and Figure~\ref{fig:eBiasP} show the energy loss for pion and proton interactions on argon, respectively. The energy loss is measured at kinetic energies relevant for neutrino interactions at the Short-Baseline Neutrino (SBN) program and for neutrino interactions at the higher energies predicted for DUNE~\cite{MicroBooNE:2015bmn,DUNE:2020jqi}. For pions and nucleons, pion absorption and nucleon knockout are two inelastic channels that lead to a significant loss in energy deposited in the detector. A pion absorption interaction is a pion-nucleus interaction that has no pions in the final state. The nucleon knockout channel refers to interactions with three or more nucleons produced from nucleon-nucleus scattering. Understanding the differences in energy loss due to the interactions between various particle models is vital for providing model flexibility when fitting neutrino data for cross section and oscillation measurements. 

This work employs the \textsc{GENIE} particle transport simulations, which include \textsc{GENIE}'s own particle transport models and implementations of \textsc{Geant4} BC and \textsc{INCL++} models. A more detailed explanation comparing and contrasting these models within \textsc{GENIE} can be found in Refs.~\cite{Yinrui,Dytman}. Additionally, as neutrino experiments widely utilize \textsc{GENIE}'s interaction simulation, this study can directly benefit ongoing experiments such as NOvA~\cite{NOvA:2025tmb}, KM3NeT~\cite{KM3NeT:2024ecf}, IceCube~\cite{IceCubeCollaborationP:2025rpl}, and the SBN detectors of ICARUS~\cite{ICARUS:2023gpo}, SBND~\cite{SBND:2025pjc}, and MicroBooNE~\cite{MicroBooNE:2025nll}.

We use \textsc{GENIE} v3.4.2 to parameterize the number of nucleons produced from pion absorption and nucleon knockout across energy ranges and nuclear targets relevant for neutrino scattering experiments. In the future, these tools should allow experiments to fit a wide variety of particle scattering data, including data from DUET~\cite{DUET}, LArIAT~\cite{elena}, and DUNE's ProtoDUNE hadron analysis program~\cite{pdunePion,pduneProton}, to significantly reduce pion and proton cross section uncertainties in neutrino oscillation experiments. While these constraints have not been applied to nucleon multiplicities,  T2K has shown positive results for these tuning efforts by optimizing their exclusive and inclusive pion scattering predictions using DUET data to nearly eliminate pion reinteractions as an uncertainty in their oscillation measurements~\cite{DUETTune,T2KOsc}.

The paper first summarizes how hadron interaction models compare to hadron scattering data in Sec.~\ref{sec:hadron-nucleus}. It then discusses how the popular \textsc{GENIE} model describes nucleon final states for pion absorption and nucleon knockout in Sec.~\ref{sec:GENIEhA2018}. Section~\ref{sec:results} compares and contrasts the performance on nucleon multiplicity predictions of the four particle transport models included in the \textsc{GENIE} simulation framework. Section~\ref{sec:reweight} then outlines the reweighting protocol for altering the \textsc{GENIE} hA2018 model to mimic the results of simulating hadron-nucleus interactions with \textsc{Geant4} BC and \textsc{INCL++}, demonstrating how these alterations reduce model spread and impact comparisons to recent neutrino datasets. Finally, Sec.~\ref{sec:conclusion} concludes the paper and discusses future prospects for constraining additional hadron scattering channels.

\section{Comparisons of Models to Historical Hadron Scattering Data} \label{sec:hadron-nucleus}

A wide range of historical datasets can be used to evaluate pion absorption and nucleon knockout in modern hadron-nucleus interactions. Traditionally, total reaction cross sections are the most common comparison~\cite{Dytman}. This provides a measure of the mean free path but does not give information about the dynamics of the final state. Fortunately, there is a large body of inclusive data available, and GENIE has a large repository of it. Inclusive measurements place a detector at various angles relative to the beam direction and sample a range of the final state nucleon energies.

Simulations of nucleon-nucleus interactions are a series of nucleon-nucleon scattering events usually called an intranuclear cascade (INC). Even though semi-classical approximations are involved, INC models provide excellent agreement with data. Models such as hN2018, \textsc{INCL++}, and \textsc{Geant4} BC simulate the full cascade. hA2018 attempts to collect all that action into a single event that can be easily reweighted.

The nucleon-nucleon (proton or neutron) cross section is forward-peaked for elastic, quasielastic, or knockout interactions~\cite{said}. The energy of the leading nucleon drops rapidly with angle. If the incident nucleon transferred significant momentum to nucleons inside the nucleus, additional nucleons of various energies end up in the final state. These events are labeled knockout and inclusive data for nucleon-nucleus scattering at backward angles is dominated by them.

\begin{figure*}
    \centering
    \includegraphics[width=0.49\linewidth]{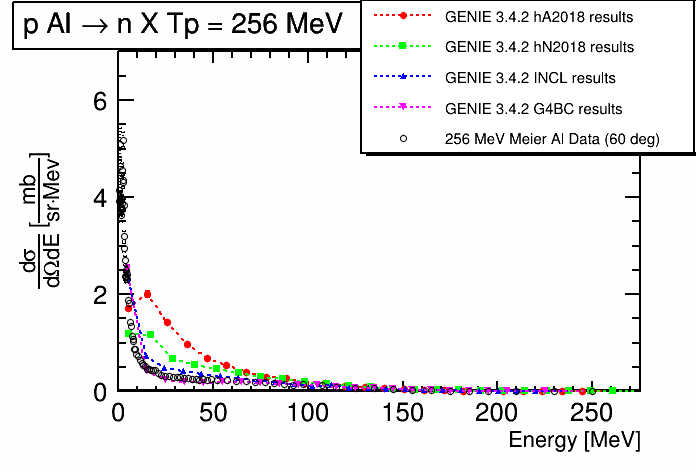}
    \includegraphics[width=0.49\linewidth]{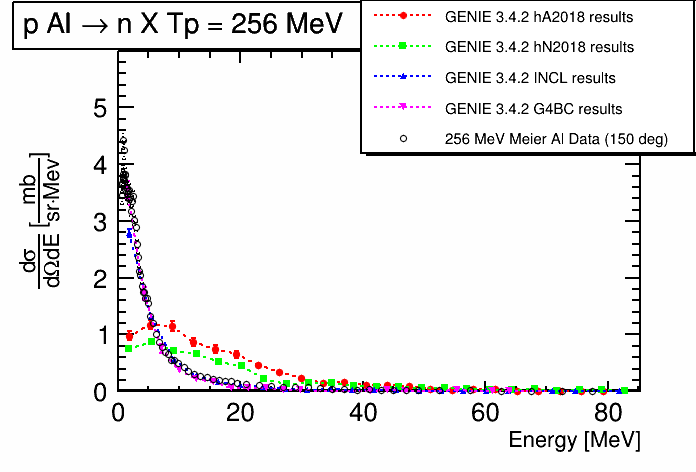}
    \caption{Inclusive cross section for $\text{Al}(p,n)$ interaction~\cite{Meier:1992anx} for a 0.256 GeV incident proton compared with all four models in this study. The left plot is for 60$^\circ$ scattering angle, and the right plot is for 150$^\circ$ scattering angle. The horizontal axis is the neutron kinetic energy. }
    \label{fig:pn-al-256}
\end{figure*}

To demonstrate differences among models and against data, we show inclusive scattering predictions across various targets and energies with either a proton or a neutron in the final state. Figure~\ref{fig:pn-al-256} shows neutron kinetic energy spectra at two detection angles for the inclusive $\text{Al}(p,n)X$ interaction at 256~MeV~\cite{Meier:1992anx}, where $p$ denotes the incident proton and $n$ denotes the detected neutron while all other final state particles are integrated over. 

In Fig.~\ref{fig:pn-al-256}, there is a large peak below about 20 MeV with a tail to higher energies. The large peak is due to compound nucleus interactions, where a group of low-energy nucleons is emitted. The tail is due to the emission of a small number of nucleons, still more than two. There is almost no cross section for neutrons above half of the beam energy. Because most of the leading neutron energy is significantly below the incident nucleon energy, these studies show that all of the data in these figures comes from knockout interactions. 

\textsc{INCL++} and \textsc{Geant4} BC have compound nucleus mechanisms, while hA2018 and hN2018 do not have them. As a result, \textsc{INCL++} and \textsc{Geant4} BC have a good match to the data for all neutron energies. hA2018 and hN2018 predict the normalization of the cross section, but the shape of the distribution is wrong. The average neutron energy for hA2018 and hN2018 is larger than for the more sophisticated models. This discrepancy is explored further in the following figures. 

\begin{figure*}
    \centering
    \includegraphics[width=0.49\linewidth]{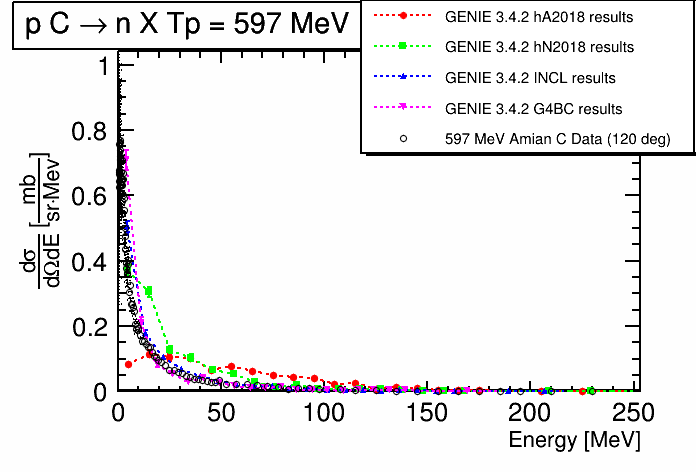}
    \includegraphics[width=0.49\linewidth]{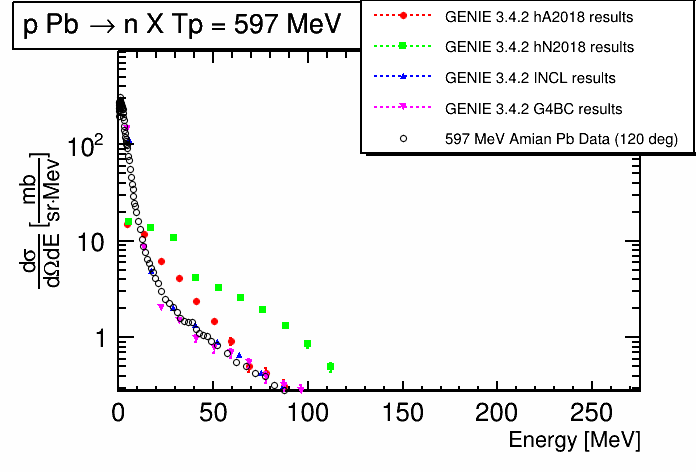}
    \caption{Inclusive cross section for $\text{C}(p,n)$ and $\text{Pb}(p,n)$ interaction~\cite{Amian:1993dhv} for a 0.597 GeV incident proton compared with all four models in this study. The left plot is for carbon, and the right plot is for lead (in log-scale). The horizontal axis is neutron kinetic energy. }
    \label{fig:pn-cpb-597}
\end{figure*}

\noindent Figure~\ref{fig:pn-cpb-597} shows inclusive scattering for heavier targets at different energies. The same effect seen for the 0.256 GeV $(p,n)$ interactions is also observed for the $(p,n)$ interactions with a 0.597 GeV incident proton on carbon and lead targets. 

\begin{figure*}
    \centering
    \includegraphics[width=0.49\linewidth]{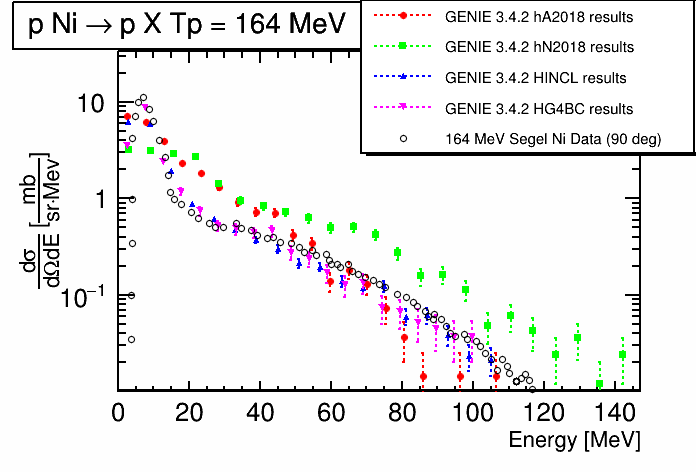}
    \includegraphics[width=0.49\linewidth]{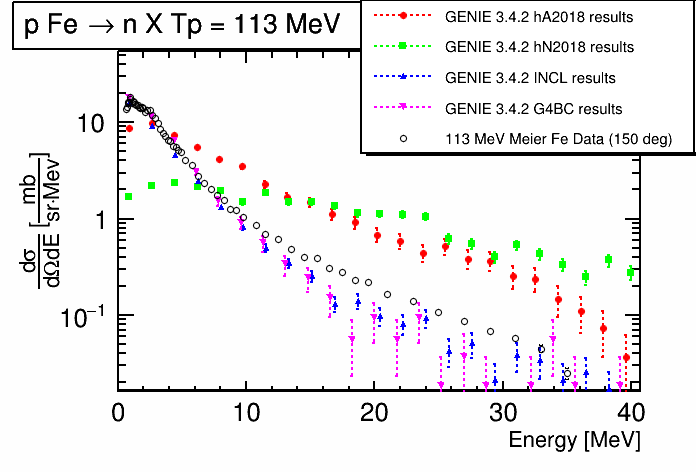}
    \caption{Inclusive cross section for $\text{Ni}(p,p')$~\cite{Segel:1982uf} and $\text{Fe}(p,n)$ interaction~\cite{Meier:1989tjq} for 164 MeV and 113 MeV proton beams, respectively, compared with all four models in this study. The left plot is for nickel, and the right plot is for iron (both in log-scale). The horizontal axis is the proton (left) and neutron (right) kinetic energy. }
    \label{fig:pp-pn-cfe-150}
\end{figure*}

At lower nucleon beam energies, the quasielastic cross section is larger. However, it is still very forward-peaked, and so the knockout cross section can be studied at angles of 60$^\circ$ or larger. Figure~\ref{fig:pp-pn-cfe-150} shows both $(p,p')$ and $(p,n)$ interactions for 164 MeV protons on nickel and 113 MeV protons on iron. Again, a dominant peak is seen for scattered nucleons with energies less than about 20 MeV. The tail to higher nucleon energy is much larger than was seen for higher energy proton beams. \textsc{INCL++} and \textsc{Geant4} BC models describe both features of the data better, suggesting that model alterations to the number of nucleons in knockout and energy distribution could address the model spread observed in this phase space.

\begin{figure*}
    \centering
    \includegraphics[width=0.49\linewidth]{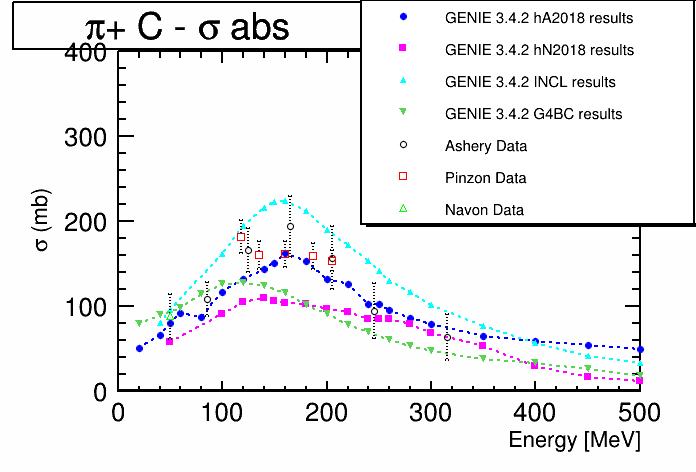}
    \includegraphics[width=0.49\linewidth]{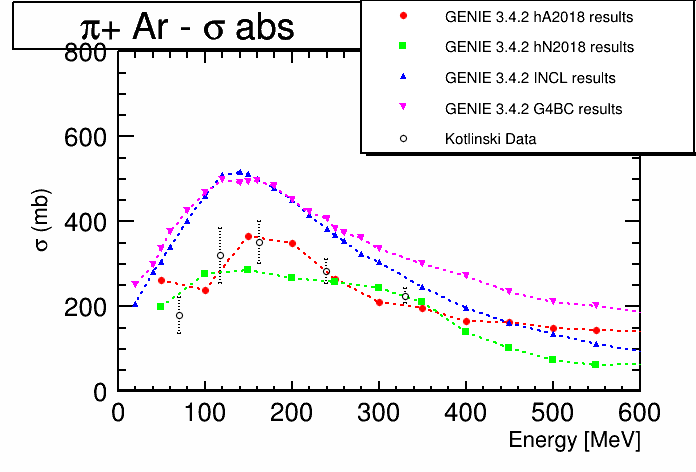}
    \caption{Total absorption cross section for $\pi^+$ using carbon (left) and argon (right) targets compared to the four models used in this work. The horizontal axis is pion kinetic energy.}
    \label{fig:pip-abs-c-ar}
\end{figure*}

A broad view of pion absorption comes from the total exclusive cross section model for that process. The total absorption cross section using carbon and argon targets is shown in Fig.~\ref{fig:pip-abs-c-ar} with comparisons to data on carbon~\cite{Ashery:1981tq,Navon:1983xj} and argon~\cite{LADS:2000pzh}. A recent experiment~\cite{DUET} is seen to be in good agreement with older carbon data. hA2018 is in the best agreement with these datasets, in part because Ashery data~\cite{Ashery:1981tq} is used as an input dataset to the model. \textsc{INCL++} has moderate agreement and \textsc{Geant4} BC has the worst agreement.

A more focused examination can be made using the LADS pion absorption data~\cite{LADS:2000pzh}. They also published total cross sections for specific channels with thresholds of 30 MeV for protons, deuterons, and neutrons. Table~\ref{tab:lads_mult} shows the 2p0n, 3p0n, 4p0n, 5p0n and summed cross sections for 239 MeV absorption events $\pi{^+}$ $^{40}\text{Ar}$, including the detection threshold of 30 MeV for protons, neutrons, and deuterons. We note that  \textsc{Geant4} BC has a somewhat larger 2p0n (i.e. quasideuteron final state) cross sections than the data. Both \textsc{INCL++} and \textsc{Geant4} BC have the best description of the higher multiplicities. \textsc{GENIE} hA2018 describes the lower multiplicities well but is significantly larger than data for the higher multiplicities. Table~\ref{tab:lads_mult2} explores the multiplicity of neutrons for the case of 2p in the final state. LADS data for 2p1n is almost as large as data for 2p0n. Only the\textsc{Geant4} BC model has this feature. Again, hA2018 is much larger than data for the higher multiplicity (2p2n). \textsc{INCL++} has the best description of these detailed data.

\begin{table}[tp]
\centering
\caption{Total cross sections as a function of proton multiplicity for 239 MeV $\pi{^+}$ absorption interactions on $^{40}\text{Ar}$. Results from the four models in this study are compared with LADS data~\cite{LADS:2000pzh}. All cross section units are millibarns (mb). This table includes a 30 MeV threshold for data and models for protons, deuterons, and neutrons.}
\label{tab:lads_mult}
\begin{tabular}{ |c|c|c|c|c|c| }
 \hline
 Model & 2p0n & 3p0n & 4p0n & 5p0n & Sum \\
 \hline
 hA2018 & 72.1 & 25.0 & 16.5 & 4.4 & 118.0 \\
 \hline
 hN2018 & 38.3 & 21.7 & 7.4 & 1.8 & 69.2 \\
 \hline
 \textsc{INCL++} &  79.1 & 16.8  &  0.99 &   0.03 & 96.9 \\ 
 \hline
 \textsc{G4} BC & 82.8 & 34.3 & 3.7 & 0.06 & 120.8 \\
 \hline
 Data & 72.9$\pm$5.8 & 26.8$\pm$2.5 & 2.0$\pm$0.2 & 0.04$\pm$0.01 & 101.7$\pm$6.3 \\
 \hline
\end{tabular}
\end{table}

\begin{figure*}
    \centering
    \includegraphics[width=0.49\linewidth]{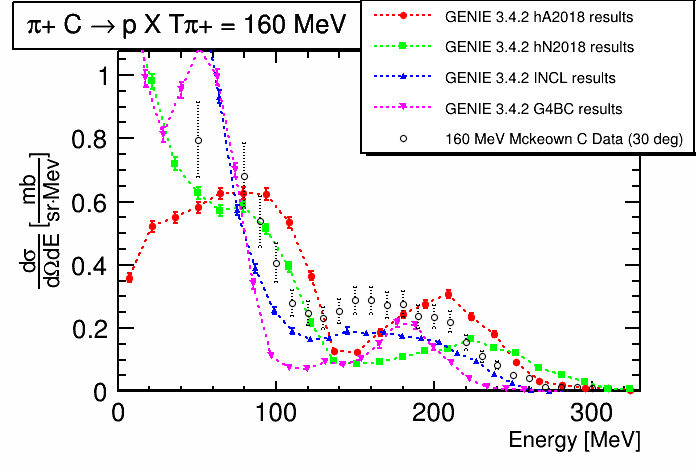}
    \includegraphics[width=0.49\linewidth]{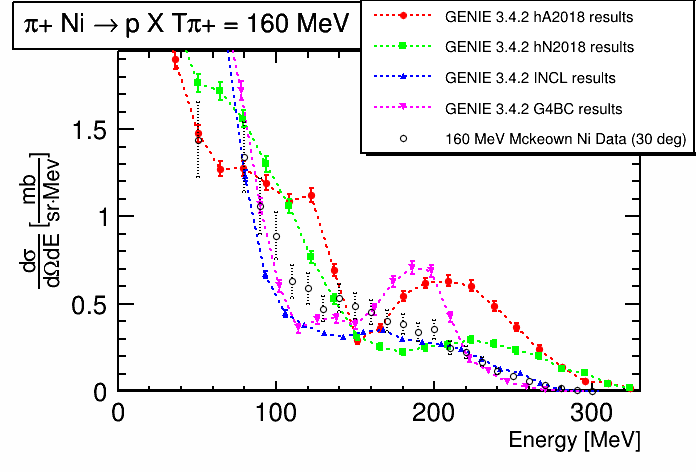}
    \caption{Inclusive cross section for $\text{C}(\pi^+,p)$ and $\text{Ni}(\pi^+,p)$~\cite{Mckeown:1981pw} for 160 MeV pion beams compared with all four models in this study. The left plot is for carbon, and the right plot is for nickel.}
    \label{fig:pip-p-cni-30}
\end{figure*}

\begin{table}[tp]
\centering
\caption{Total cross sections for channels involving neutrons for 239 MeV $\pi{^+}$ absorption on $^{40}\text{Ar}$. Results from the four models in this study are compared with LADS data~\cite{LADS:2000pzh}. All cross section units are millibarns (mb). This table includes a 30 MeV threshold for data and models for protons, deuterons, and neutrons.}
\label{tab:lads_mult2}
\begin{tabular}{ |c|c|c|c|c| }
 \hline
 Model & 2p0n & 2p1n & 2p2n & Sum\\
 \hline
 hA2018 & 72.1 & 23.8 & 17.8 & 113.7\\
 \hline
 hN2018 & 38.3 & 36.7 & 23.0 & 98.0\\
 \hline
 \textsc{INCL++} & 79.1 & 35.0 & 4.0 & 118.1 \\
 \hline
\textsc{G4} BC & 82.8 & 58.2 & 6.2 & 147.2 \\
 \hline
 Data & 72.9$\pm$5.8 & 62.9$\pm$6.6 & 5.6$\pm$1.0 & 141.4$\pm$8.8 \\
 \hline
\end{tabular}
\end{table}
The signature of pion absorption in pion interactions for inclusive data is more complicated~\cite{McKeown}. The $(\pi^+,p)$ interactions in Ref.~\cite{Mckeown:1981pw} show contributions from 2-body in-medium quasielastic pion scattering and absorption ($\pi^+d\rightarrow pp$) interactions. At forward angles, these basic interactions are of comparable size~\cite{said}. The inclusive data for proton emission from $\pi^+$ beams show separate peaks for protons emitted from scattering and absorption reactions~\cite{Mckeown:1981pw}. This is shown in Fig.~\ref{fig:pip-p-cni-30} (left), which shows the data for a 160 MeV $\pi^+$ beam impinging on a carbon target with protons detected at $30^\circ$. The peak at $\sim$70 MeV comes from $\pi^+ p$ scattering and the peak at $\sim$160 MeV comes from quasielastic absorption in the nuclear medium. For comparison, the right plot shows the same data for a Ni target. Here, the peaks are more smeared out by Fermi motion and FSI. None of the models provide a good match to the higher energy peak for both targets. For carbon (left), hA2018 has the right magnitude, but the other models all undershoot the data. For nickel (right), \textsc{INCL++} and hN2018 have moderate agreement with the data, while hA2018 and \textsc{Geant4} BC are about a factor of two above the data.

These comparisons demonstrate that while empirical models like hA2018 accurately constrain total reaction normalizations and broad features seen in the data, full cascade models such as \textsc{INCL++} and \textsc{Geant4} BC provide a more physically complete description of spectral shapes and nuclear de-excitation, making them compelling benchmarks for our reweighting framework.

\section{\textsc{GENIE} hA2018 Particle Transport Simulation Details}
\label{sec:GENIEhA2018}

\textsc{GENIE}, a frequent generator of choice for liquid argon neutrino experiments, defines the following mesonless final state topologies as pion absorption and nucleon knockout:

\begin{itemize}
    \item \textbf{\textsc{GENIE} pion absorption}: A pion-nucleus interaction with no pions and at least two nucleons in the final state.
    \item \textbf{\textsc{GENIE} nucleon knockout}: A nucleon-nucleus interaction with at least three nucleons in the final state. This topology differentiates it from inelastic interactions.
\end{itemize}

 A key feature of both pion absorption and nucleon knockout interactions in the GENIE hA2018 model is the definition of an intermediate step of nucleon multiplicity distributions. This is followed by the ``decay'' of the intermediate ensemble of nucleons by phase space distributions. The multiplicity distributions were defined in terms of the sum and difference of proton and neutron values. 
  
\begin{itemize}
\item Total number of protons and neutrons ($N_p+N_n$).
\item Difference between the number of protons and neutrons ($N_p-N_n$).
\end{itemize}

 This allowed simple extension of $\pi^+$ to $\pi^0$ and $\pi^-$ distributions and to neutron-rich nuclei. Fits to inclusive data and the \textsc{GENIE} hN2018 model determined the nuclear $A$ and probe energy dependence of these templates. \textsc{GENIE}'s hN2018, \textsc{Geant4} BC, and \textsc{INCL++} employ an intranuclear cascade that simulates intermediate particles with the ability to re-interact within the remnant nucleus. These functions are a useful way to compare hA2018 with the INC models and an important part of this analysis.


\textsc{GENIE}'s hA2018 particle transport simulation is empirical and data-driven, using a wide variety of hadron scattering data supported with theoretical input~\cite{Dytman,Andreopoulos:2015wxa}. The original model was based on hadron-nucleus data, as shown in Sec.~\ref{sec:hadron-nucleus}, and modeling available at the time. Because of the empirical nature of the simulation, \textsc{GENIE}'s hA2018 model is able to be altered using simple distributions, unlike simulations with a full intranuclear cascade.


\subsection{Pion Absorption}

As an empirical simulation without a full intranuclear cascade, \textsc{GENIE} hA2018 probabilistically designates an interaction fate for a pion impinging upon a nucleus based on the branching fraction of interaction types: elastic scattering, inelastic scattering, single charge exchange, double charge exchange, pion production, and absorption. For pion absorption interactions, there is a subsequent probabilistic split into two fates based on data from Ref.~\cite{McKeown}. These fates represent a 2-nucleon quasideuteron (QD) mechanism and a multinucleon (MN) mechanism. The QD mechanism covers low-energy pions, and the MN mechanism covers higher-energy pions. The fraction of interactions divided between these two mechanisms is determined by the pion kinetic energy $T_\pi$ and target nucleus nucleon multiplicity $A$ via the following formula hardcoded into \textsc{GENIE} hA2018~\cite{Andreopoulos:2015wxa}:
\begin{equation}
    P_{\text{QD}} = 1.14 \cdot (0.903 - 0.001989 \cdot A) \cdot (1.35 - 4.767 \cdot T_\pi), \label{eqn:qd}
\end{equation}
where $T_\pi$ is measured in GeV. Any probabilities below 0 are fixed to 0, and those above 1 are fixed to 1. For $A=40$ (argon), QD interactions represent 100\% of events for $T_{\pi}\lesssim 0.062\text{ GeV}$. MN events are sampled 100\% of the time for $T_{\pi}\gtrsim 0.289\text{ GeV}$, with the probability linearly interpolated between these values.
\par
Modeling of the MN mechanism is the primary topic of this paper, since its outputs are comparable to the INC models. For completeness, we discuss the simulation details of the QD mechanism here. In determining the proportion of QD absorption interactions that occur on a given nucleon pair, GENIE hA2018 utilizes pion absorption cross section ratios calculated from a BUU simulation for nucleon-nucleon pion production from Ref.~\cite{PionNucleus_1994}. The GENIE simulation extrapolates to different nuclei by multiplying the cross section ratio for that incident target pair by the fraction $\frac{Z}{A}$ or $\frac{N}{A}$ for each proton or neutron, respectively, in the incident target pair~\cite{Andreopoulos:2015wxa}. For $^{40}$Ar, the probabilities, given the charge of the probe, are provided in Table~\ref{tab:qd_probs}:

\begin{table}[tp]
\centering
\caption{Quasideuteron nucleon pair absorption probabilities in \textsc{GENIE} hA2018 for different types of pions on $^{40}\text{Ar}$~\cite{Andreopoulos:2015wxa}.}
\label{tab:qd_probs}
\begin{tabular}{ |c|c|c|c| }
 \hline
 Probe / Nucleons Impinged & $pp$ & $pn$ & $nn$ \\
 \hline
 $\pi^+$ & 0 & 95.2\% & 4.8\% \\
 \hline
 $\pi^-$ & 3.3\% & 96.7\% & 0 \\
 \hline
 $\pi^0$ & 9.8\% & 75.5\% & 14.7\% \\
 \hline
\end{tabular}
\end{table}

The preference for absorption on $pn$ pairs is important for interpretation of data for pion absorption in nuclei. The charges of the nucleons are appropriately converted given the charge of the pion when emitted into the final state. For example, if a $\pi^+$ interacts with a $pn$ pair, then two protons will be emitted.
\par Once the emitted nucleon species are determined, the simulation assigns final state kinematics. The Fermi momentum of the initial state nucleons is determined via a Local Fermi Gas nuclear model. As is universal in GENIE, a Woods-Saxon spatial density is used for nuclei heavier than oxygen, and a modified Gaussian spatial nuclear density is used for lighter nuclei. The probe energy is given to the kinetic energies of the struck nucleon(s) to provide the total outgoing nucleon energy. This includes the mass of the pion for absorption, and the outgoing nucleons are given a random energy and angle according to phase space.

The MN simulation, relevant for all pions in argon above $T_{\pi}\simeq 0.062$ GeV and guaranteed above $T_{\pi}\simeq 0.289$ GeV, models the sum and difference of proton and neutron multiplicities in the final state using Gaussian distributions, specified by the mean and standard deviation based on INC calculations. Dividing the response into sum and difference allows easy extension from $p$ and $\pi^+$ to the other charge states via isospin symmetry.
\par
\textsc{GENIE} hA2018 describes these Gaussian parameters for MN pion absorption final state multiplicities as dependent on the $A$-value of the target nucleus and the energy of the incident hadron based on INC simulations available at that time. The relationships between the target nucleus, the kinetic energy of the hadron, and the nucleon distributions are shown in Table~\ref{tab:pion}. This works well for up to 18 emitted nucleons. For larger groups, the nucleons must be divided into groups that can be effectively handled, and the maximum is set to 85. Studies presented here show that neither of these criteria of large numbers of emitted nucleons is relevant for existing neutrino experiments.
\par
Final state kinematics in the MN pion absorption are determined via phase space decay. The phase space decay module generates weights for the final state particles in the center-of-mass frame, determining the momentum fraction a given particle will carry. It simulates angles uniformly at random in $\cos\theta$ for each particle in the center-of-mass frame while demanding that four-momentum be conserved. This is an important and unjustified assumption. The simulation then boosts the particles according to the probe incident momentum to determine the final state momenta. 
\par
Detailed model comparisons of kinematic distributions of the final state for pion absorption interactions between \textsc{GENIE} hA2018, \textsc{GENIE} hN2018, \textsc{Geant4} BC and \textsc{INCL++} are included in Appendix \ref{app:pionabs}.

\begin{table}[tp]
\caption{Multinucleon (MN) pion absorption parameters for nucleon multiplicity in \textsc{GENIE} hA2018. $T$ is the pion kinetic energy (GeV), and $A$ is the target nucleon number~\cite{Andreopoulos:2015wxa}.}
\centering
\begin{tabular}{|c|c|c|}
\hline
 & $\mu$ & $\sigma$ \\
\hline
Sum & $10^{-4}(1+4\,T)(A-10)^2 + 3.5$ & $(10+16\,T)(A/250)^{0.9}$ \\
\hline
Diff. & $(1+4\,T) - (A/200)(1+8\,T)$ & $4\left(1 - e^{-0.03\,T}\right)$ \\
\hline
\end{tabular}
\label{tab:pion}
\end{table}

\subsection{Nucleon Knockout}

Description of this channel has the disadvantage of a lack of channel-specific cross section data. Therefore, INC simulations must be used and the main validation is with inclusive data such as was shown in Sect.~\ref{sec:hadron-nucleus}.
INC simulations show that a simple Gaussian function is not sufficient to describe all nucleon knockout (KO) data. Nucleon interactions that produce more than two nucleons in the final state are described by an exponential decay function for the total nucleon multiplicity and a Gaussian distribution for the difference between protons and neutrons, the same underlying distribution as MN pion absorption interactions. The decay constant ($\gamma_{\rm ns}$) is the relevant parameter for the set of exponential decay fits. Table~\ref{tab:knockout} and Table~\ref{tab:gamma} outline the parameters used by \textsc{GENIE} hA2018.
\par
The kinematics of final state nucleons in a nucleon knockout interaction follow the same phase space decay procedure as in MN pion absorption, except with a nucleon probe instead of a pion probe. The other key difference is that the final state proton and neutron multiplicities include the probe (whereas an absorbed pion does not appear in the final state).

\begin{table}[tp]
\centering
\caption{Nucleon knockout modeling parameters for \textsc{GENIE} hA2018 given the kinetic energy ($T$) of the incident nucleon and the target nucleus number of nucleons ($A$) and number of protons ($Z$). The coefficients for the sum of nucleons are shown in Table~\ref{tab:gamma}. \textsc{GENIE} mandates that $\gamma_{\rm ns} \ge 0.002$~\cite{Andreopoulos:2015wxa}.}
\begin{tabular}{|c|c|c|}
\hline
 & $\mu$ & $\sigma$ \\
\hline
Sum & \multicolumn{2}{c|}{$\gamma_{\rm ns} = c_1 e^{c_2 A} + c_3$} \\
\hline
Diff. & 
$\begin{cases}
135.2\, e^{-7.1 (A-Z)/A} - 2.8 & A\ge2Z \\[2pt]
-135.2\, e^{-7.1 Z/A} + 4.9 & A \leq 2Z
\end{cases}$ 
& $2.0 + 0.0078\, A$ \\
\hline
\end{tabular}

\label{tab:knockout}
\end{table}

\begin{table}[tp]
\centering
\caption{Coefficients for $\gamma_{\rm ns}$ as a function of kinetic energy ($T$) of the incident particle in GeV~\cite{Andreopoulos:2015wxa}.}
\begin{tabular}{|c|c|}
\hline
Coefficient & Expression \\
\hline
$c_1$ & $0.041 + 0.1525\,T$ \\
$c_2$ & $-0.003444 - 0.02324\,T$ \\
$c_3$ & $0.064 - 0.015\,T$ \\
\hline
\end{tabular}

\label{tab:gamma}
\end{table}

\section{Nucleon Multiplicities from Pion-Nucleus and Proton-Nucleus Scattering}
\label{sec:results}

A library of hadron scattering events was produced using \textsc{GENIE}'s hadron scattering software across all four base models of hA2018, hN2018, \textsc{Geant4} BC v4.10.6.p01g, and \textsc{INCL++} v5.2.9.5g implemented in \textsc{GENIE} v3.4.2. Pions and nucleons with kinetic energies ranging from 0.05 GeV to 2.00 GeV were simulated. In total, approximately 4 million interactions were simulated per particle per target. The particles included in the sample are positively-charged pions, negatively-charged pions, neutral pions, protons, and neutrons. The targets chosen are carbon, oxygen, argon, and iron. These targets reflect those of a wide range of popular detector media for neutrino experiments, such as water (Super-Kamiokande~\cite{T2KOsc}, Hyper-Kamiokande~\cite{Hyper-Kamiokande:2025fci}, KM3NeT~\cite{KM3NeT:2024ecf}, IceCube~\cite{IceCubeCollaborationP:2025rpl}), scintillator (NOvA~\cite{NOvA:2025tmb}, T2K's ND280~\cite{T2K:2020sbd}), and argon (DUNE~\cite{DUNE:2020jqi}, SBN detectors of ICARUS~\cite{ICARUS:2023gpo}, SBND~\cite{SBND:2025pjc}, and MicroBooNE~\cite{MicroBooNE:2025nll}). 

Fortunately, the basic parameterizations chosen for hA2018 are valid for the general features of all GENIE FSI models. However, \textsc{INCL++} and \textsc{Geant4} BC models contain additional physics in two ways. The emission of nuclear fragments from nuclear evaporation, such as deuterium and helium is modeled. Here, each multinucleon state is counted according to the component nucleons. In addition, the evaporated nucleons coming from compound nucleus excitation distort the sum and difference spectra in important ways. 
A minimum kinetic energy threshold was applied to account for differences between models. 
The kinetic energy threshold for a particle was set to 5 MeV per nucleon, meaning 5 MeV for a proton or neutron, 10 MeV for a deuteron, etc. In the context of applications to experiments, this threshold is intended to match the kinetic energy threshold of DUNE's Phase II Near Detector~\cite{DUNE:2021tad,DUNE:2024wvj}, which has the lowest threshold of current and next-generation detectors. The analysis uses the following definitions for pion absorption and nucleon knockout:

\begin{itemize}
    \item \textbf{Pion absorption}: A pion-nucleus interaction with no pions and at least two nucleons with kinetic energies above 5 MeV in the final state.
\item \textbf{Nucleon knockout}: A nucleon-nucleus interaction with at least three nucleons with kinetic energies above 5 MeV in the final state. \textsc{GENIE} classifies events with two or fewer nucleons in the final state as inelastic interactions, which describe the quasielastic nucleon-nucleon process.
\end{itemize}

\begin{figure*}
    \centering
    \includegraphics[width=0.49\linewidth]{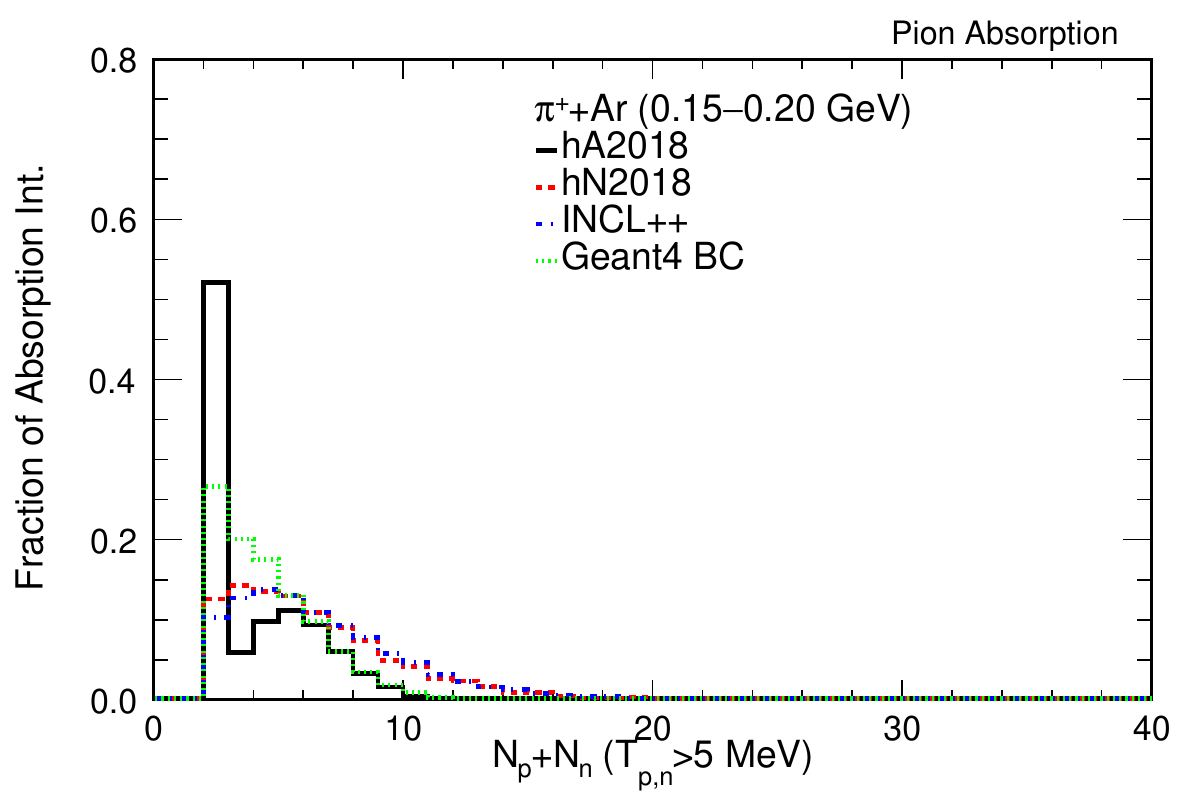}
    \includegraphics[width=0.49\linewidth]{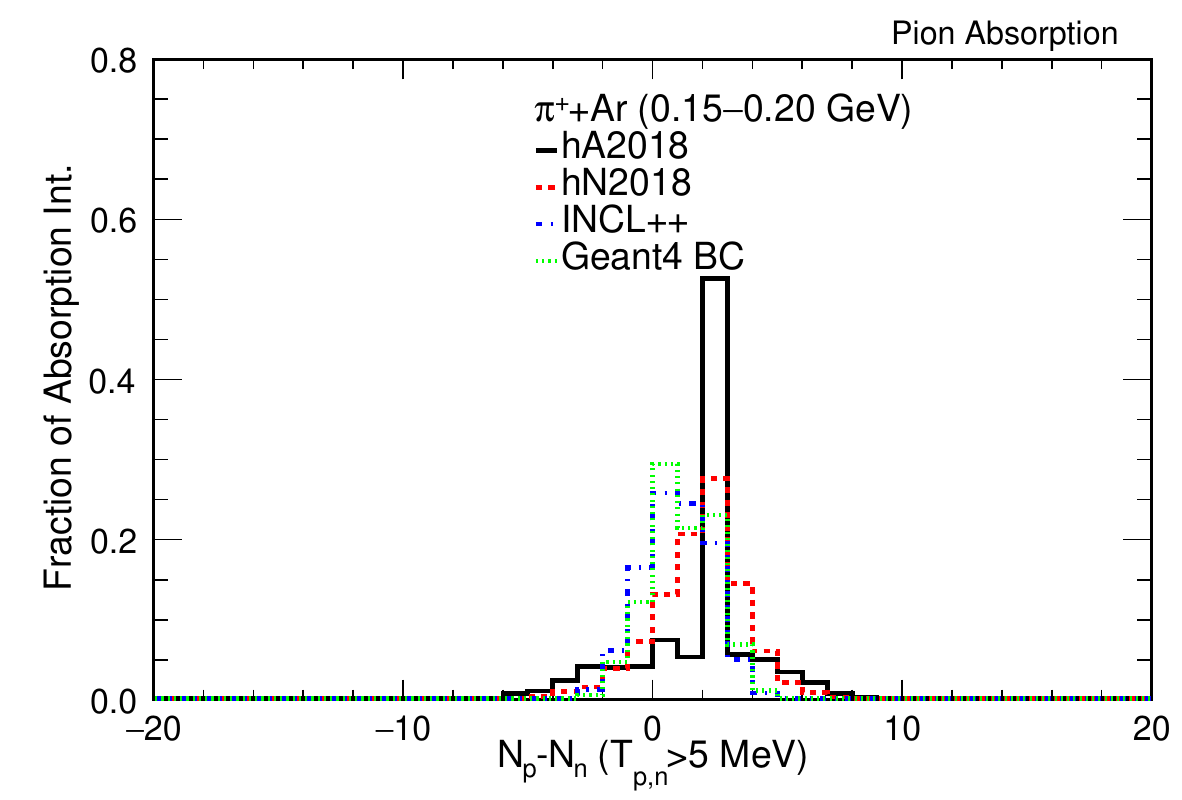}
    \includegraphics[width=0.49\linewidth]{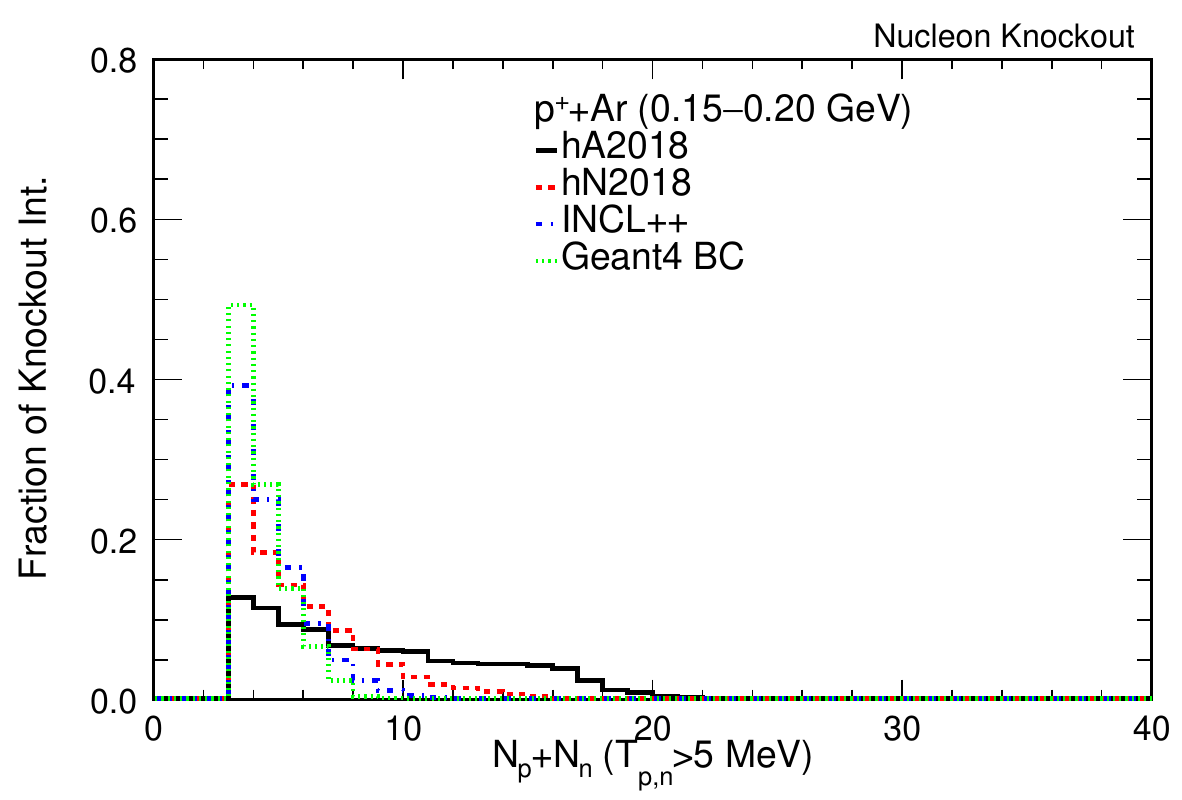}
    \includegraphics[width=0.49\linewidth]{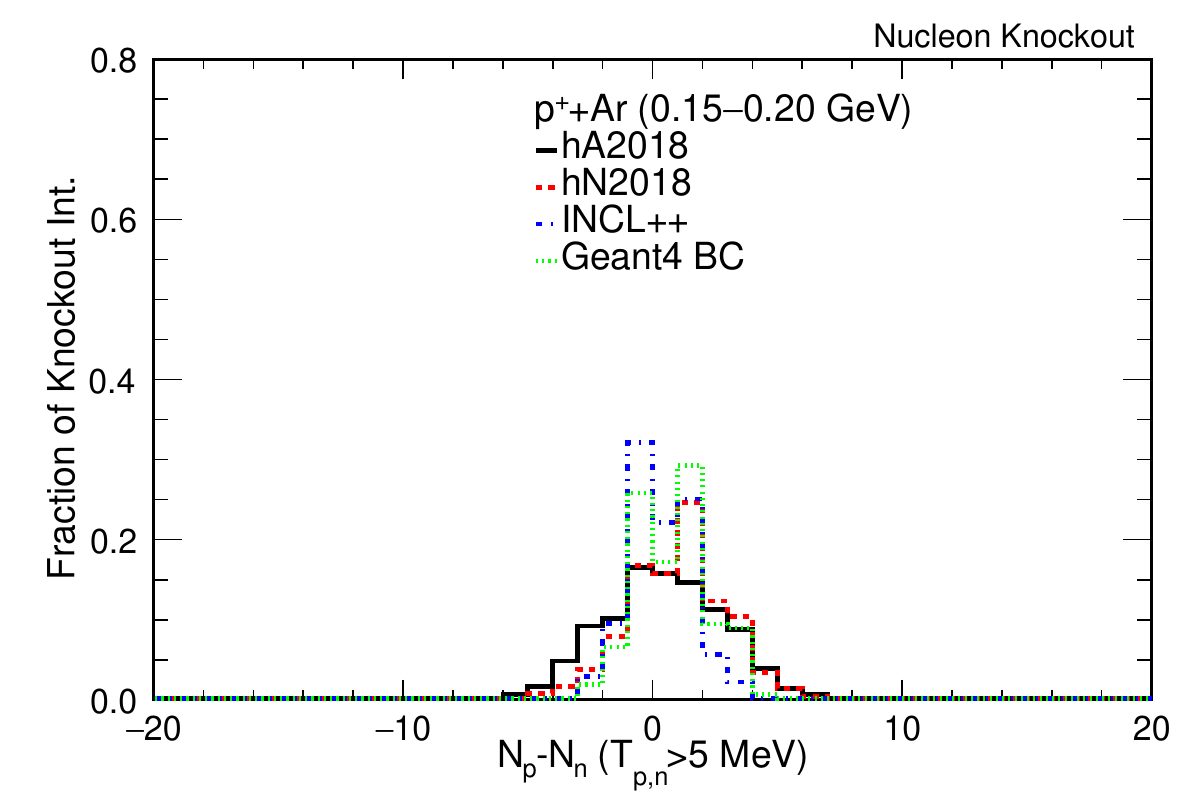}
    \caption{Total number of nucleons (left) and the difference in the number of protons and neutrons (right) at an average incident kinetic energy of 0.175 GeV from pion absorption (top) and nucleon knockout interactions on argon.}
    \label{fig:ke125}
\end{figure*}

\begin{figure*}
    \centering
    \includegraphics[width=0.49\linewidth]{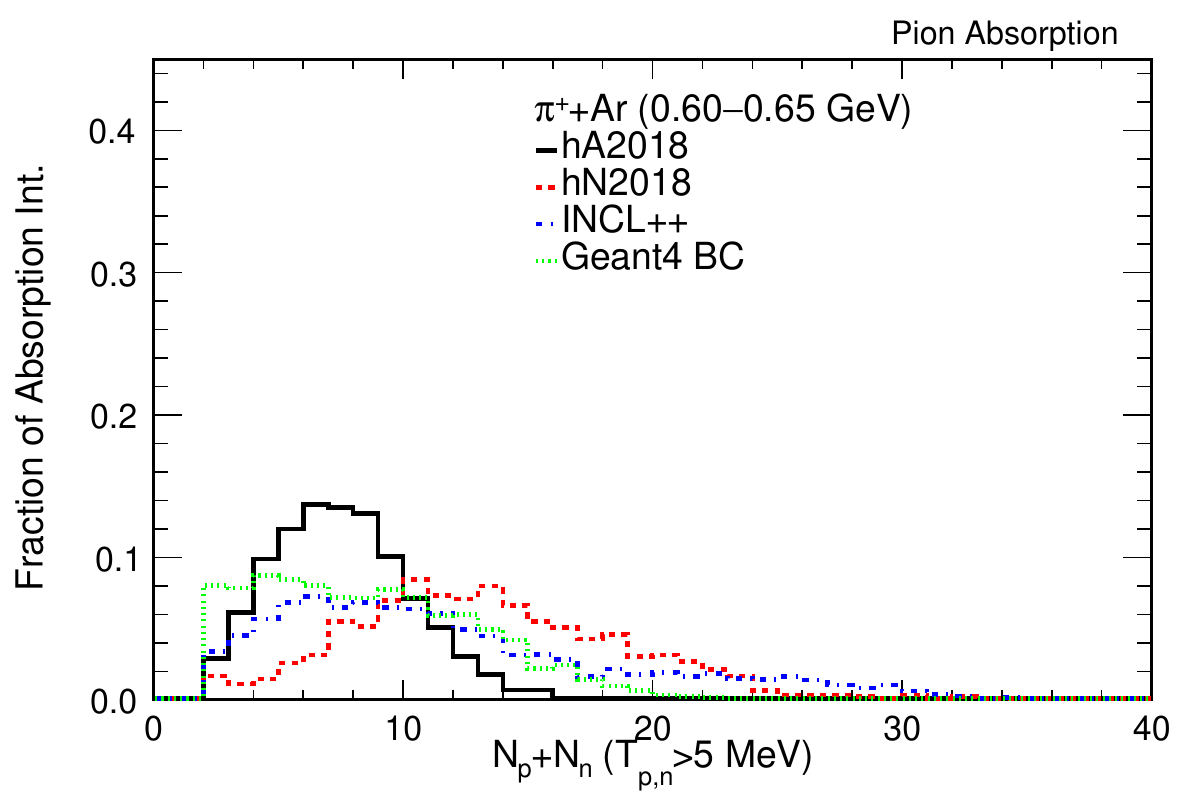}
    \includegraphics[width=0.49\linewidth]{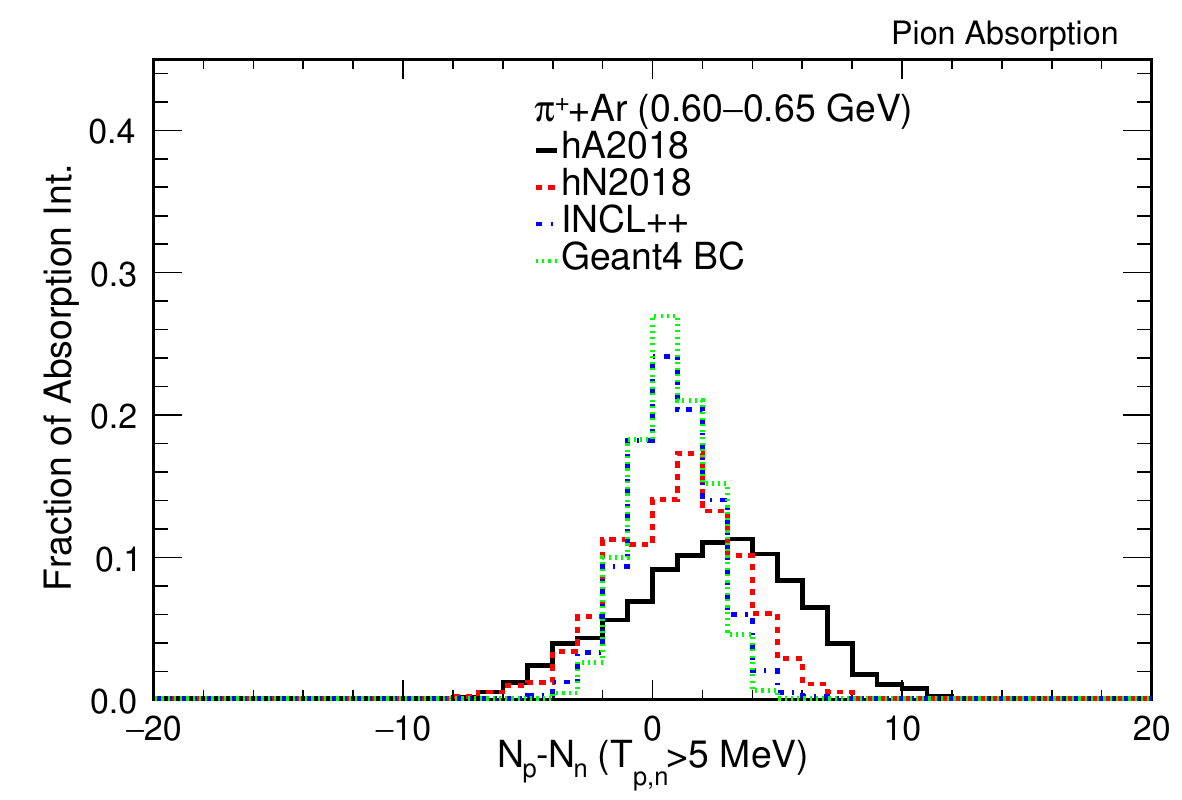}
    \includegraphics[width=0.49\linewidth]{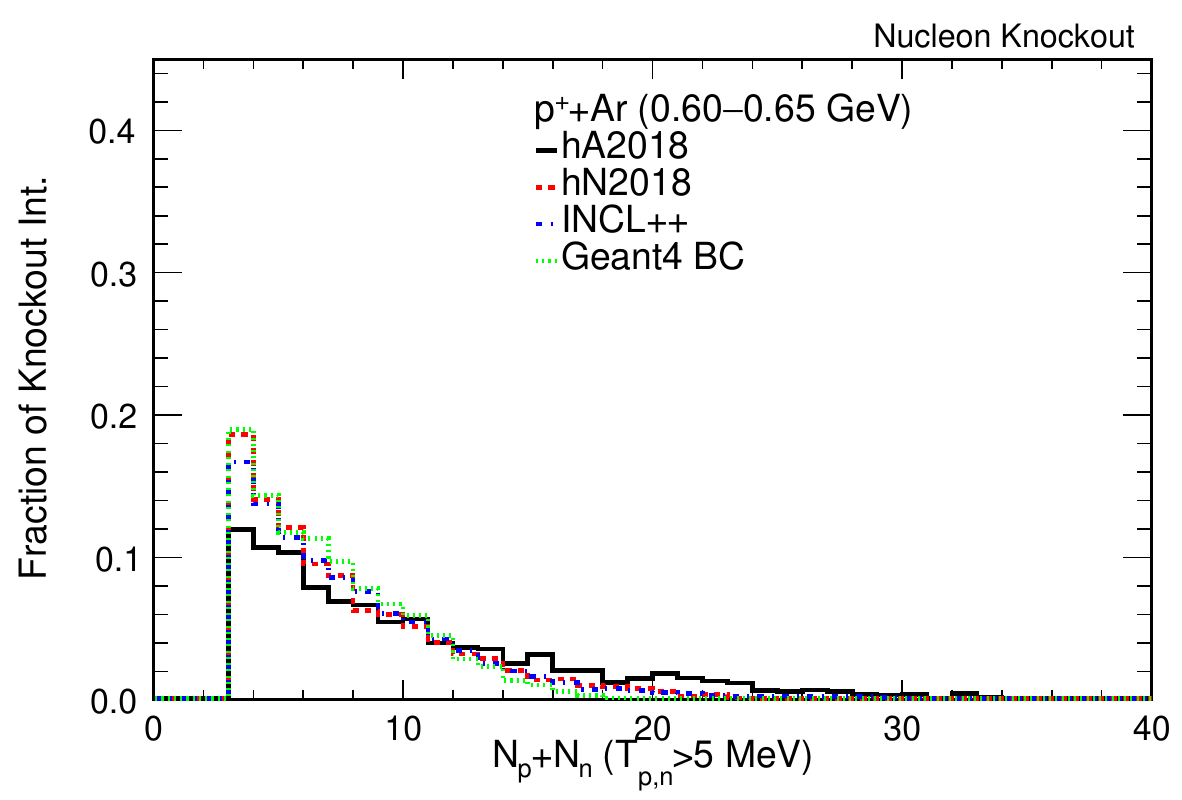}
    \includegraphics[width=0.49\linewidth]{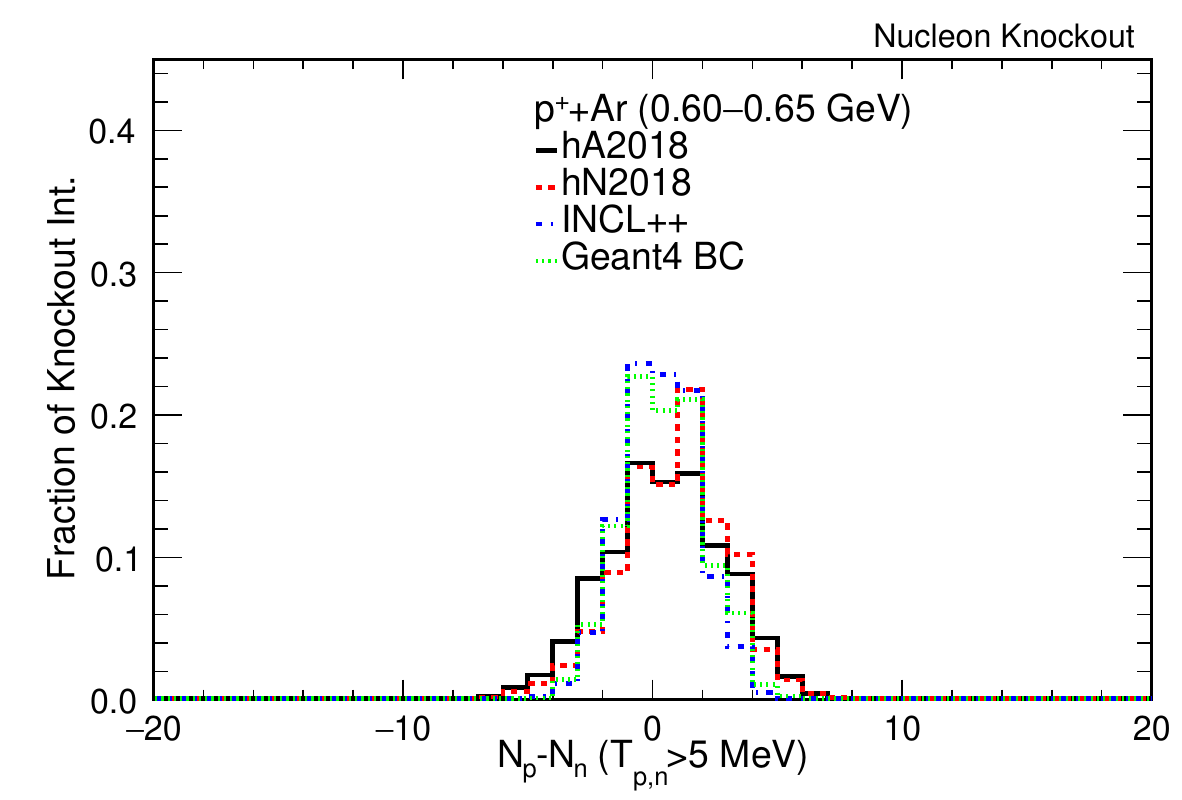}
    \caption{Total number of nucleons (left) and the difference in the number of protons and neutrons (right) at an average incident kinetic energy of 0.625 GeV from pion absorption (top) and nucleon knockout interactions on argon.}
    \label{fig:ke625}
\end{figure*}

The total number and the difference between the numbers of protons and neutrons in the final state were ascertained for pion absorption and nucleon knockout interactions. Example distributions for argon at the average kinetic energy for SBN are shown in Figure~\ref{fig:ke125} and at the average kinetic energy for DUNE are shown in Figure~\ref{fig:ke625}. In all cases, the \textsc{GENIE} hA2018 model differs significantly from models with a full intranuclear cascade. Thus, this is an interesting way to characterize these models.

\begin{figure*}
    \centering
    \includegraphics[width=0.45\linewidth]{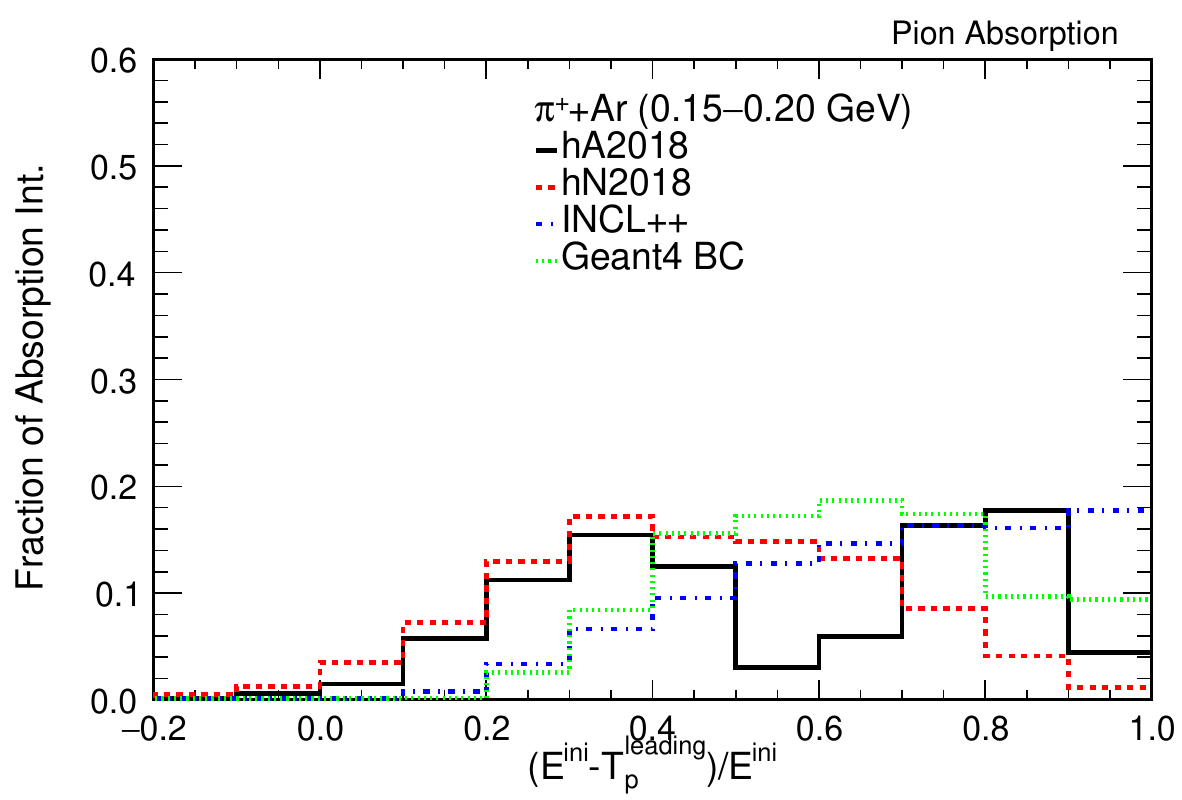}
    \includegraphics[width=0.45\linewidth]{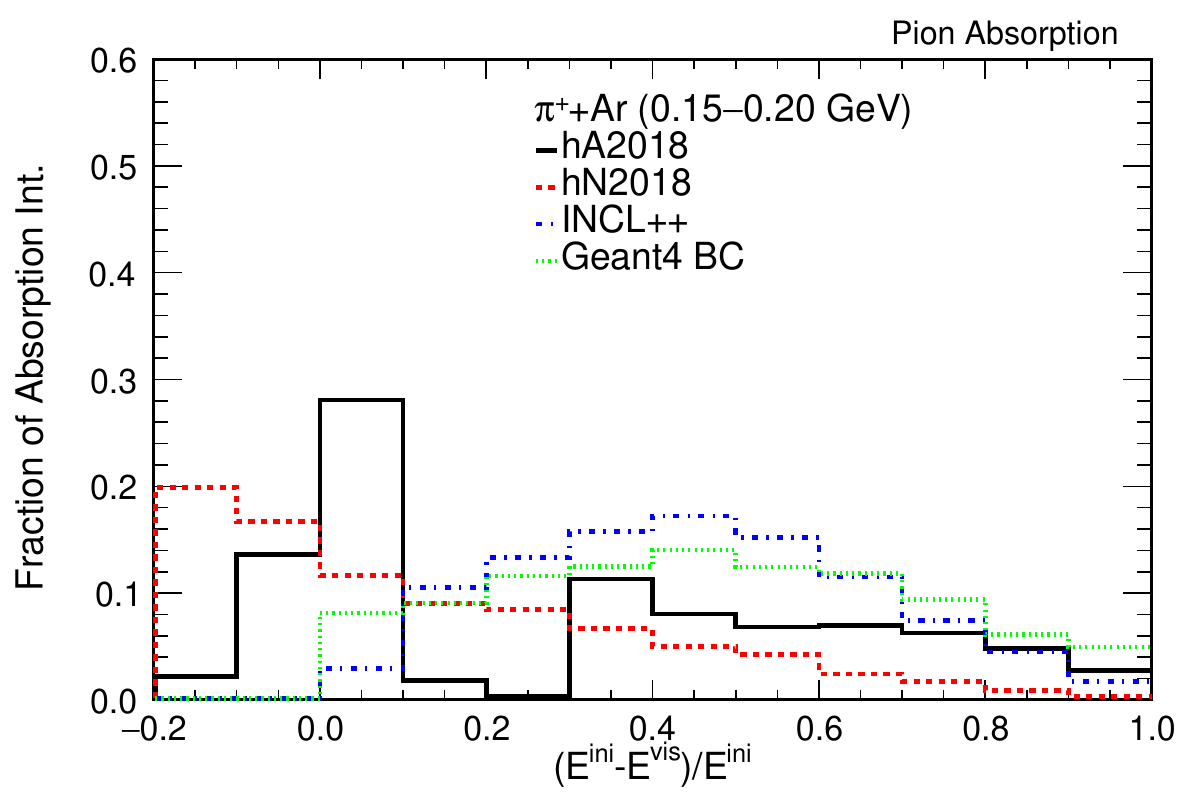}

    \includegraphics[width=0.45\linewidth]{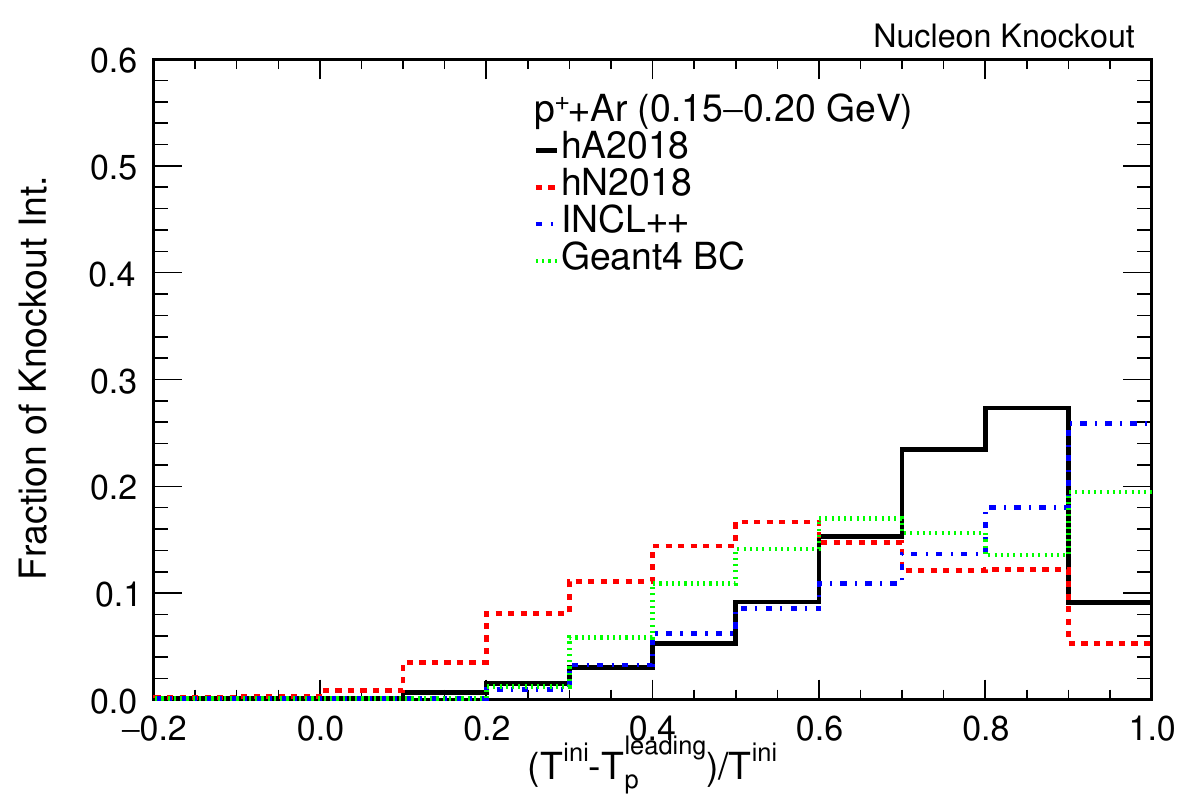}
    \includegraphics[width=0.45\linewidth]{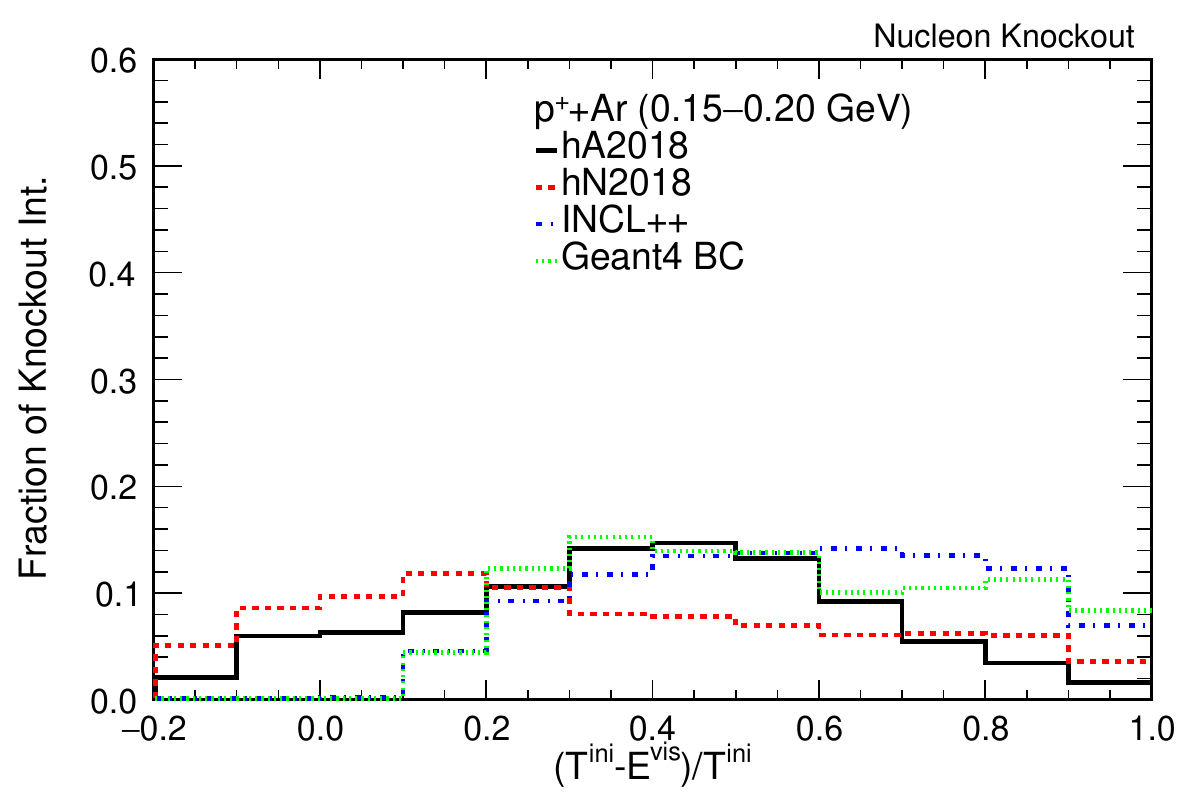}
    \caption{Ratios of the total leading proton kinetic energy (left) and the total visible energy (right) for pion absorption (top) and nucleon knockout (bottom) on argon. At low incident hadron energies, \textsc{GENIE} hA2018 and hN2018 may have interactions with an energy bias below -0.2.}
    \label{fig:Evis}
\end{figure*}

For pion absorption, at higher energies relevant for DUNE (Fig.~\ref{fig:ke625}) where the MN mechanism dominates, both $N_p+N_n$ and $N_p-N_n$ distributions can be reasonably approximated by Gaussian functions, whereas at lower energies (Fig.~\ref{fig:ke125}), the two-nucleon QD process introduces a prominent non-Gaussian peak at two final state nucleons. Conversely, for nucleon knockout, the total number of nucleons is well described by an exponential decay function, alongside a Gaussian function for the distribution of $N_p-N_n$. The energy distribution of nucleons is also relevant in understanding modeling differences. Figure~\ref{fig:Evis} shows the leading proton kinetic energy and visible energy for pion absorption and nucleon knockout interactions. \textsc{INCL++} and \textsc{Geant4} BC typically behave similarly, but \textsc{GENIE} hadron scattering models prefer more energy to go to the leading nucleon and more visible energy. Due to a lack of Pauli blocking in \textsc{GENIE} models, there is also an unphysical population of interactions with more visible energy than the energy of the incident hadron.

Because the QD subchannel of pion absorption in \textsc{GENIE} hA2018 produces non-Gaussian multiplicity distributions and is altogether separate from the MN model, the linear fits of how the Gaussian parameters evolve as a function of pion kinetic energy are limited to $T_{\pi}\geq 0.3\text{ GeV}$.

A series of these distributions is made across the energy range simulated for each target. The Gaussian and exponential decay parameters for each kinetic energy bin are then plotted and fitted using \textsc{MINUIT}~\cite{James:1975dr} to either a linear or an exponential decay function. As examples, the mean of the total number of nucleons for pion absorption and the exponential decay constant are shown below in Figure~\ref{fig:par} for argon, with comparisons directly made between \textsc{GENIE} hA2018, \textsc{Geant4} BC, and \textsc{INCL++}. Because of the definition of pion absorption, the mean number of nucleons in the distribution must be at least two nucleons.

\begin{figure*}[h]

    \centering
\includegraphics[width=0.445\textwidth]{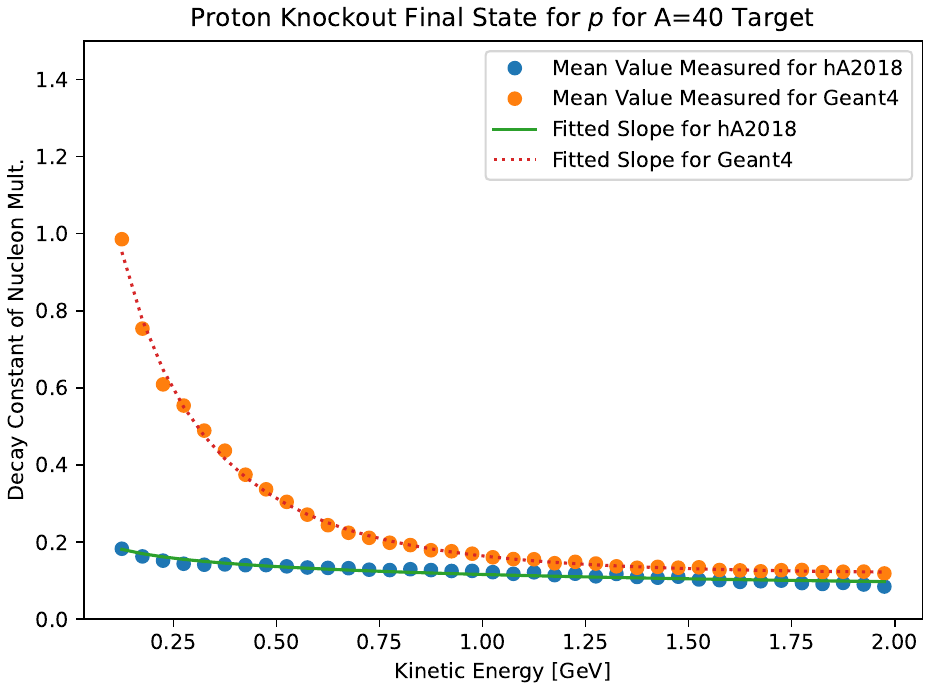}
\includegraphics[width=0.45\textwidth]{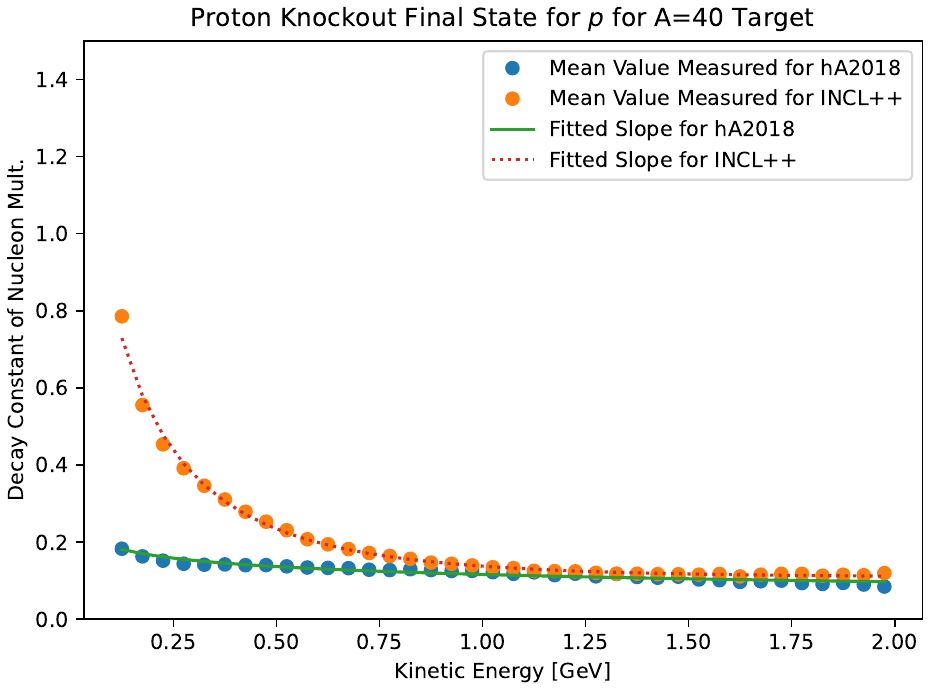}

\includegraphics[width=0.45\textwidth]{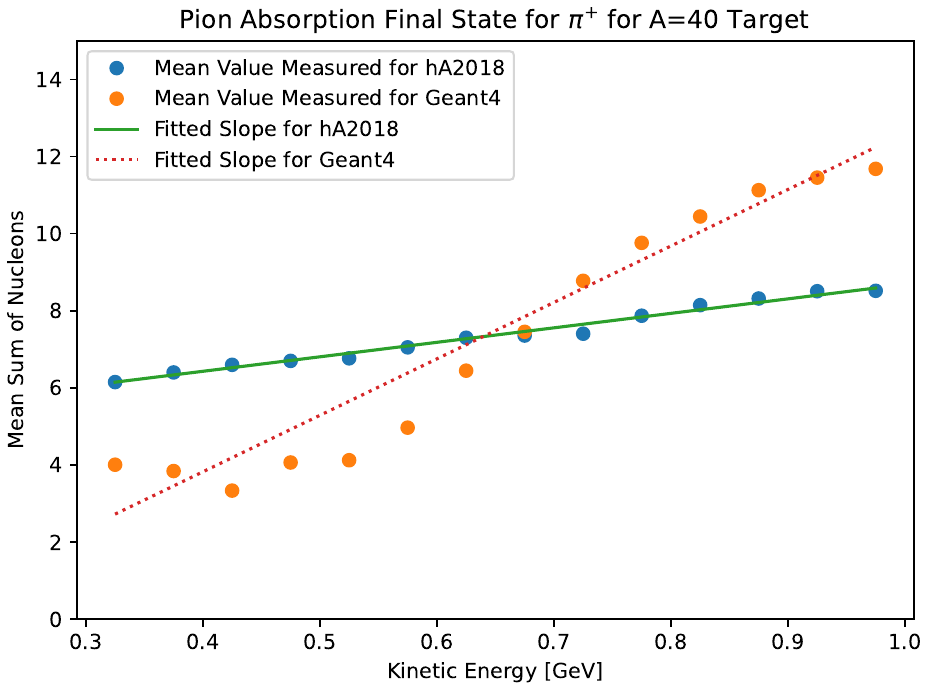}
\includegraphics[width=0.45\textwidth]{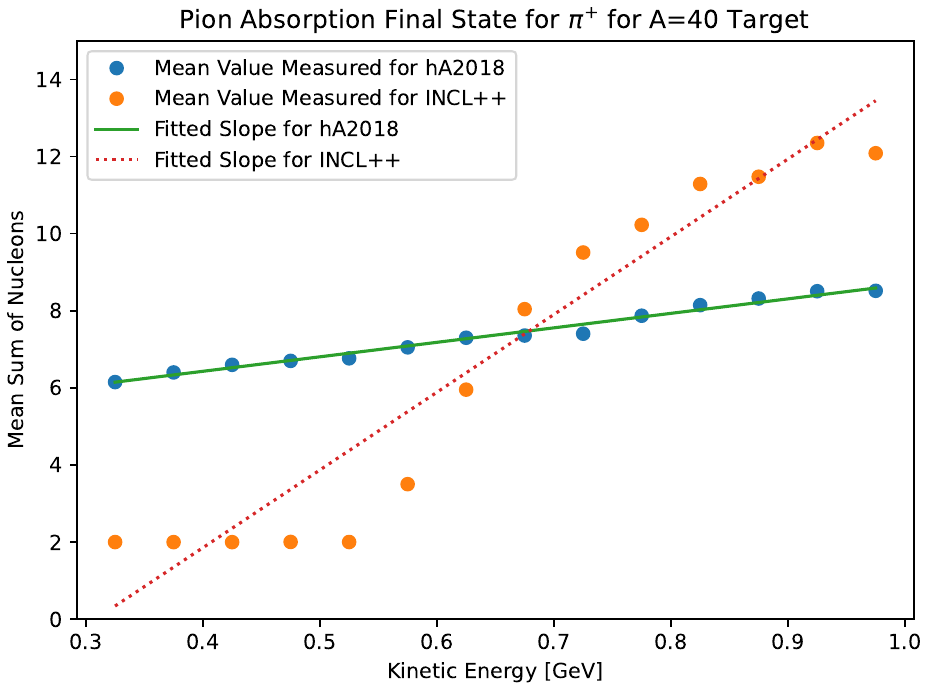}
    \caption{Evolution of multiplicity fit parameters on $^{40}\text{Ar}$ versus incident hadron kinetic energy. (Top) Knockout exponential decay constant $\Gamma$ (Eq.~\ref{eqn:gamma}) for $p$-Ar. (Bottom) Pion absorption Gaussian mean $\mu$ (Eq.~\ref{eqn:lin}) for $\pi^+$-Ar in the MN regime ($T_\pi \ge 0.3\text{ GeV}$). Markers denote bin-by-bin fits; lines indicate the fitted curves.}
        \label{fig:par}

\end{figure*}

The most basic function is the Gaussian where the mean and width are fit to the multiplicity distribution for each probe energy. The energy dependence is then
fit using a linear function and an exponential decay function for both multiplicity sum and difference for pions and the multiplicity difference for nucleons. The linear function is written as:

\begin{equation}
\mu = j + i \cdot T \label{eqn:lin}
\end{equation}

\noindent The exponential decay function is used for the nucleon multiplicity sum. It needs additional parameters to handle the kinetic energy ($T$) dependence of the fast-falling edge for simulations with a full intranuclear cascade and is written as:

\begin{equation}
\Gamma = j \cdot \exp(T^{k} \cdot i) + l \label{eqn:gamma}
\end{equation}

\noindent with $i$, $j$, $k$, and $l$ as fit parameters of $\Gamma$. This emphasis on solely the kinetic energy dependence for each nuclear target is different from the approach taken by the \textsc{GENIE} authors, as shown in Tables \ref{tab:pion} and \ref{tab:knockout}.
The extracted fit parameters for carbon, oxygen, argon, and iron targets across all four models are compiled in Tables~\ref{tab:fitparams_c12}--\ref{tab:fitparams_fe56} in Appendix \ref{app:tables}.

\section{Using Nucleon Fitting Results to Modify and Optimize Simulations}
\label{sec:reweight}

The parameterizations of the various particle scattering models can be used to alter the output of the simulations and allow them to cover the model spread or data-simulation discrepancies. The linear and exponential fits of how the distributions evolve per target are recorded for each model. 

The alteration of the \textsc{GENIE} empirical model hA2018 involves two sequential steps. First, the total number of nucleons and the difference in nucleons are altered using the tables of parameters. The ratio of the probability of an interaction having that total number of nucleons and difference in nucleons is calculated and used to reweight the probability of that interaction occurring. For Gaussian-based parameters, the weight of an interaction ($w$) to alter the nominal simulation to an alternative one is:

\begin{equation}
    w(N)=\frac{G_{\alt}(N, \mu_{\alt}, \sigma_{\alt})}{G_{\nom}(N, \mu_{\nom}, \sigma_{\nom})} \label{eqn:gauss}
\end{equation}
\noindent where $G$ represents the total fraction of interactions with multiplicity of integer $N$ within a Gaussian distribution of mean ($\mu$) and standard deviation ($\sigma$). The mean and standard deviation are determined from Eq.~\ref{eqn:lin}. 

The exponential decay function for the total number of nucleons in nucleon knockout follows a similar weighting scheme. The decay parameter ($\gamma$) modifies an exponential decay distribution dependent on the number of nucleons ($e^{-\gamma N}$), and is calculated using Eq.~\ref{eqn:gamma}. The total weight for the exponential decay is:

\begin{equation}
    w=\frac{E_{\alt}(N, \gamma_{\alt})}{E_{\nom}(N, \gamma_{\nom})} \label{eqn:exp}
\end{equation}

\noindent where the weight is the ratio of the total number of interactions $E$ of a specific integer number of nucleons $N$ for an exponential decay function with a decay rate of $\gamma$ defined by the model parameterization in Eq.~\ref{eqn:gamma}.

To ensure QD scattering does not bias results, the QD component of \textsc{GENIE} hA2018 is suppressed during the reweighting of nucleons using Gaussian and exponential decay functions. Using Eq.~\ref{eqn:qd}, the probability of QD scattering is calculated and then altered, so the probability of QD-like interactions with only two nucleons matches the Gaussian weights calculated in the denominator of Eq.~\ref{eqn:gauss}. 

The second component is to alter the visible energy using templates of the visible energy from nucleon knockout and pion absorption interactions for each model. These templates for \textsc{GENIE} hA2018 are produced after nucleon parameter reweighting. These templates are intended to accommodate the fact that \textsc{GENIE} hA2018 and hN2018 distribute the available energy more uniformly across final state nucleons than \textsc{Geant4} BC and \textsc{INCL++}. Without this second alteration, the nucleon reweighting would significantly alter the visible energy to a point that is not representative of the model being used to alter the nominal distributions. The visible energy templates are only derived and applied to interactions that fit the definitions of nucleon knockout and pion absorption outlined in Sec.~\ref{sec:results}. The weight is applied as a ratio of the fraction ($F$) of interactions within a particular bin of visible energy, and is defined as 
\begin{equation}
    w_{\text{energy}} = \frac{F_{\text{alt}}(E_{\text{vis}})}{F_{\text{nom}}(E_{\text{vis}})}
\end{equation}

Occasionally, the integral of the phase space of \textsc{GENIE} hA2018 in terms of nucleon multiplicity and visible energy is different between the two models being used to generate weights. This behavior is most often observed with \textsc{GENIE} hN2018, which, unlike other models, has interactions with negative energy bias and the highest average number of nucleons produced at energies greater than 0.50 GeV. To accommodate this, the phase space and integral of the nominal simulation are used to scale the normalization of the weights of the alternative simulation being utilized. Combined, the total weight applied to interactions, assuming the distributions are normalized to the same phase space, is:

\begin{equation}
    w_{\text{tot}} = w_{\text{sum}}(N_p+N_n)\cdot w_{\text{diff}}(N_p-N_n)\cdot w_{\text{energy}}(E_{\text{vis}}) \label{eqn:total}
\end{equation}

\noindent where $w_{\text{sum}}$ and $w_{\text{diff}}$ are the multiplicity weight ratios. The nucleon multiplicity weights are evaluated for the total number of protons and neutrons ($N_p+N_n$) and the difference of protons and neutrons ($N_p-N_n$) as previously defined in Eq.~\ref{eqn:gauss} and Eq.~\ref{eqn:exp} for parameters defined by Gaussian or exponential decay functions. 

\begin{figure*}
    \centering
    \includegraphics[width=0.45\linewidth]{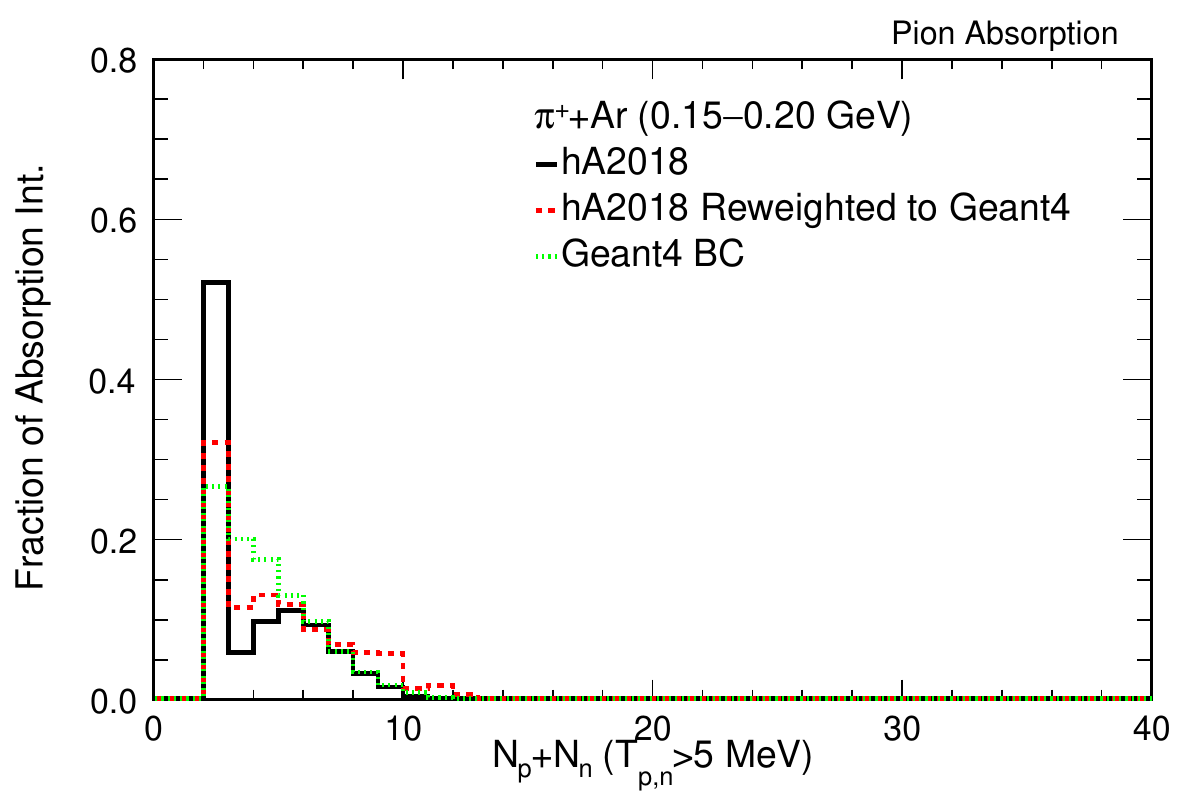}
        \includegraphics[width=0.45\linewidth]{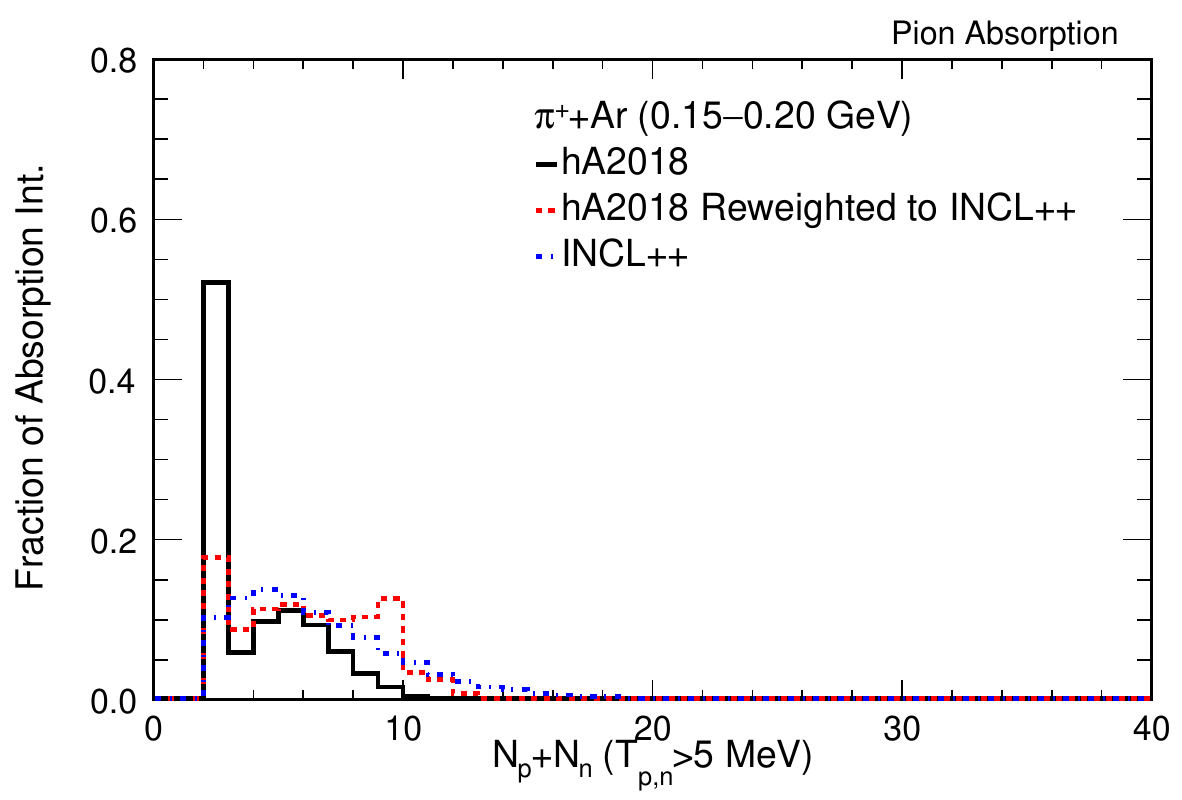}

        \includegraphics[width=0.45\linewidth]{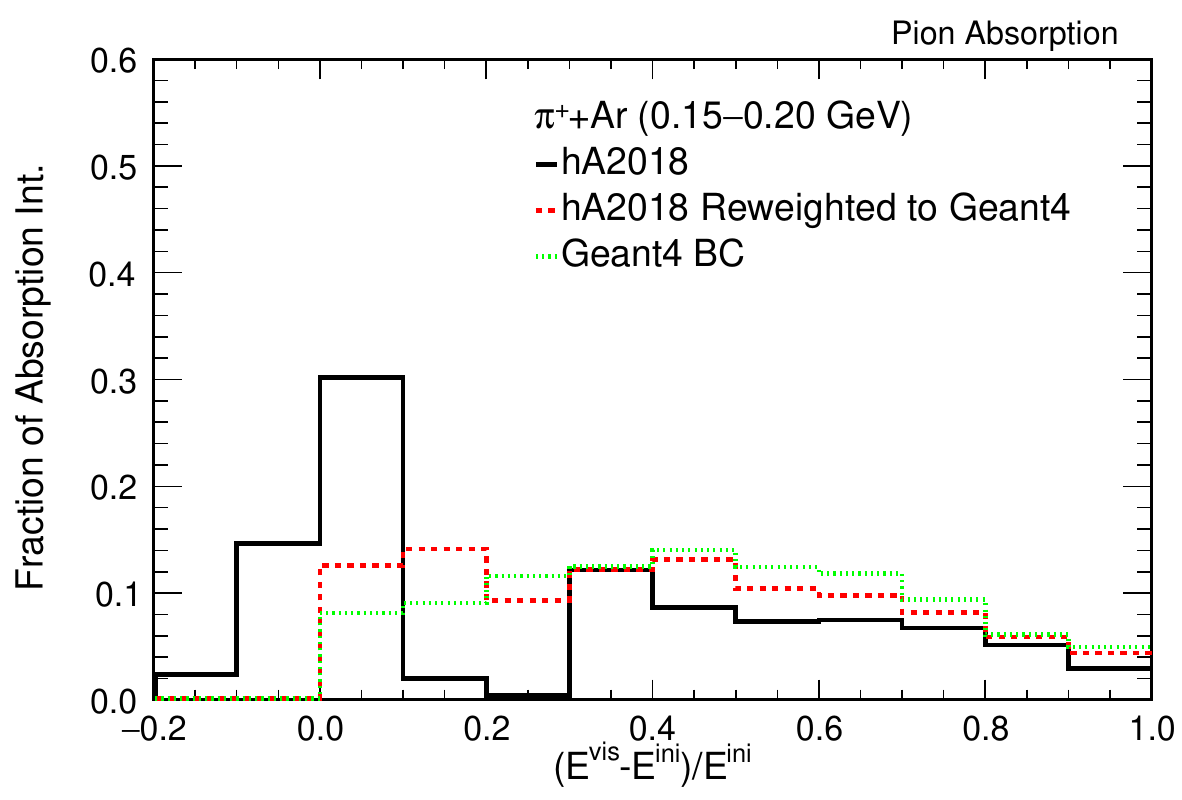}
        \includegraphics[width=0.45\linewidth]{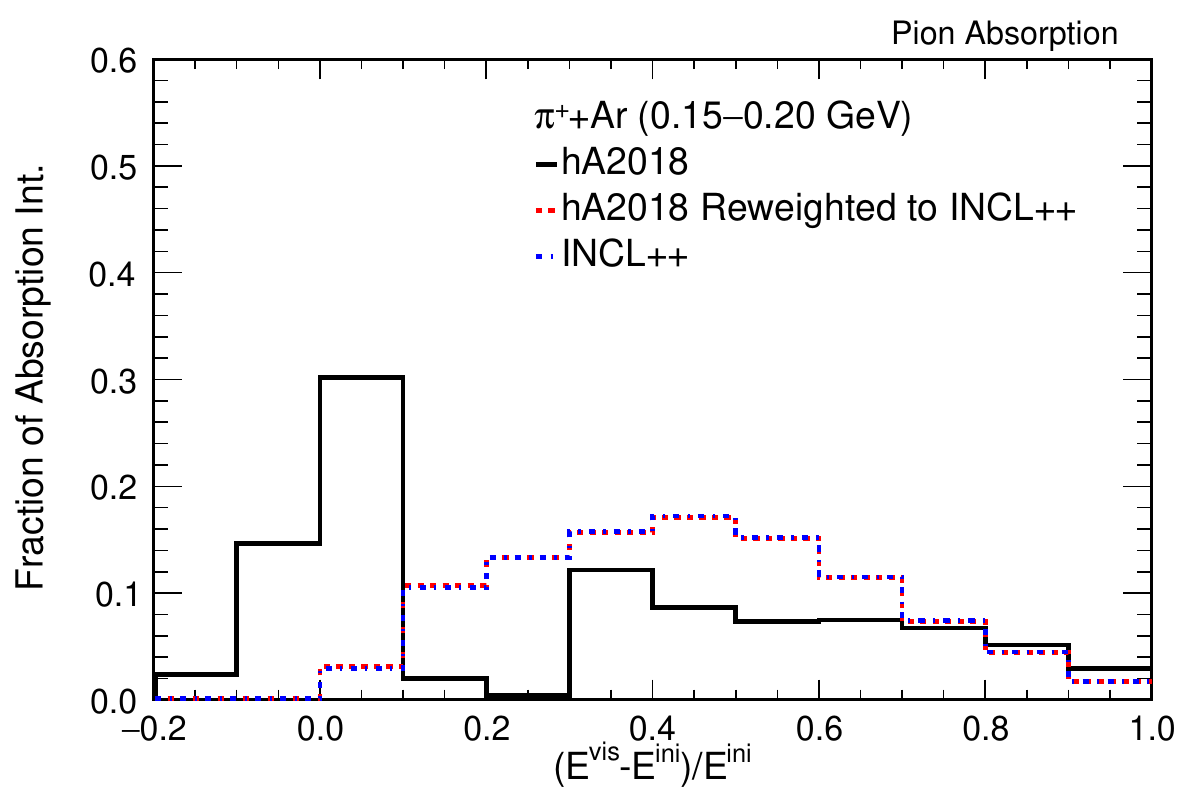}
    \caption{Distributions on the total number of nucleons and the visible energy available for various hadron scattering models of pion absorption interactions on argon with a mean incident pion energy of 0.175 GeV. Alterations of the base \textsc{GENIE} hA2018 to \textsc{Geant4} BC and \textsc{INCL++} are shown on the left and right, respectively.}
    \label{fig:rwPiAbs}
\end{figure*}

\begin{figure*}
    \centering
        \includegraphics[width=0.45\linewidth]{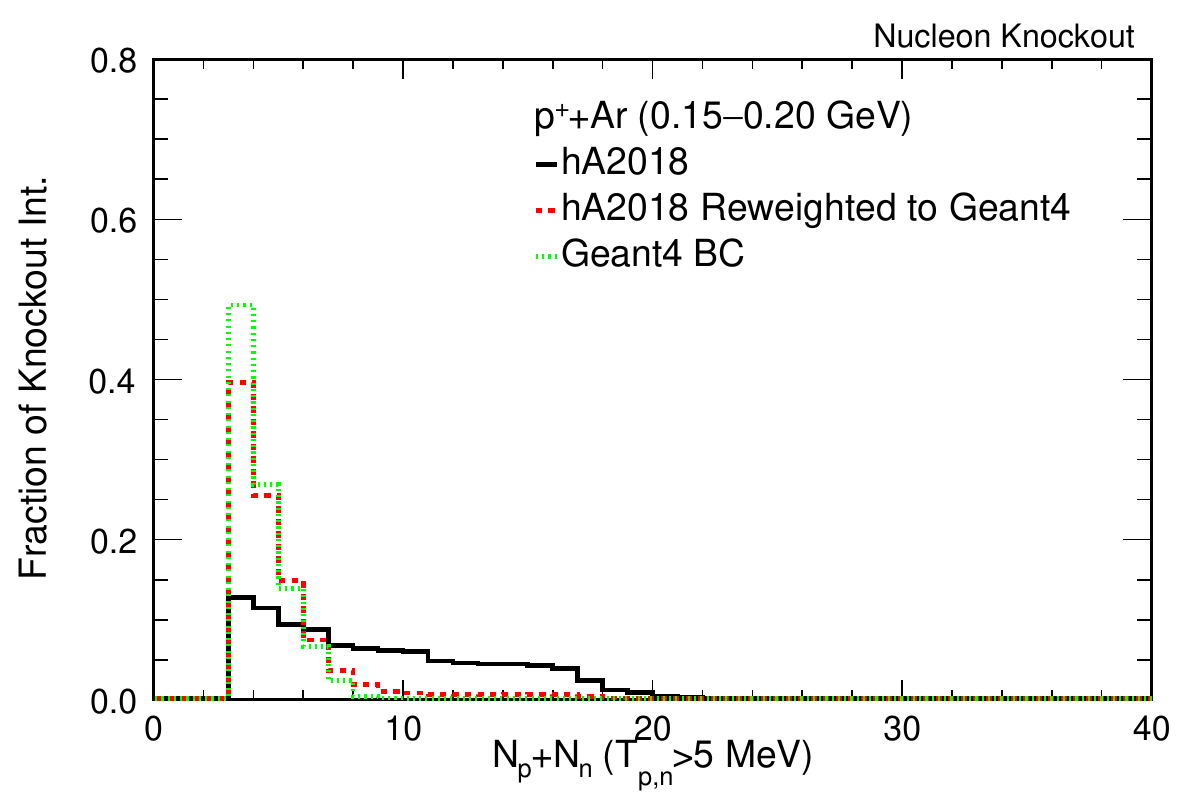}
        \includegraphics[width=0.45\linewidth]{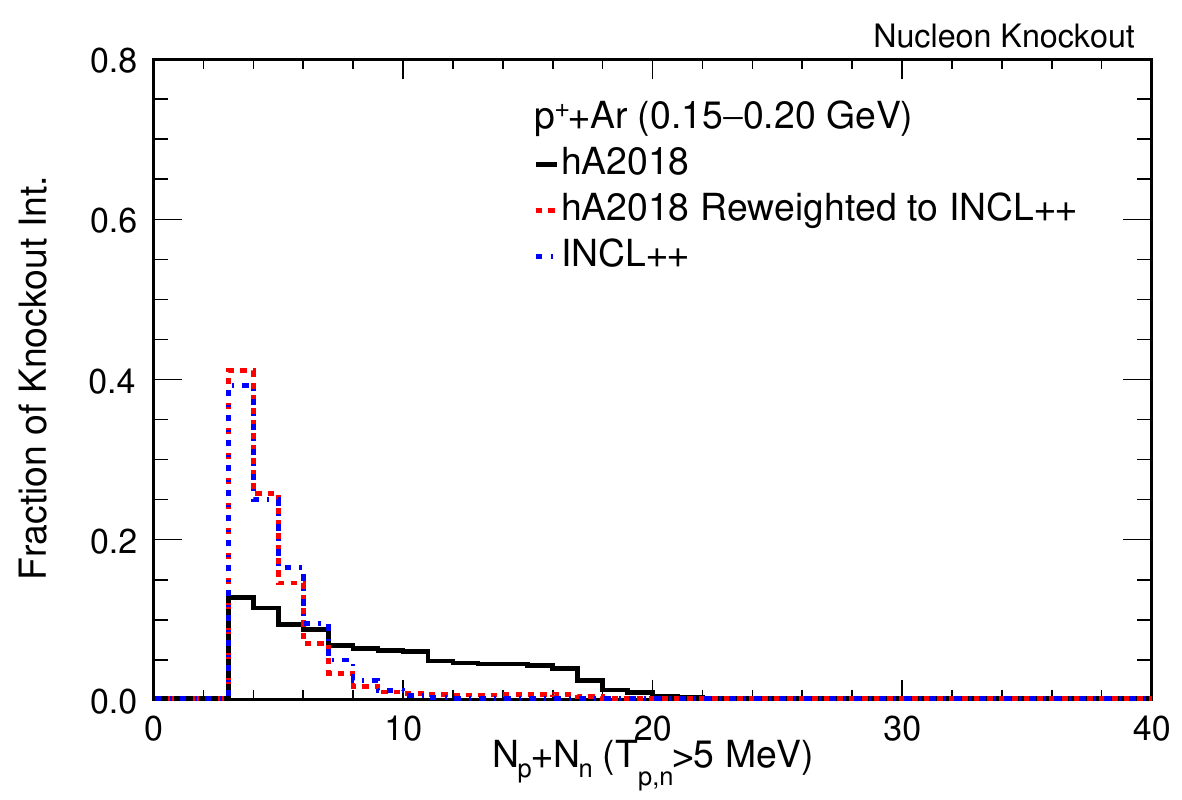}

        \includegraphics[width=0.45\linewidth]{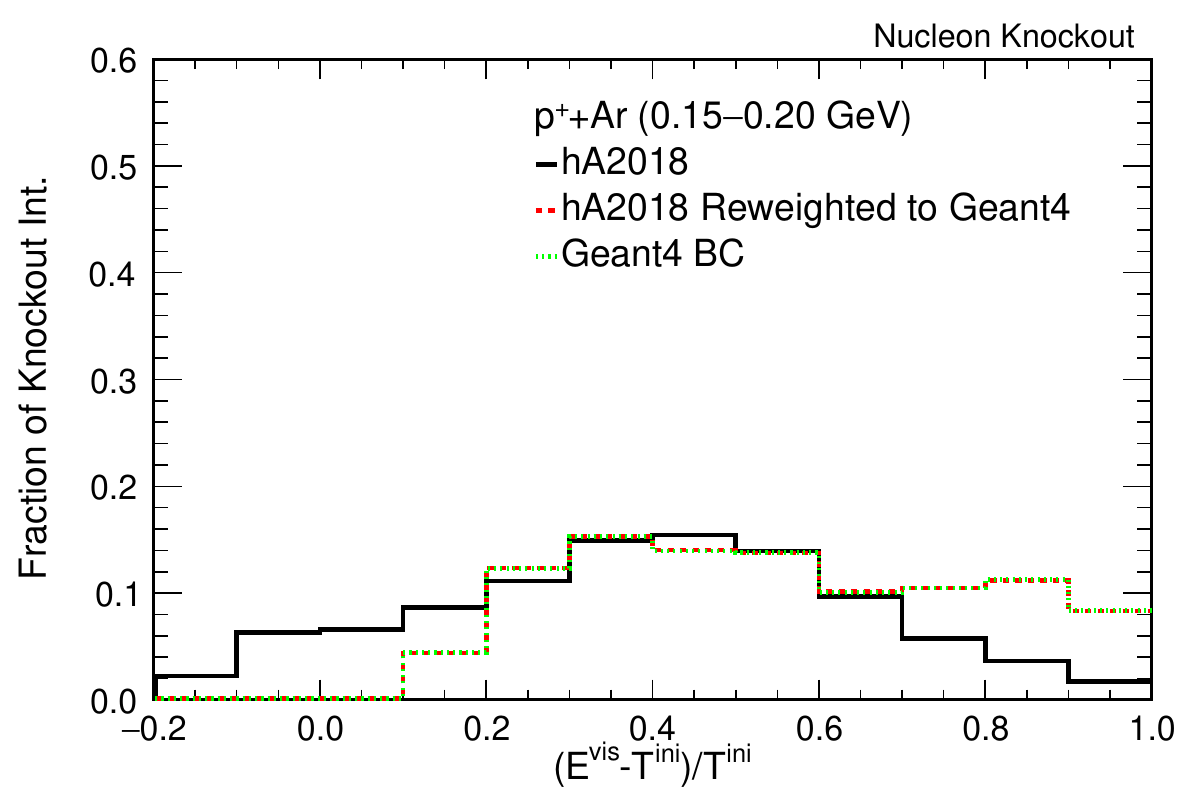}
        \includegraphics[width=0.45\linewidth]{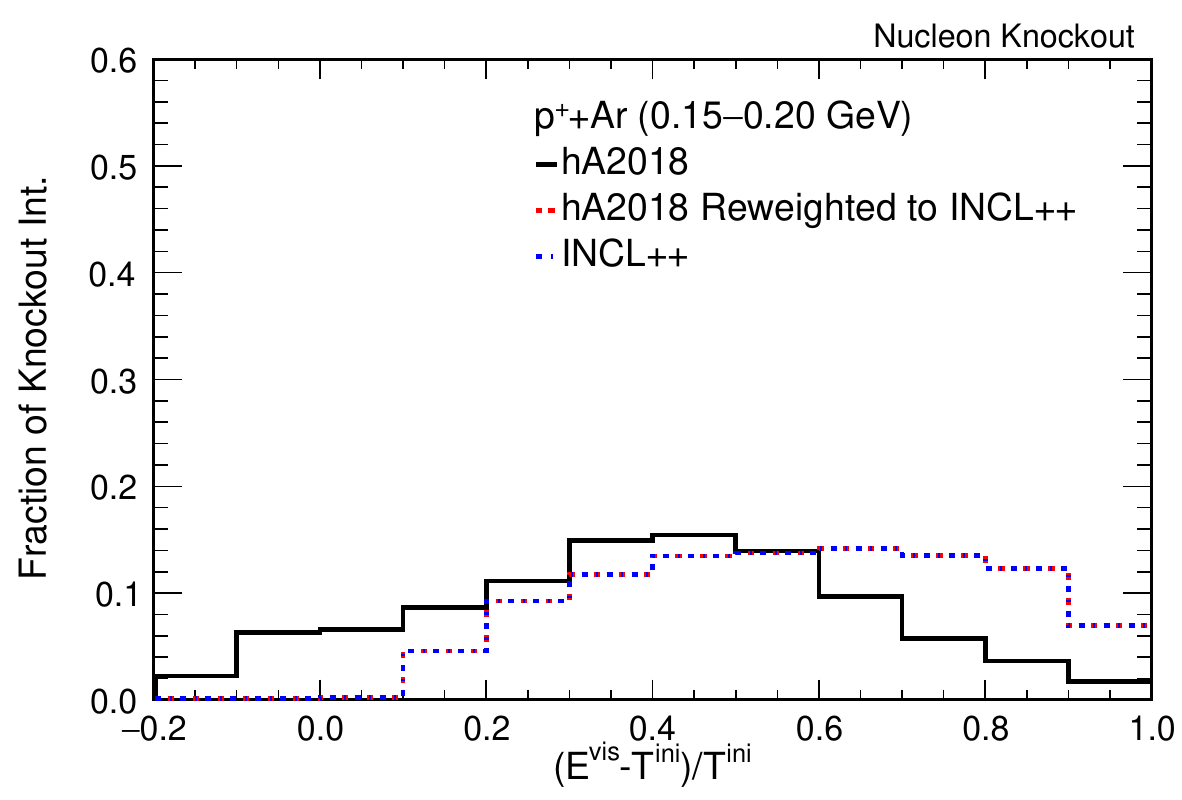}
    \caption{Distributions on the total number of nucleons and the visible energy available for various hadron scattering models for proton knockout interactions on argon with a mean incident proton energy of 0.175 GeV. Alterations of the base \textsc{GENIE} hA2018 to \textsc{Geant4} BC and \textsc{INCL++} are shown on the left and right, respectively.}
    \label{fig:rwKO}
\end{figure*}

Examples of altering \textsc{GENIE} hA2018 to \textsc{Geant4} BC and \textsc{INCL++} are shown in Figure~\ref{fig:rwPiAbs} and Figure~\ref{fig:rwKO}. As the nucleon multiplicity alterations are calculated from linear fits of Gaussian and exponential decay parameters, it is expected that there is not a complete mapping to the alternative simulation, unlike with the visible energy templates that are implemented at the final step.

The model spread dials have been included in the popular neutrino reweighting software \textsc{nusystematics} and have been used with the neutrino cross section comparison software \textsc{NUISANCE} to test simulation-to-data performance~\cite{NUISANCE}. We compared the different particle transport simulations to MicroBooNE~\cite{MicroBooNE:2023cmw}, T2K~\cite{T2K:2018rnz}, and MINERvA~\cite{MINERvA:2025tem} muon neutrino data, specifically data of charged-current interactions without mesons in the final state (CC0$\pi$). These three experiments provide muon neutrino datasets across different GeV-scale beams and on different targets. Generally, pion absorption and nucleon knockout mainly impact the leading proton kinematics, as the leading protons should have smaller kinetic energies for interactions with more nucleons in the final state. Because muon neutrino CC0$\pi$ interactions have only nucleons in the hadronic final state, they allow for a better study of leading proton kinematics and transverse kinematic imbalances than datasets with primarily resonant and deep inelastic interactions.

As a base model, we use the AR23 model configuration in \textsc{GENIE} which ensures the same neutrino cross sections are used for all particle transport model comparisons. This model is shared between DUNE and SBN. Comparisons will be made with the AR23 configuration between the hA2018 model and the two external models implemented in \textsc{GENIE}, namely \textsc{INCL++} and \textsc{Geant4} BC. A third prediction will be shown of hA2018, altered using the methods described in this work, to demonstrate the impact of the knockout and absorption channels on the data-to-simulation agreement and model spread. The total and exclusive hadron scattering cross sections are unaltered to focus on model predictions of nucleon multiplicities and visible energies of final state interactions.

\begin{figure*}[h]
    \centering
    \includegraphics[width=0.45\linewidth]{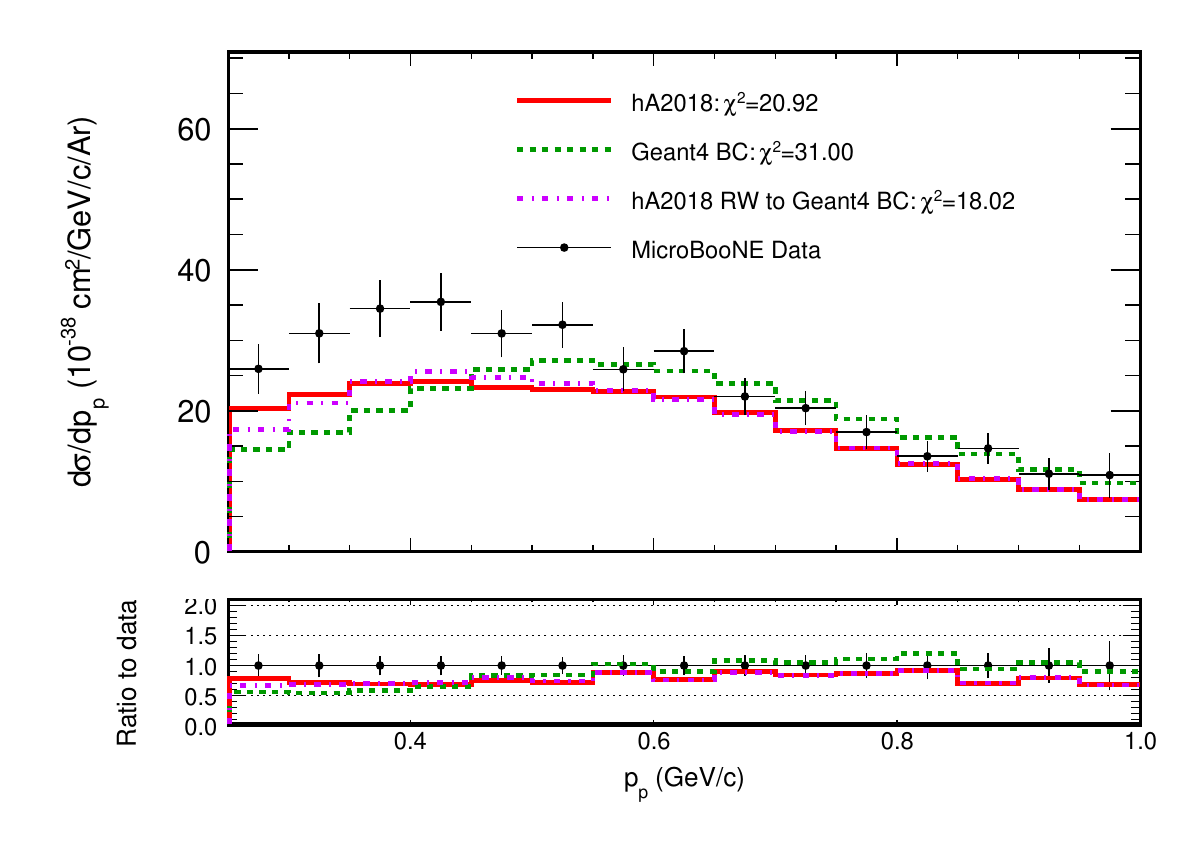}
    \includegraphics[width=0.45\linewidth]{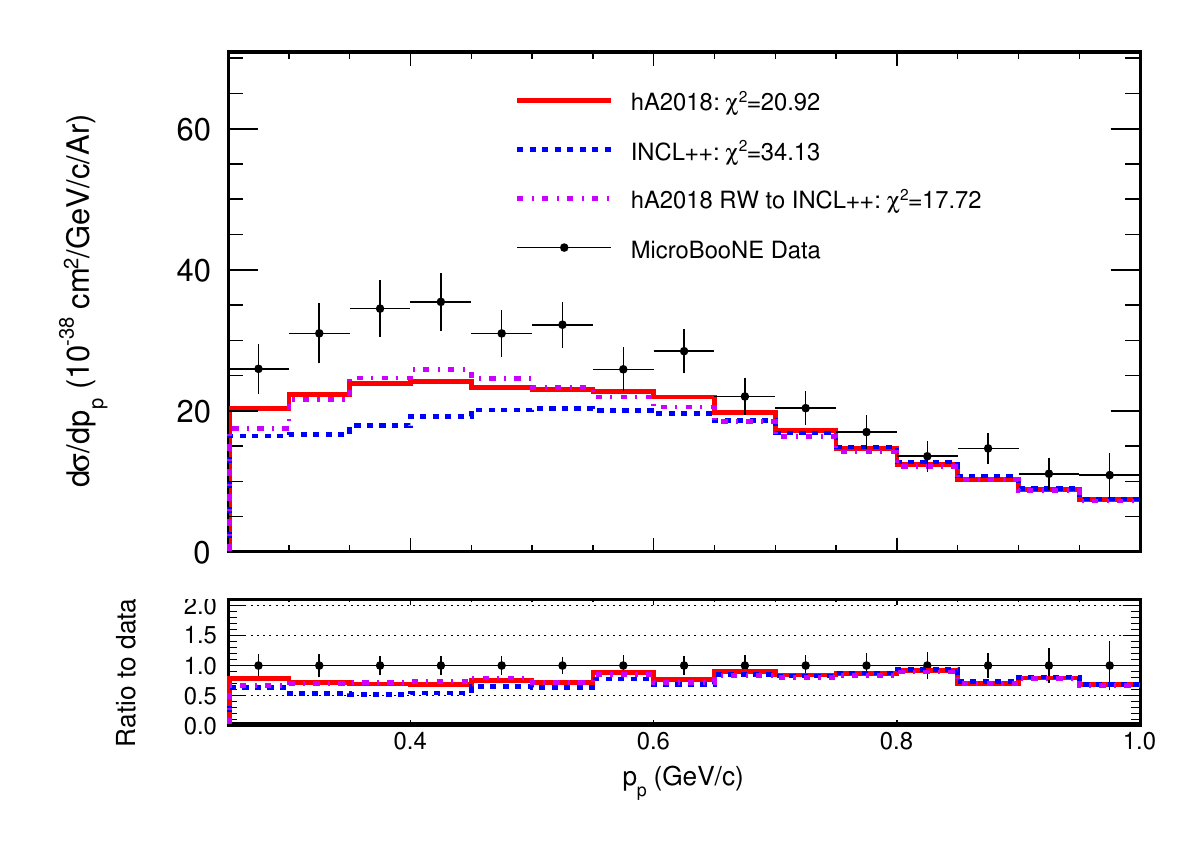}

    \includegraphics[width=0.45\linewidth]{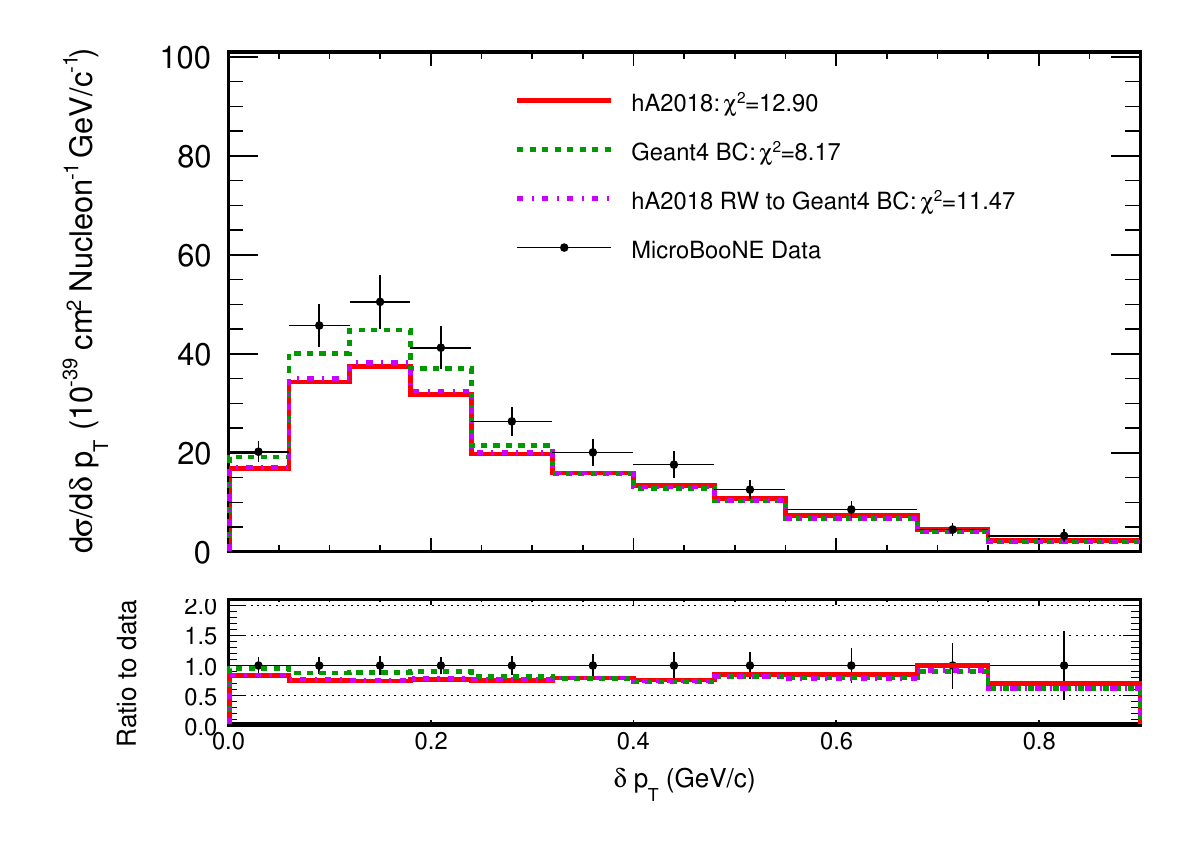}
    \includegraphics[width=0.45\linewidth]{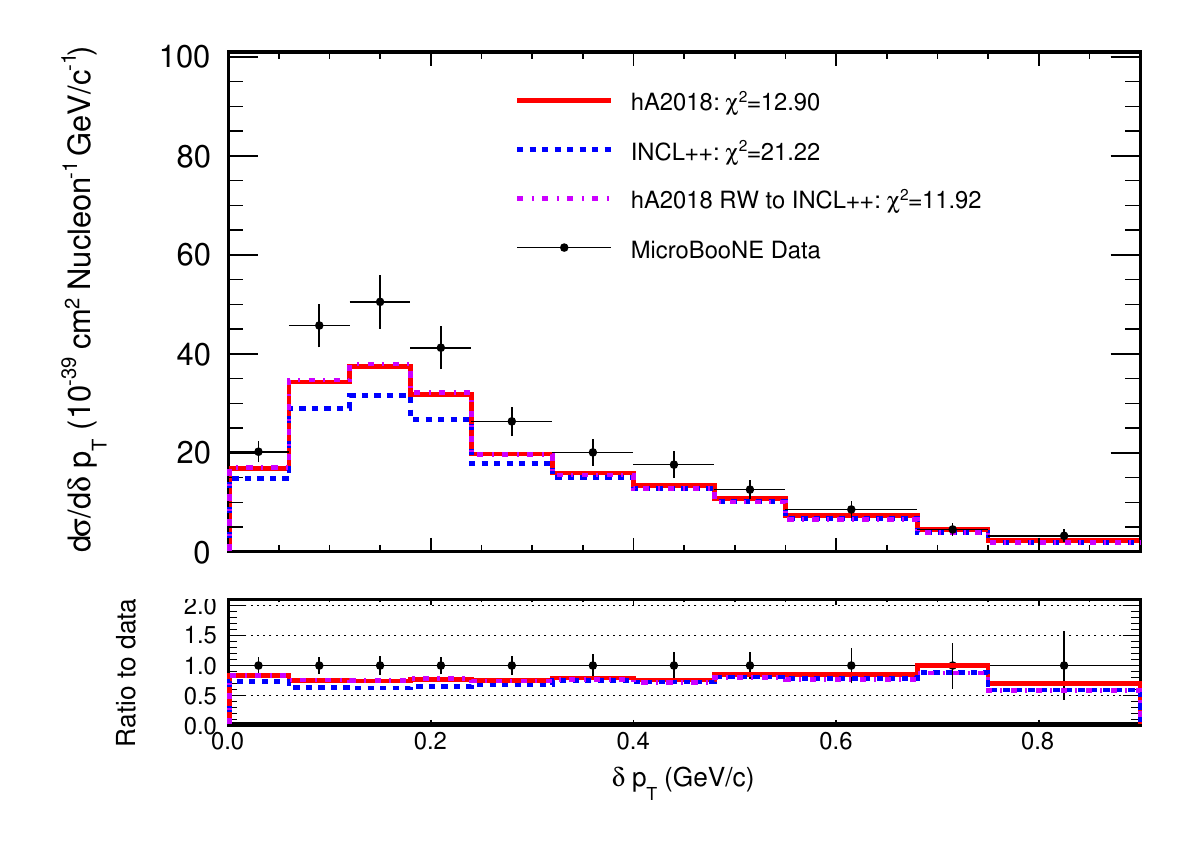}
    \caption{Comparisons of the model predictions on argon and MicroBooNE CC0$\pi$ data for the leading proton momentum (top) and transverse kinematic imbalance (bottom)~\cite{MicroBooNE:2023cmw}.}
    \label{fig:uBCC0pi}
\end{figure*}

Figure~\ref{fig:uBCC0pi} shows comparisons of the various hadron scattering models usable with the AR23 neutrino interaction model to MicroBooNE data~\cite{MicroBooNE:2023cmw}. All interactions without mesons are selected with this dataset. Generally, \textsc{INCL++} shows the best agreement, with \textsc{Geant4} BC having the worst agreement by consistently underestimating the neutrino cross section when paired with the AR23 neutrino model. The actual model spread does not significantly decrease when the predictions are altered using both the numerical multiplicity distribution parameters and the visible energy templates. There are slight improvements in the model spread for the leading proton momentum, but none are observed for the transverse kinematic imbalance. We redid the comparisons with alterations to the inclusive and exclusive hadron cross section rates to be the same between the reweighted model and the alternative model. We obtained identical results that appear to point to other effects in particle transport outside of pion absorption and nucleon knockout modeling.

\begin{figure*}[h]
    \centering
    \includegraphics[width=0.45\linewidth]{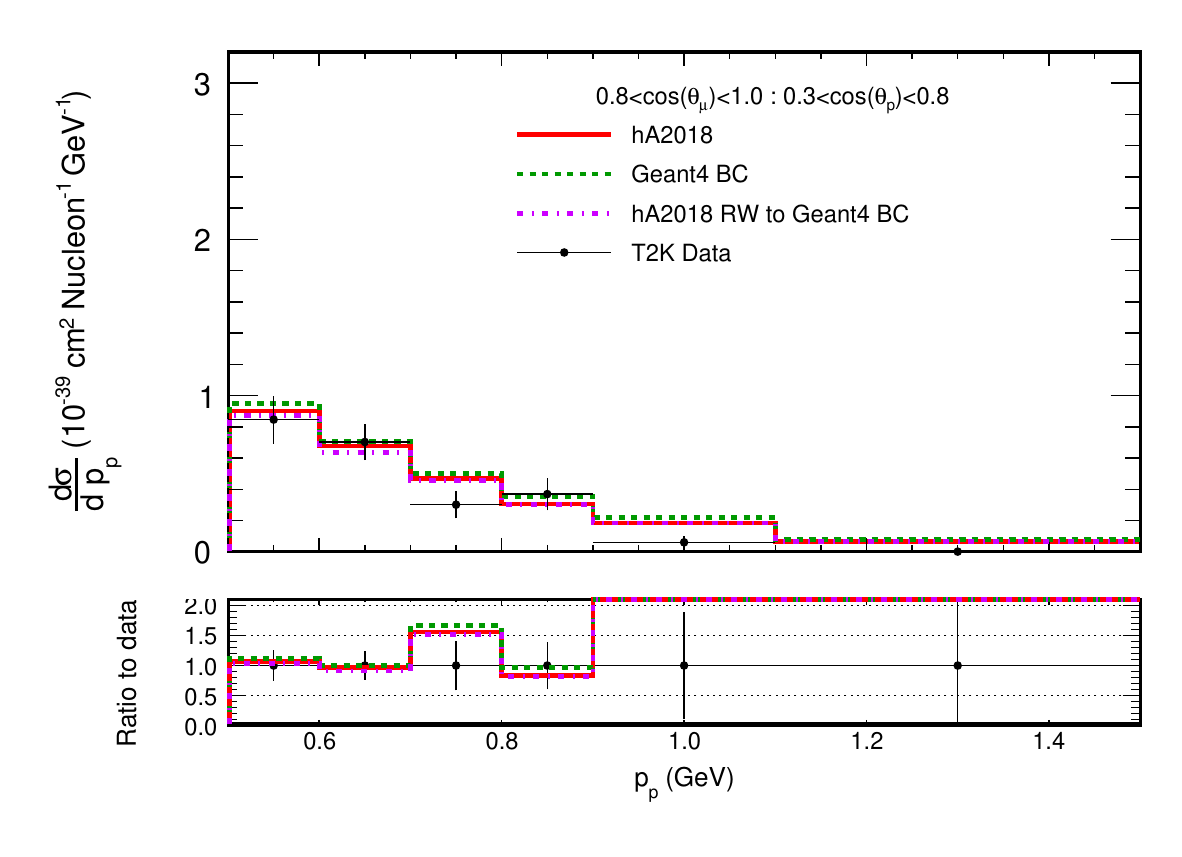}
    \includegraphics[width=0.45\linewidth]{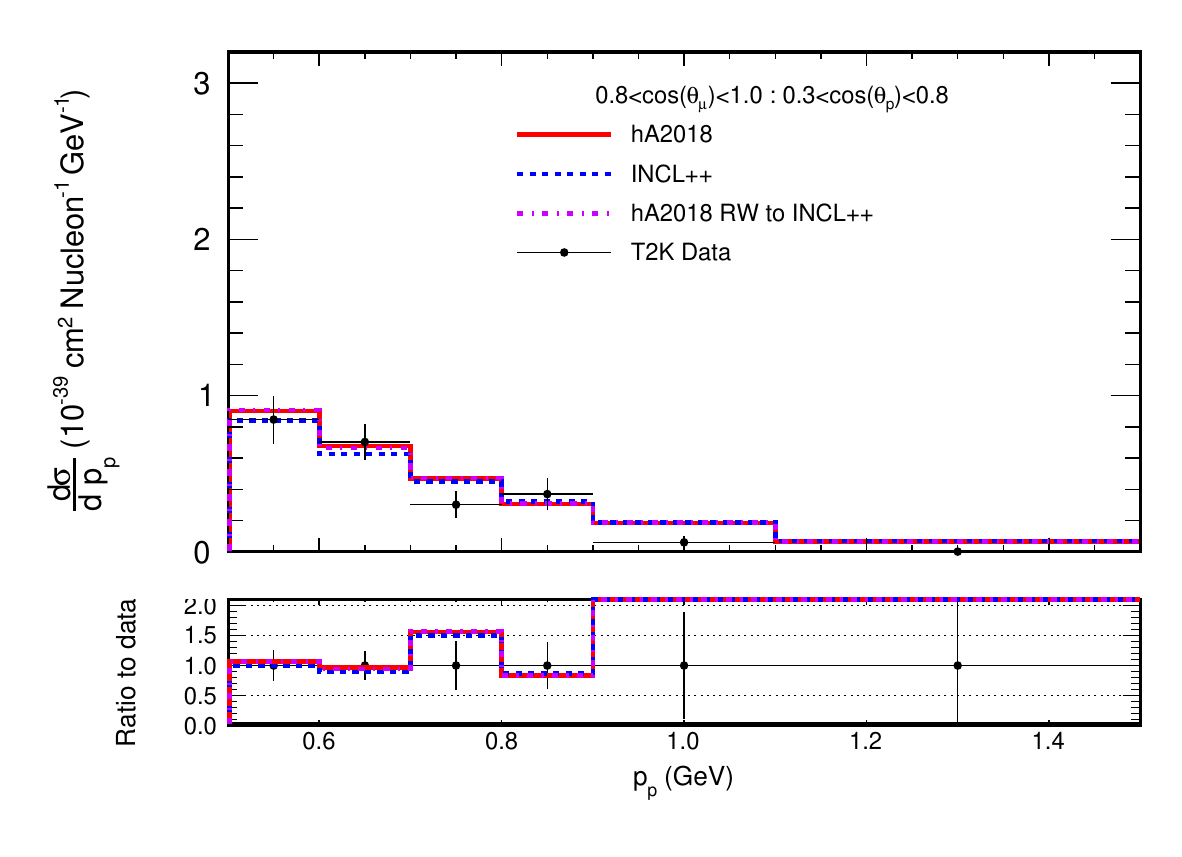}
    \caption{Comparisons of T2K CC1p0$\pi$ data and model predictions on hydrocarbon at the ND280 detector for proton momentum in a specific phase space of proton angle and muon angle~\cite{T2K:2018rnz}. A chi-squared metric is not presented as this differential cross section is a subset of the phase space of the full dataset.}
    \label{fig:t2kCC0pi}
\end{figure*}

T2K published a similar analysis to MicroBooNE in 2018~\cite{T2K:2018rnz}. The analysis divides the differential cross section into specific slices of lepton and proton kinematics. The dataset studied is mainly of CC1p0$\pi$ interactions. Figure~\ref{fig:t2kCC0pi} presents the predictions and reveals almost no model spread for this phase space.

\begin{figure*}[h]
    \centering
    \includegraphics[width=0.45\linewidth]{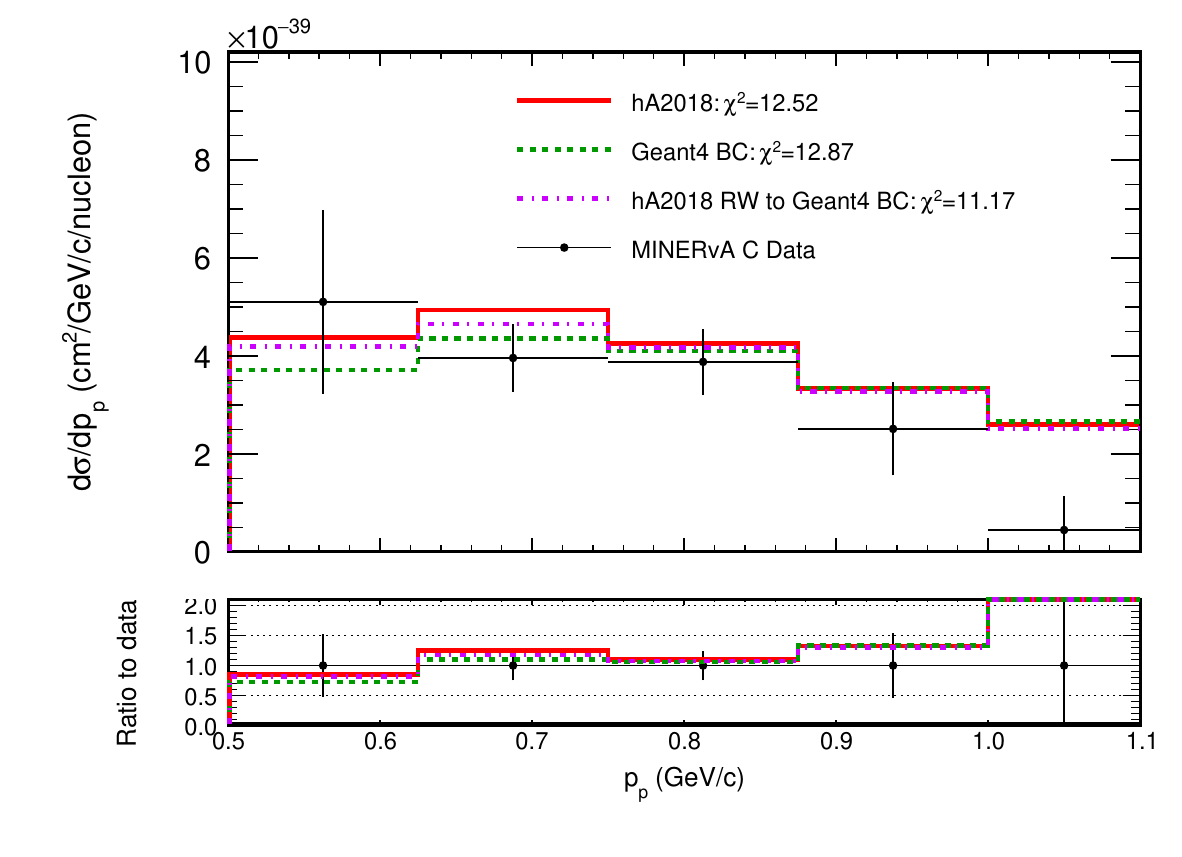}
    \includegraphics[width=0.45\linewidth]{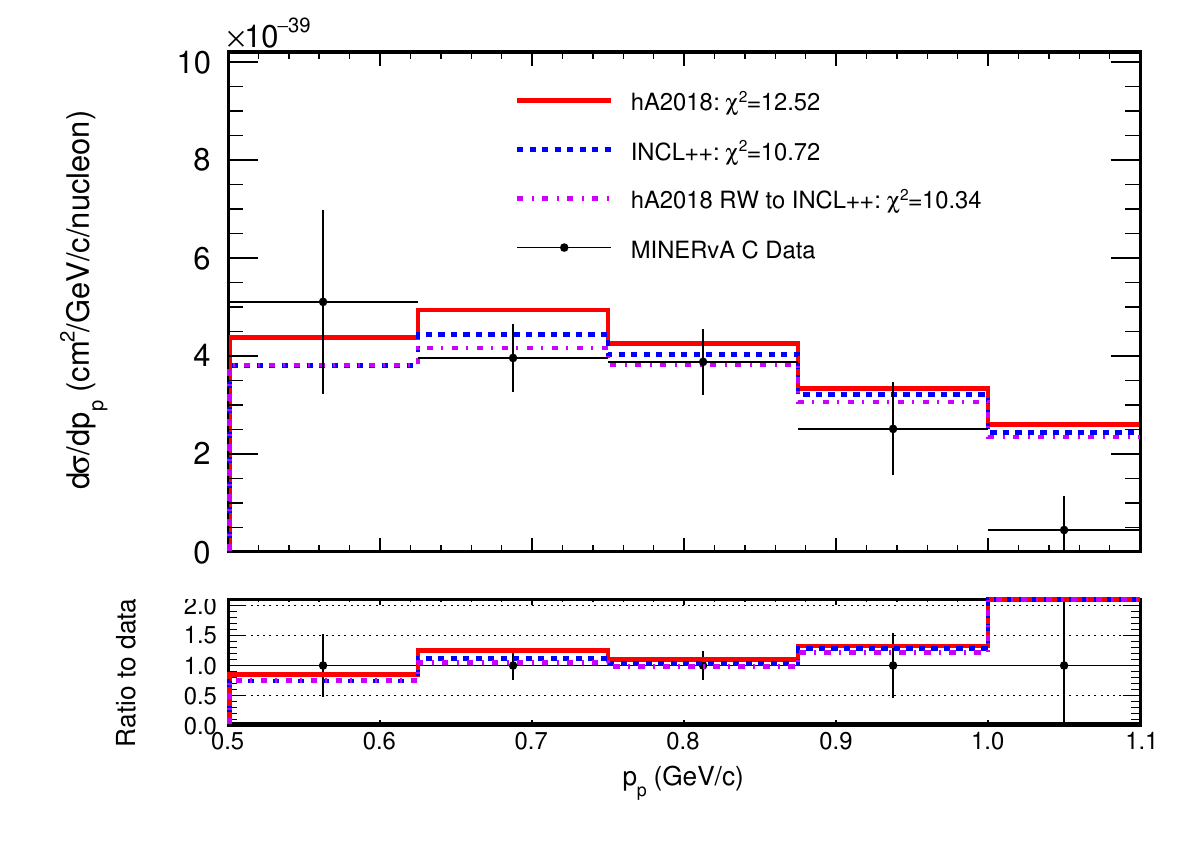}

    \includegraphics[width=0.45\linewidth]{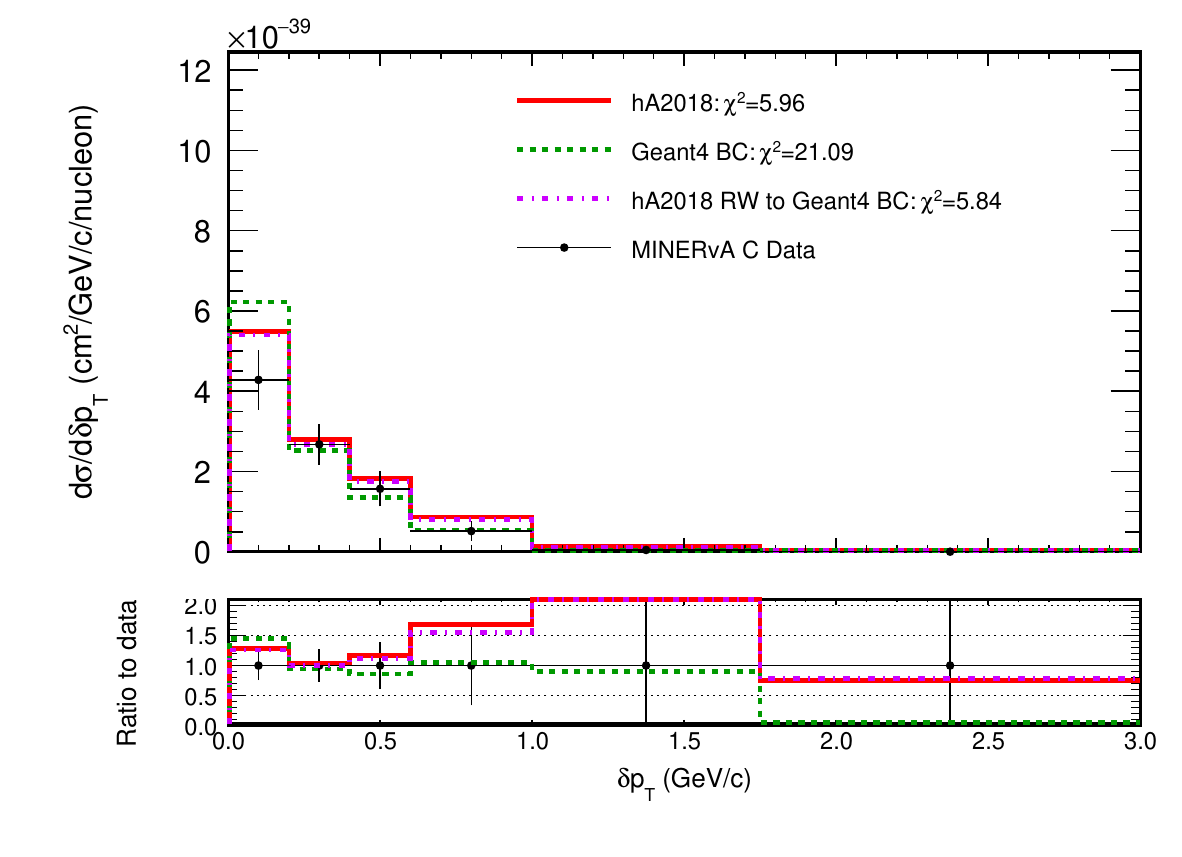}
    \includegraphics[width=0.45\linewidth]{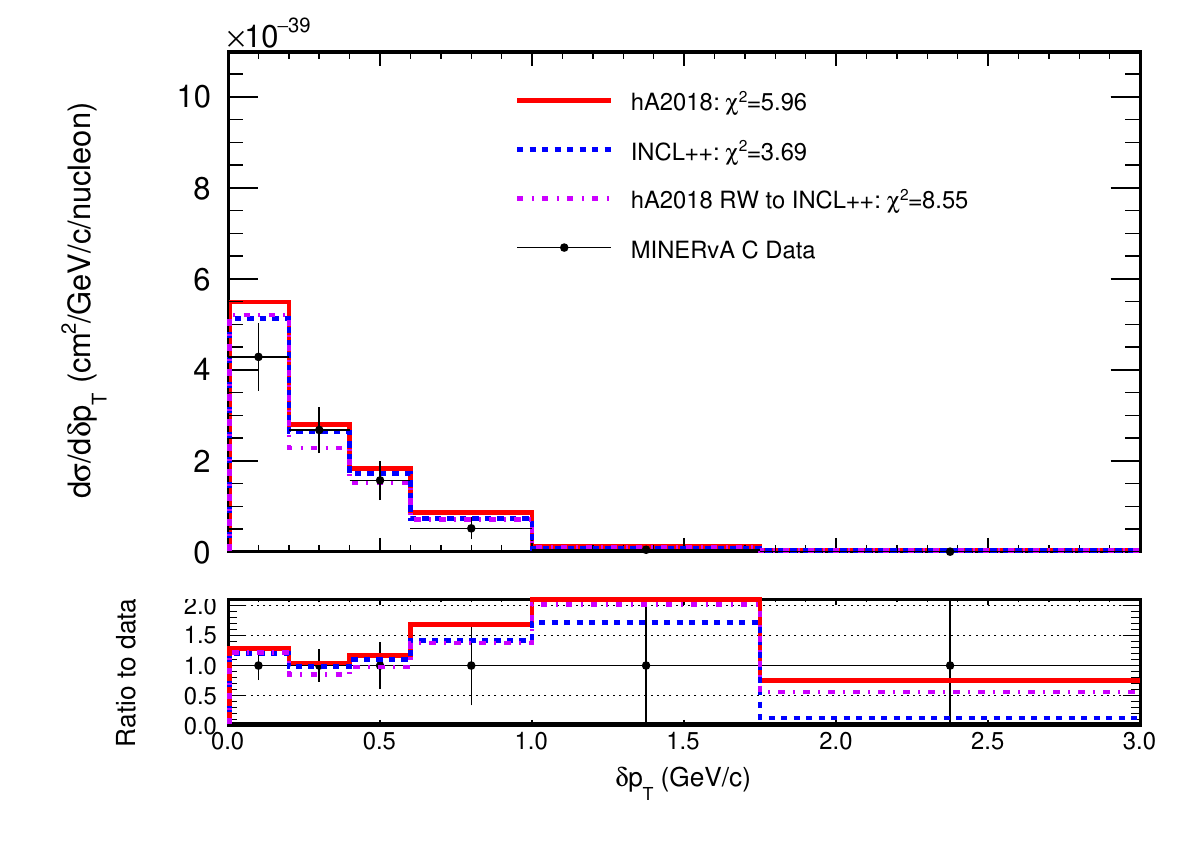}
    \caption{Results on the model predictions compared to MINERvA CC0$\pi$ on carbon~\cite{MINERvA:2025tem}. The differential neutrino cross sections as a function of leading proton momentum (top) and transverse kinematic imbalance (bottom) are shown.}
    \label{fig:mnvC}
\end{figure*}

\begin{figure*}[h]
    \centering
    \includegraphics[width=0.45\linewidth]{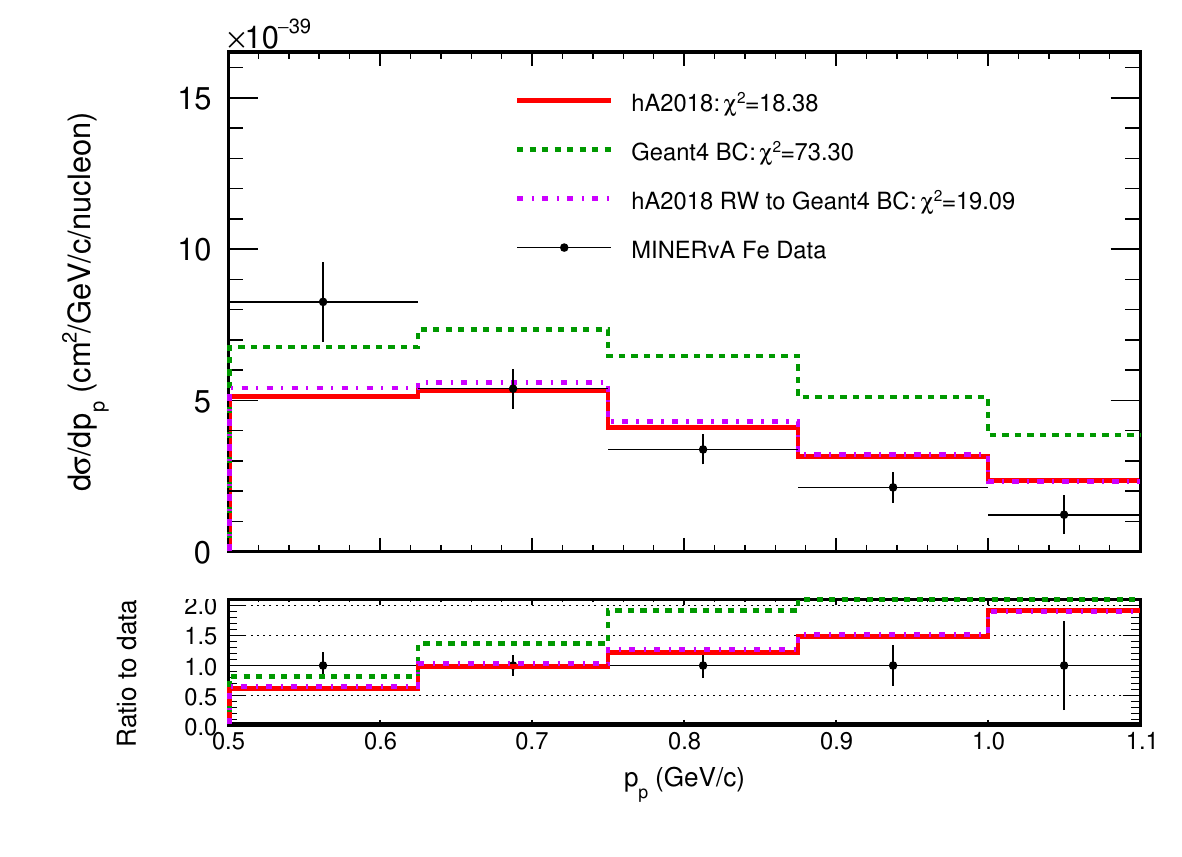}
    \includegraphics[width=0.45\linewidth]{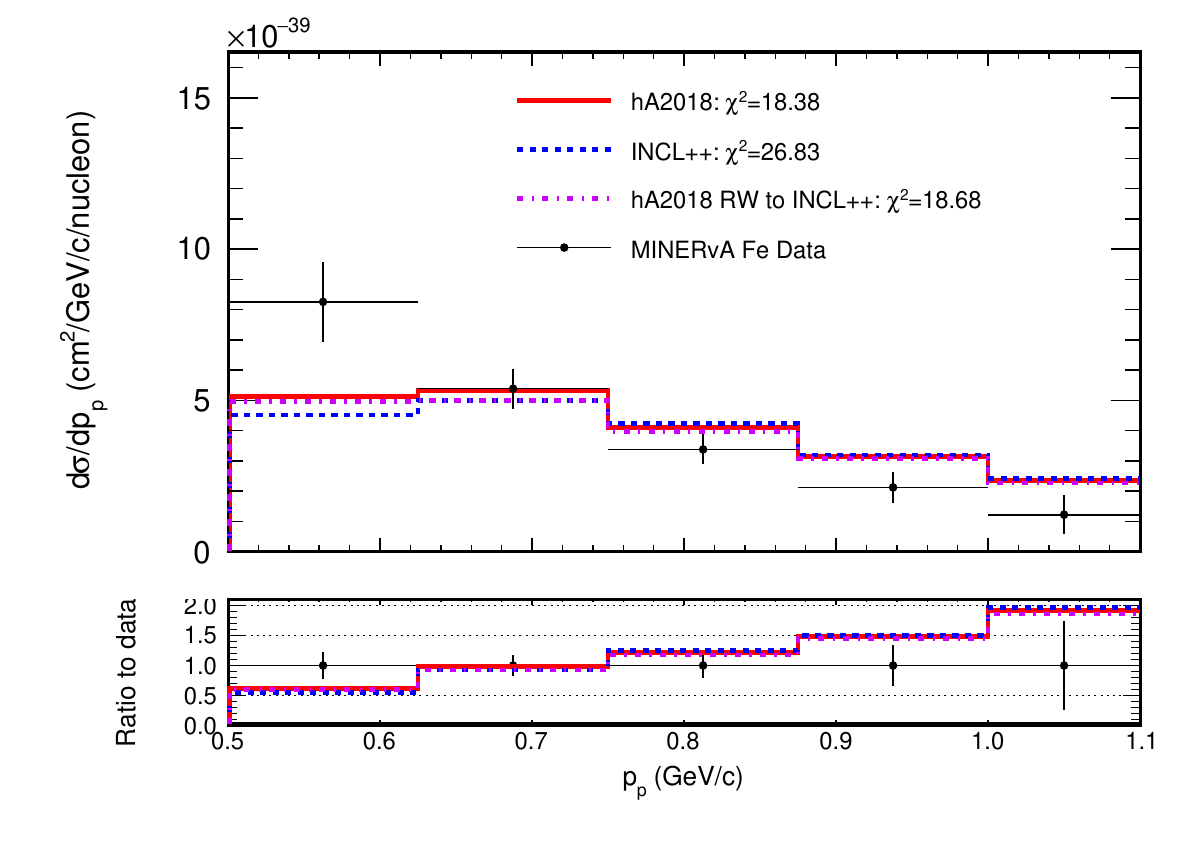}

    \includegraphics[width=0.45\linewidth]{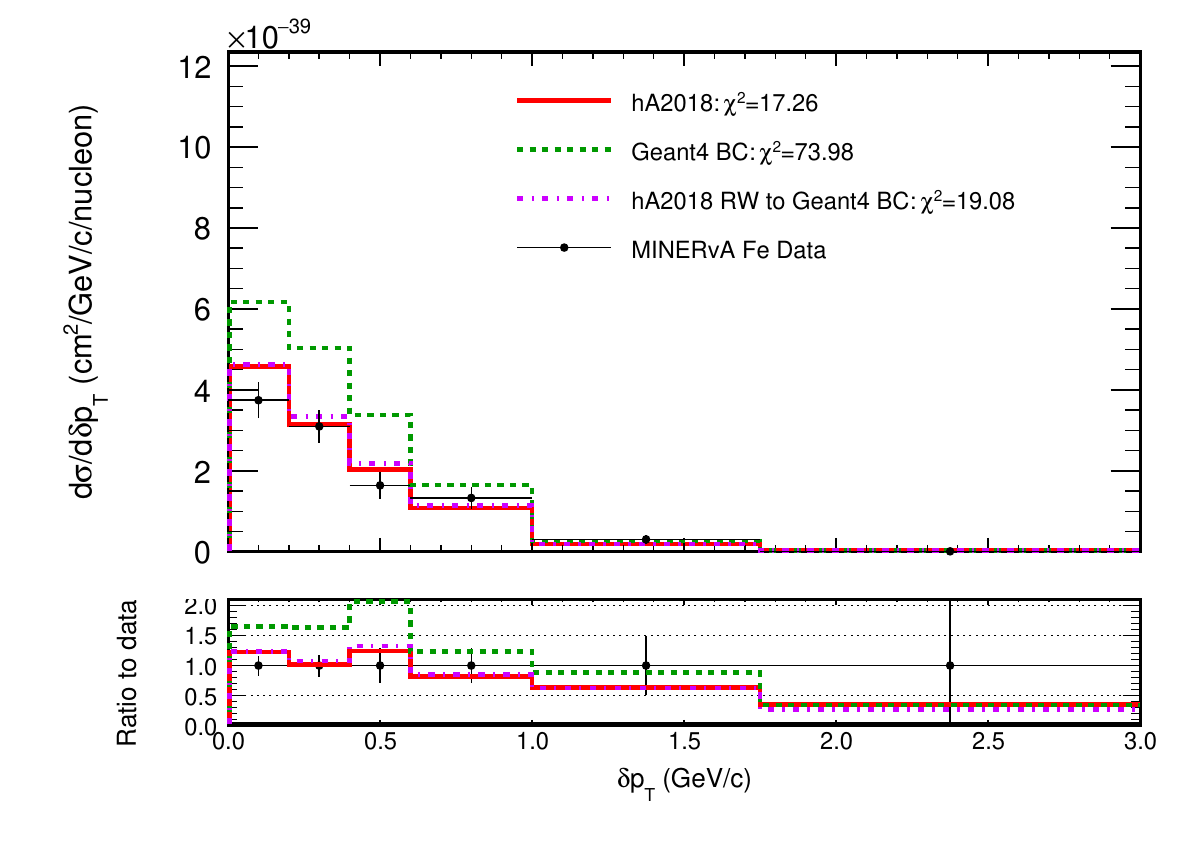}
    \includegraphics[width=0.45\linewidth]{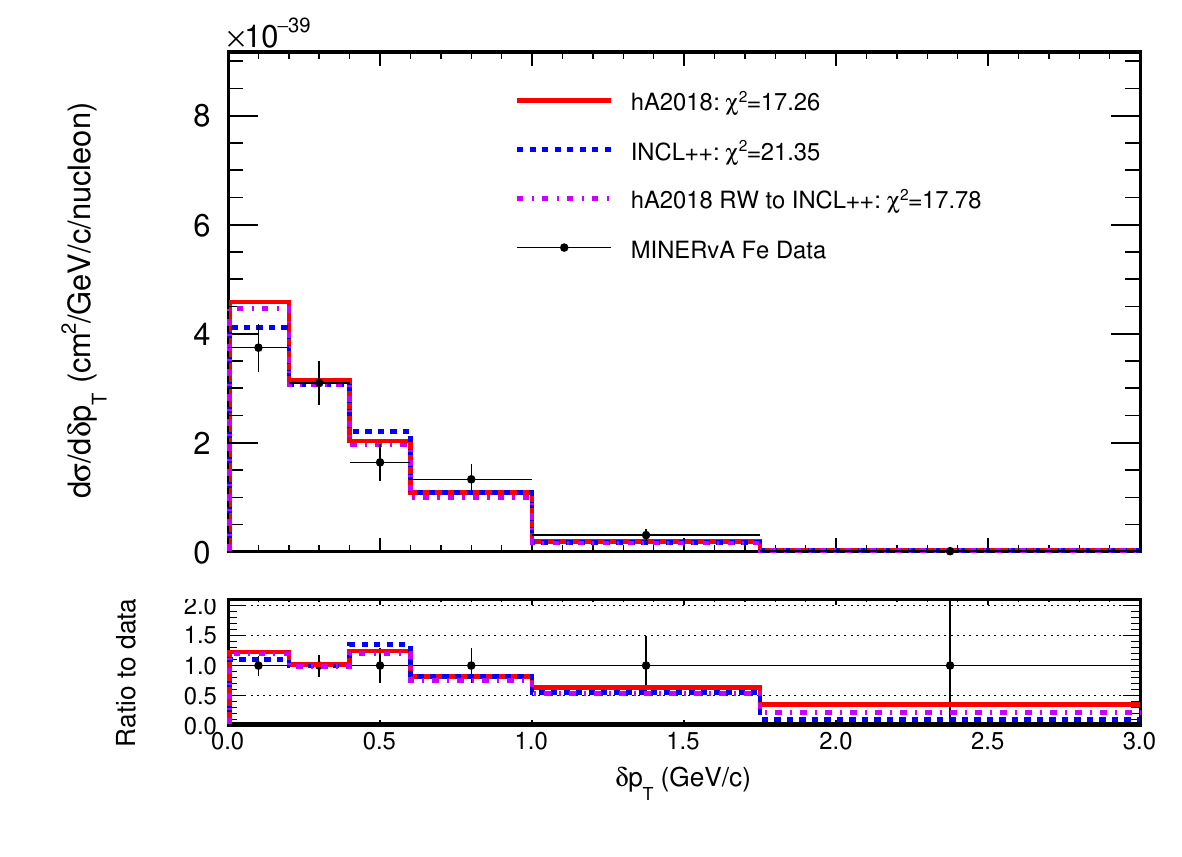}
    \caption{Results on the model predictions compared to MINERvA CC0$\pi$ on iron~\cite{MINERvA:2025tem}. Similar to comparisons on carbon, the differential neutrino cross sections for leading proton momentum (top) and transverse kinematic imbalance (bottom) are shown.}
    \label{fig:mnvFe}
\end{figure*}

The minimal model spread observed between the T2K and MicroBooNE datasets leads to inferences that the overall impact of nucleon multiplicity in intranuclear hadronic interactions does not substantially alter simulation-to-data comparisons. As a further test, MINERvA data was also used for comparisons. MINERvA uses the NuMI beam, which has an average neutrino energy of 6 GeV, a significant increase in neutrino energy compared to the average neutrino energies of 0.6 GeV and 0.8 GeV for the T2K and MicroBooNE experiments, respectively. The neutrino flux of the NuMI beam is more similar to the wide-band beam planned for DUNE than those of MicroBooNE and T2K. We show comparisons to two nuclear targets from MINERvA CC0$\pi$ datasets on carbon and iron~\cite{MINERvA:2025tem}. Predictions from \textsc{GENIE} AR23 compared to MINERvA data are shown in Figure~\ref{fig:mnvC} and Figure~\ref{fig:mnvFe}. Similar to before, all AR23 predictions with different hadronic models have similar agreement with data. However, the model spread reduces for both the proton momentum and the transverse kinematic imbalance for \textsc{INCL++} on both targets and for \textsc{Geant4} BC on carbon, suggesting the nucleon multiplicity and visible energy dials do reduce the difference between model predictions for datasets with neutrino energies comparable to the NuMI beam.

In this section, a method to alter the number of neutrons and protons based on Gaussian and exponential decay fits and the total visible energy was demonstrated. The impact these hadronic interaction predictions have on the final neutrino simulations was shown using the \textsc{GENIE} AR23 model with benchmarks to MicroBooNE, T2K, and MINERvA data. The hadronic models do not significantly alter simulation-to-data agreement. Still, the model spread does decrease when predictions are applied to the signal definition of the MINERvA CC0$\pi$ dataset on carbon and iron.

\section{Conclusion}\label{sec:conclusion}

FSI sensitivity remains an important unknown for any neutrino interaction or oscillation analysis. GENIE provides four models, and various references provide a first picture of the variations between these models. The hA2018 FSI model in GENIE is often used in analyses. It has a significantly different basis from the other models (hN2018, INCL++, and \textsc{Geant4} BC), and understanding this difference is the major goal of this paper. We identify the multiplicity of nucleons resulting from the FSI handling of pion absorption and nucleon knockout as a major intermediate step. This is an input to hA2018 and an output from the other codes, thus an important consideration. The hA2018 model identified the importance of the sum and difference of nucleon multiplicities and based that model on fits to available INC models. Here, the Gaussian parameterization was recognized for pion absorption multiplicity sum and difference and the nucleon knockout difference. On the other hand, the nucleon KO sum was described by an exponential decay function. These same forms were used in the present work, but had to be modified in some ways.

Sec.~\ref{sec:hadron-nucleus} shows comparisons to historical datasets that are strongly influenced by the nucleon knockout or pion absorption mechanisms. Double differential cross sections~\cite{Meier:1989tjq,Meier:1992anx,Mckeown:1981pw} are used here for validation of FSI codes. For proton or neutron knockout, the backward-going baryon production angle data are extremely sensitive to baryon knockout interactions. Here, the GENIE external codes (INCL++, \textsc{Geant4} BC) provide a better match to the data. The higher energy loss peak in $(\pi^+,p)$ interactions is very sensitive to pion absorption in the nuclear medium. Here, we find that \textsc{Geant4} BC is in poor agreement with the data.

These features are examined at the more basic level on nucleon multiplicities. The model behavior seen for hadron data is largely reproduced in the multiplicity distributions and visible energy distributions. 

To examine these differences, the paper introduces a way to parameterize the final state of pion absorption and nucleon knockout interactions using simple Gaussian and exponential decay functions. To be consistent across models that include or do not include compound nucleus processes, a cutoff in nucleon kinetic energy of 5 MeV is applied to both the nucleon multiplicity and the total visible energy distributions. Separate reweighting functions to adjust hA2018 distributions to match either \textsc{INCL++} or \textsc{Geant4} BC are presented.

A library of particle scattering simulations that approximate or simulate the entire cascade was produced using \textsc{GENIE}, which has its own particle transport models as well as implementations of \textsc{INCL++} and the \textsc{Geant4} BC model. Agreement between intranuclear cascade models improves due to alterations of the nucleon multiplicities via Gaussian and exponential decay reweighting functions. These parameterizations complement other procedures to handle model spread in hadron scattering, namely, template fitting the visible energy detectable through protons and pions in the final state.

Section~\ref{sec:reweight} demonstrates that simple alterations by Gaussian and exponential decay functions can be implemented for \textsc{GENIE} hA2018 simulations and decrease model spread in hadron-nucleus interactions. While the model spread decreases on MINERvA neutrino data by parameterizing nucleon multiplicities of the various models, the overall simulation-to-data differences in comparisons to MINERvA, MicroBooNE, and T2K datasets remain largely unchanged. In summary, the model spread and simulation-to-data discrepancies cannot be explained solely by the visible energy and final state hadron multiplicity of pion absorption and nucleon knockout FSI.

This is the major conclusion of this work. Although discrepancies are observed for hA2018 when compared to hadron-nucleus data, these differences are overwhelmed by the features of neutrino-nucleus interactions in existing measurements. There, wide ranges of neutrino energies are sampled, and many distributions in the final state are semi-inclusive functions, e.g., leading proton momentum distributions integrated over all angles. If statistical uncertainties were reduced, then improved sensitivities through double-differential distributions, such as those common with electron or hadron beams, would become possible.

Another benefit from this work is the careful examination in Appendix~\ref{app:pionabs} of the details inside these models. They show problems with energy balance for the hA2018 and hN2018 models which are now fixed in the upcoming GENIE v3.8 release.

The capability introduced here allows neutrino experiments to use the tools for fake data studies and systematic uncertainty evaluations addressing differences in particle transport models. This is much simpler than making multiple simulations using the different FSI codes, where care is needed to make sure every other aspect of the interaction is kept constant.

Future studies can broaden the impact of kinematic and particle multiplicity tuning for all hadron scattering channels, such as pion production, pion charge exchange, and quasielastic interactions, which have been omitted from this study. While quasielastic hadronic interactions have only two particles in the final state and can be altered with a simple visible energy template, the other hadronic scattering channels include mesons in the final state and require new parameters to describe these more complicated interactions.

\section*{Acknowledgments}
The authors thank Jeremy Wolcott, Gray Putnam, Rik Gran, Jaesung Kim, Joshua Barrow, Francisco Martinez Lopez, Laura Munteanu, and Stephen Dolan for their discussions regarding particle scattering models and analysis methods. Matthew King was supported by the National Science Foundation Graduate Research Fellowship Program under
NSF Award No. 2140001. 

\bibliographystyle{apsrev4-2}
\bibliography{references}
\appendix
\section{Software Used to Parameterize Hadron Scattering Simulations and Tables of Results}
\label{app:tables}

This section includes four tables of the parameters determined for the four models for carbon, oxygen, argon, and iron targets. The full simulation and analysis software is archived at \url{github.com/rdiurba/genieFSIStudies}. This includes scripts to simulate hadron scattering and produce Gaussian and exponential decay fits using a combination of Python and C++. For convenience, the tools for reweighting and the measurements made in this work are in the \textsc{hadDataRelease} directory.
\begin{table*}[htbp]
\centering
\caption{Fit parameters for hadronic rescattering models on $^{12}$C. Linear fits use $y = j + iT$. Nucleon sum distributions use $\Gamma = j \times \exp(i \times T^{k}) + l$.}
\label{tab:fitparams_c12}
\begin{tabular}{|>{\centering\arraybackslash}p{1.2cm}|l|cccc|cccc|cccc|cccc|}
\hline
\multicolumn{2}{c|}{} & \multicolumn{4}{c|}{\textbf{hA2018}} & \multicolumn{4}{c|}{\textbf{hN2018}} & \multicolumn{4}{c|}{\textbf{INCL++}} & \multicolumn{4}{c|}{\textbf{Geant4}} \\
\hline
 &  & $i$ & $j$ & $k$ & $l$ & $i$ & $j$ & $k$ & $l$ & $i$ & $j$ & $k$ & $l$ & $i$ & $j$ & $k$ & $l$ \\
\hline\hline
\multirow{3}{*}{$p$} & Diff.\ $\mu$ & 0.040 & 1.589 &  &  & 0.007 & 1.467 &  &  & 0.103 & 1.343 &  &  & -0.092 & 1.681 &  &  \\
\cline{2-18}
 & Diff.\ $\sigma$ & 0.026 & 2.076 &  &  & -0.024 & 1.565 &  &  & -0.067 & 0.992 &  &  & -0.040 & 1.131 &  &  \\
\cline{2-18}
 & Sum $\Gamma$ & -0.000 & 0.210 & 4.999 & 0.003 & -1.599 & 1.759 & 0.103 & 0.000 & -6.480 & 8.060 & 0.379 & 0.194 & -4.862 & 0.587 & 1.850 & 0.293 \\
\hline
\multirow{3}{*}{$n$} & Diff.\ $\mu$ & -0.048 & -0.403 &  &  & -0.003 & -0.480 &  &  & -0.118 & -0.199 &  &  & 0.066 & -0.454 &  &  \\
\cline{2-18}
 & Diff.\ $\sigma$ & 0.024 & 2.050 &  &  & -0.015 & 1.558 &  &  & -0.067 & 1.001 &  &  & -0.027 & 1.108 &  &  \\
\cline{2-18}
 & Sum $\Gamma$ & -0.566 & 0.000 & 2.788 & 0.212 & -1.401 & 1.513 & 0.100 & 0.000 & -5.438 & 0.701 & 1.065 & 0.197 & -4.615 & 0.559 & 1.893 & 0.287 \\
\hline
\multirow{4}{*}{$\pi^+$} & Diff.\ $\mu$ & 2.971 & 1.719 &  &  & -0.421 & 2.504 &  &  & 0.080 & 2.333 &  &  & -0.119 & 2.140 &  &  \\
\cline{2-18}
 & Diff.\ $\sigma$ & -0.068 & 3.709 &  &  & -0.003 & 1.320 &  &  & -0.126 & 0.880 &  &  & 0.031 & 0.989 &  &  \\
\cline{2-18}
 & Sum $\mu$ & 1.012 & 4.133 &  &  & 3.477 & 4.052 &  &  & 3.627 & 2.828 &  &  & 2.063 & 0.791 &  &  \\
\cline{2-18}
 & Sum $\sigma$ & 0.641 & 0.823 &  &  & 0.646 & 3.013 &  &  & 0.500 & 5.588 &  &  & 4.561 & 1.947 &  &  \\
\hline
\multirow{4}{*}{$\pi^0$} & Diff.\ $\mu$ & 2.903 & -0.137 &  &  & 0.066 & 0.462 &  &  & -0.105 & 0.610 &  &  & 0.011 & 0.538 &  &  \\
\cline{2-18}
 & Diff.\ $\sigma$ & 0.090 & 3.655 &  &  & -0.037 & 1.381 &  &  & -0.211 & 0.979 &  &  & -0.197 & 1.321 &  &  \\
\cline{2-18}
 & Sum $\mu$ & 0.786 & 4.168 &  &  & 2.254 & 5.126 &  &  & -1.084 & 5.927 &  &  & 0.002 & 1.999 &  &  \\
\cline{2-18}
 & Sum $\sigma$ & 0.678 & 0.775 &  &  & 0.010 & 3.080 &  &  & 10.101 & 0.574 &  &  & 3.901 & 2.043 &  &  \\
\hline
\multirow{4}{*}{$\pi^-$} & Diff.\ $\mu$ & 2.546 & -1.715 &  &  & 0.299 & -1.422 &  &  & -0.173 & -1.180 &  &  & 0.149 & -1.057 &  &  \\
\cline{2-18}
 & Diff.\ $\sigma$ & 0.352 & 3.402 &  &  & 0.014 & 1.349 &  &  & -0.154 & 0.917 &  &  & 0.098 & 0.981 &  &  \\
\cline{2-18}
 & Sum $\mu$ & 0.642 & 4.248 &  &  & 2.718 & 4.594 &  &  & 4.308 & 2.289 &  &  & 2.376 & 0.954 &  &  \\
\cline{2-18}
 & Sum $\sigma$ & 0.644 & 0.805 &  &  & -0.172 & 3.092 &  &  & 15.149 & -1.838 &  &  & 2.292 & 3.233 &  &  \\
\hline
\end{tabular}
\end{table*}

\begin{table*}[htbp]
\centering
\caption{Fit parameters for hadronic rescattering models on $^{16}$O. Linear fits use $y = j + iT$. Nucleon sum distributions use $\Gamma = j \times \exp(i \times T^{k}) + l$.}
\label{tab:fitparams_o16}
\begin{tabular}{|>{\centering\arraybackslash}p{1.2cm}|l|cccc|cccc|cccc|cccc|}
\hline
\multicolumn{2}{c|}{} & \multicolumn{4}{c|}{\textbf{hA2018}} & \multicolumn{4}{c|}{\textbf{hN2018}} & \multicolumn{4}{c|}{\textbf{INCL++}} & \multicolumn{4}{c|}{\textbf{Geant4}} \\
\hline
 &  & $i$ & $j$ & $k$ & $l$ & $i$ & $j$ & $k$ & $l$ & $i$ & $j$ & $k$ & $l$ & $i$ & $j$ & $k$ & $l$ \\
\hline\hline
\multirow{3}{*}{$p$} & Diff.\ $\mu$ & 0.019 & 1.599 &  &  & -0.003 & 1.480 &  &  & 0.118 & 1.338 &  &  & -0.139 & 1.813 &  &  \\
\cline{2-18}
 & Diff.\ $\sigma$ & 0.001 & 2.167 &  &  & -0.009 & 1.680 &  &  & -0.035 & 1.082 &  &  & -0.014 & 1.188 &  &  \\
\cline{2-18}
 & Sum $\Gamma$ & -0.000 & 0.187 & 4.889 & 0.000 & -2.254 & 2.831 & 0.115 & 0.000 & -4.971 & 0.843 & 1.076 & 0.211 & -4.408 & 0.804 & 1.677 & 0.267 \\
\hline
\multirow{3}{*}{$n$} & Diff.\ $\mu$ & -0.036 & -0.404 &  &  & 0.015 & -0.502 &  &  & -0.141 & -0.070 &  &  & 0.039 & -0.319 &  &  \\
\cline{2-18}
 & Diff.\ $\sigma$ & -0.002 & 2.143 &  &  & -0.008 & 1.685 &  &  & -0.036 & 1.095 &  &  & -0.010 & 1.183 &  &  \\
\cline{2-18}
 & Sum $\Gamma$ & -34.832 & 0.006 & 1.344 & 0.186 & -5.009 & 8.397 & 0.188 & 0.251 & -5.070 & 0.898 & 1.072 & 0.210 & -4.792 & 0.756 & 1.798 & 0.275 \\
\hline
\multirow{4}{*}{$\pi^+$} & Diff.\ $\mu$ & 2.845 & 1.900 &  &  & -0.230 & 2.293 &  &  & 0.200 & 2.217 &  &  & -0.072 & 2.222 &  &  \\
\cline{2-18}
 & Diff.\ $\sigma$ & -0.119 & 3.929 &  &  & 0.117 & 1.508 &  &  & -0.106 & 1.027 &  &  & 0.039 & 1.088 &  &  \\
\cline{2-18}
 & Sum $\mu$ & 1.386 & 4.263 &  &  & 3.627 & 5.164 &  &  & 6.738 & -0.854 &  &  & 4.987 & -0.404 &  &  \\
\cline{2-18}
 & Sum $\sigma$ & 0.958 & 0.920 &  &  & -0.058 & 3.962 &  &  & 4.777 & 4.435 &  &  & 5.093 & 2.396 &  &  \\
\hline
\multirow{4}{*}{$\pi^0$} & Diff.\ $\mu$ & 3.244 & -0.418 &  &  & 0.115 & 0.411 &  &  & 0.045 & 0.578 &  &  & -0.123 & 0.707 &  &  \\
\cline{2-18}
 & Diff.\ $\sigma$ & 0.223 & 3.624 &  &  & 0.006 & 1.618 &  &  & -0.173 & 1.141 &  &  & -0.141 & 1.362 &  &  \\
\cline{2-18}
 & Sum $\mu$ & 1.191 & 4.245 &  &  & 3.372 & 5.534 &  &  & 3.606 & 1.053 &  &  & 4.071 & -0.071 &  &  \\
\cline{2-18}
 & Sum $\sigma$ & 0.938 & 0.905 &  &  & 0.653 & 3.195 &  &  & 4.791 & 4.481 &  &  & 4.548 & 2.242 &  &  \\
\hline
\multirow{4}{*}{$\pi^-$} & Diff.\ $\mu$ & 2.474 & -1.801 &  &  & 0.082 & -1.218 &  &  & -0.178 & -1.021 &  &  & -0.020 & -0.895 &  &  \\
\cline{2-18}
 & Diff.\ $\sigma$ & 0.190 & 3.538 &  &  & 0.004 & 1.583 &  &  & -0.139 & 1.057 &  &  & 0.146 & 1.051 &  &  \\
\cline{2-18}
 & Sum $\mu$ & 0.951 & 4.388 &  &  & 4.206 & 4.812 &  &  & 3.501 & 0.742 &  &  & 3.479 & 0.357 &  &  \\
\cline{2-18}
 & Sum $\sigma$ & 1.018 & 0.872 &  &  & 0.159 & 3.551 &  &  & 4.999 & 4.300 &  &  & 5.611 & 2.139 &  &  \\
\hline
\end{tabular}
\end{table*}

\begin{table*}[htbp]
\centering
\caption{Fit parameters for hadronic rescattering models on $^{40}$Ar. Linear fits use $y = j + iT$. Nucleon sum distributions use $\Gamma = j \times \exp(i \times T^{k}) + l$.}
\label{tab:fitparams_ar40}
\begin{tabular}{|>{\centering\arraybackslash}p{1.2cm}|l|cccc|cccc|cccc|cccc|}
\hline
\multicolumn{2}{c|}{} & \multicolumn{4}{c|}{\textbf{hA2018}} & \multicolumn{4}{c|}{\textbf{hN2018}} & \multicolumn{4}{c|}{\textbf{INCL++}} & \multicolumn{4}{c|}{\textbf{Geant4}} \\
\hline
 &  & $i$ & $j$ & $k$ & $l$ & $i$ & $j$ & $k$ & $l$ & $i$ & $j$ & $k$ & $l$ & $i$ & $j$ & $k$ & $l$ \\
\hline\hline
\multirow{3}{*}{$p$} & Diff.\ $\mu$ & -0.009 & 0.425 &  &  & -0.196 & 0.992 &  &  & -0.099 & 0.369 &  &  & -0.518 & 0.834 &  &  \\
\cline{2-18}
 & Diff.\ $\sigma$ & -0.023 & 2.405 &  &  & 0.204 & 2.049 &  &  & 0.260 & 1.337 &  &  & 0.127 & 1.546 &  &  \\
\cline{2-18}
 & Sum $\Gamma$ & -1.854 & 0.742 & 0.132 & 0.000 & -3.097 & 0.198 & 2.118 & 0.146 & -4.285 & 2.119 & 0.601 & 0.109 & -3.807 & 2.108 & 0.680 & 0.117 \\
\hline
\multirow{3}{*}{$n$} & Diff.\ $\mu$ & -0.098 & -1.390 &  &  & -0.162 & -0.972 &  &  & -0.415 & -0.807 &  &  & -0.348 & -1.174 &  &  \\
\cline{2-18}
 & Diff.\ $\sigma$ & -0.018 & 2.326 &  &  & 0.172 & 2.053 &  &  & 0.207 & 1.402 &  &  & 0.191 & 1.464 &  &  \\
\cline{2-18}
 & Sum $\Gamma$ & -1.779 & 0.691 & 0.133 & 0.000 & -2.952 & 0.213 & 1.836 & 0.154 & -4.342 & 2.273 & 0.595 & 0.110 & -3.716 & 1.930 & 0.711 & 0.118 \\
\hline
\multirow{4}{*}{$\pi^+$} & Diff.\ $\mu$ & 2.315 & 0.884 &  &  & -1.014 & 1.676 &  &  & -0.653 & 1.171 &  &  & -1.279 & 1.542 &  &  \\
\cline{2-18}
 & Diff.\ $\sigma$ & 0.342 & 3.232 &  &  & 0.523 & 2.167 &  &  & 0.691 & 1.272 &  &  & 0.298 & 1.355 &  &  \\
\cline{2-18}
 & Sum $\mu$ & 3.756 & 4.928 &  &  & 8.701 & 6.191 &  &  & 20.160 & -6.209 &  &  & 14.641 & -2.032 &  &  \\
\cline{2-18}
 & Sum $\sigma$ & 2.380 & 1.544 &  &  & 1.794 & 4.722 &  &  & 5.250 & 6.592 &  &  & 1.937 & 4.014 &  &  \\
\hline
\multirow{4}{*}{$\pi^0$} & Diff.\ $\mu$ & 1.911 & -0.451 &  &  & -0.762 & 0.022 &  &  & -0.750 & -0.282 &  &  & -1.509 & 0.205 &  &  \\
\cline{2-18}
 & Diff.\ $\sigma$ & 0.408 & 3.235 &  &  & 0.574 & 2.143 &  &  & 0.568 & 1.382 &  &  & 0.083 & 1.698 &  &  \\
\cline{2-18}
 & Sum $\mu$ & 3.465 & 4.996 &  &  & 7.666 & 6.804 &  &  & 20.225 & -6.551 &  &  & 16.991 & -4.710 &  &  \\
\cline{2-18}
 & Sum $\sigma$ & 2.527 & 1.518 &  &  & 2.105 & 4.402 &  &  & 4.463 & 6.766 &  &  & 1.791 & 4.457 &  &  \\
\hline
\multirow{4}{*}{$\pi^-$} & Diff.\ $\mu$ & 1.505 & -1.735 &  &  & -0.141 & -1.916 &  &  & -0.971 & -1.622 &  &  & -1.708 & -1.206 &  &  \\
\cline{2-18}
 & Diff.\ $\sigma$ & 0.562 & 3.142 &  &  & 0.588 & 2.119 &  &  & 0.638 & 1.328 &  &  & 0.344 & 1.557 &  &  \\
\cline{2-18}
 & Sum $\mu$ & 3.673 & 4.950 &  &  & 8.477 & 5.736 &  &  & 19.748 & -6.569 &  &  & 13.665 & -0.886 &  &  \\
\cline{2-18}
 & Sum $\sigma$ & 2.495 & 1.497 &  &  & 1.861 & 4.554 &  &  & 4.453 & 6.378 &  &  & 1.956 & 3.830 &  &  \\
\hline
\end{tabular}
\end{table*}

\begin{table*}[htbp]
\centering
\caption{Fit parameters for hadronic rescattering models on $^{56}$Fe. Linear fits use $y = j + iT$. Nucleon sum distributions use $\Gamma = j \times \exp(i \times T^{k}) + l$.}
\label{tab:fitparams_fe56}
\begin{tabular}{|>{\centering\arraybackslash}p{1.2cm}|l|cccc|cccc|cccc|cccc|}
\hline
\multicolumn{2}{c|}{} & \multicolumn{4}{c|}{\textbf{hA2018}} & \multicolumn{4}{c|}{\textbf{hN2018}} & \multicolumn{4}{c|}{\textbf{INCL++}} & \multicolumn{4}{c|}{\textbf{Geant4}} \\
\hline
 &  & $i$ & $j$ & $k$ & $l$ & $i$ & $j$ & $k$ & $l$ & $i$ & $j$ & $k$ & $l$ & $i$ & $j$ & $k$ & $l$ \\
\hline\hline
\multirow{3}{*}{$p$} & Diff.\ $\mu$ & 0.000 & 0.707 &  &  & -0.231 & 1.103 &  &  & -0.141 & 1.119 &  &  & -0.544 & 1.861 &  &  \\
\cline{2-18}
 & Diff.\ $\sigma$ & -0.003 & 2.520 &  &  & 0.320 & 2.161 &  &  & 0.310 & 1.340 &  &  & 0.236 & 1.482 &  &  \\
\cline{2-18}
 & Sum $\Gamma$ & -0.726 & 0.182 & 0.693 & 0.004 & -2.676 & 0.222 & 1.744 & 0.105 & -4.573 & 3.824 & 0.451 & 0.083 & -3.970 & 4.141 & 0.432 & 0.059 \\
\hline
\multirow{3}{*}{$n$} & Diff.\ $\mu$ & -0.086 & -1.098 &  &  & -0.175 & -0.860 &  &  & -0.443 & -0.039 &  &  & -0.357 & -0.125 &  &  \\
\cline{2-18}
 & Diff.\ $\sigma$ & -0.001 & 2.481 &  &  & 0.274 & 2.174 &  &  & 0.291 & 1.373 &  &  & 0.228 & 1.510 &  &  \\
\cline{2-18}
 & Sum $\Gamma$ & -0.724 & 0.178 & 0.692 & 0.006 & -2.747 & 0.225 & 1.661 & 0.115 & -4.513 & 3.399 & 0.490 & 0.085 & -3.923 & 3.723 & 0.451 & 0.064 \\
\hline
\multirow{4}{*}{$\pi^+$} & Diff.\ $\mu$ & 1.774 & 0.967 &  &  & -1.103 & 1.855 &  &  & -0.830 & 2.020 &  &  & -0.733 & 2.272 &  &  \\
\cline{2-18}
 & Diff.\ $\sigma$ & 0.438 & 3.356 &  &  & 0.877 & 2.260 &  &  & 0.825 & 1.234 &  &  & 0.412 & 1.408 &  &  \\
\cline{2-18}
 & Sum $\mu$ & 5.260 & 5.594 &  &  & 12.106 & 5.878 &  &  & 14.739 & -0.306 &  &  & 15.239 & -1.212 &  &  \\
\cline{2-18}
 & Sum $\sigma$ & 3.247 & 1.939 &  &  & 2.366 & 5.478 &  &  & 9.244 & 3.102 &  &  & 3.210 & 3.181 &  &  \\
\hline
\multirow{4}{*}{$\pi^0$} & Diff.\ $\mu$ & 1.347 & -0.469 &  &  & -0.686 & 0.070 &  &  & -0.865 & 0.578 &  &  & -0.996 & 0.966 &  &  \\
\cline{2-18}
 & Diff.\ $\sigma$ & 0.437 & 3.378 &  &  & 0.626 & 2.420 &  &  & 0.720 & 1.349 &  &  & 0.240 & 1.650 &  &  \\
\cline{2-18}
 & Sum $\mu$ & 5.203 & 5.498 &  &  & 10.850 & 6.579 &  &  & 14.063 & -0.056 &  &  & 17.539 & -3.721 &  &  \\
\cline{2-18}
 & Sum $\sigma$ & 3.353 & 1.927 &  &  & 2.920 & 4.934 &  &  & 9.600 & 2.949 &  &  & 2.265 & 4.023 &  &  \\
\hline
\multirow{4}{*}{$\pi^-$} & Diff.\ $\mu$ & 1.033 & -1.957 &  &  & -0.080 & -1.739 &  &  & -1.011 & -0.777 &  &  & -1.259 & -0.323 &  &  \\
\cline{2-18}
 & Diff.\ $\sigma$ & 0.558 & 3.268 &  &  & 0.766 & 2.293 &  &  & 0.754 & 1.324 &  &  & 0.562 & 1.444 &  &  \\
\cline{2-18}
 & Sum $\mu$ & 5.181 & 5.717 &  &  & 11.911 & 5.342 &  &  & 15.224 & -0.971 &  &  & 15.780 & -1.533 &  &  \\
\cline{2-18}
 & Sum $\sigma$ & 2.944 & 2.108 &  &  & 2.752 & 4.853 &  &  & 9.183 & 3.260 &  &  & 2.349 & 3.807 &  &  \\
\hline
\end{tabular}
\end{table*}

\section{Pion Absorption on Argon Final State Kinetic Energy Distributions}
\label{app:pionabs}
\begin{figure*}[h]
    \centering
    \includegraphics[width=0.45\linewidth]{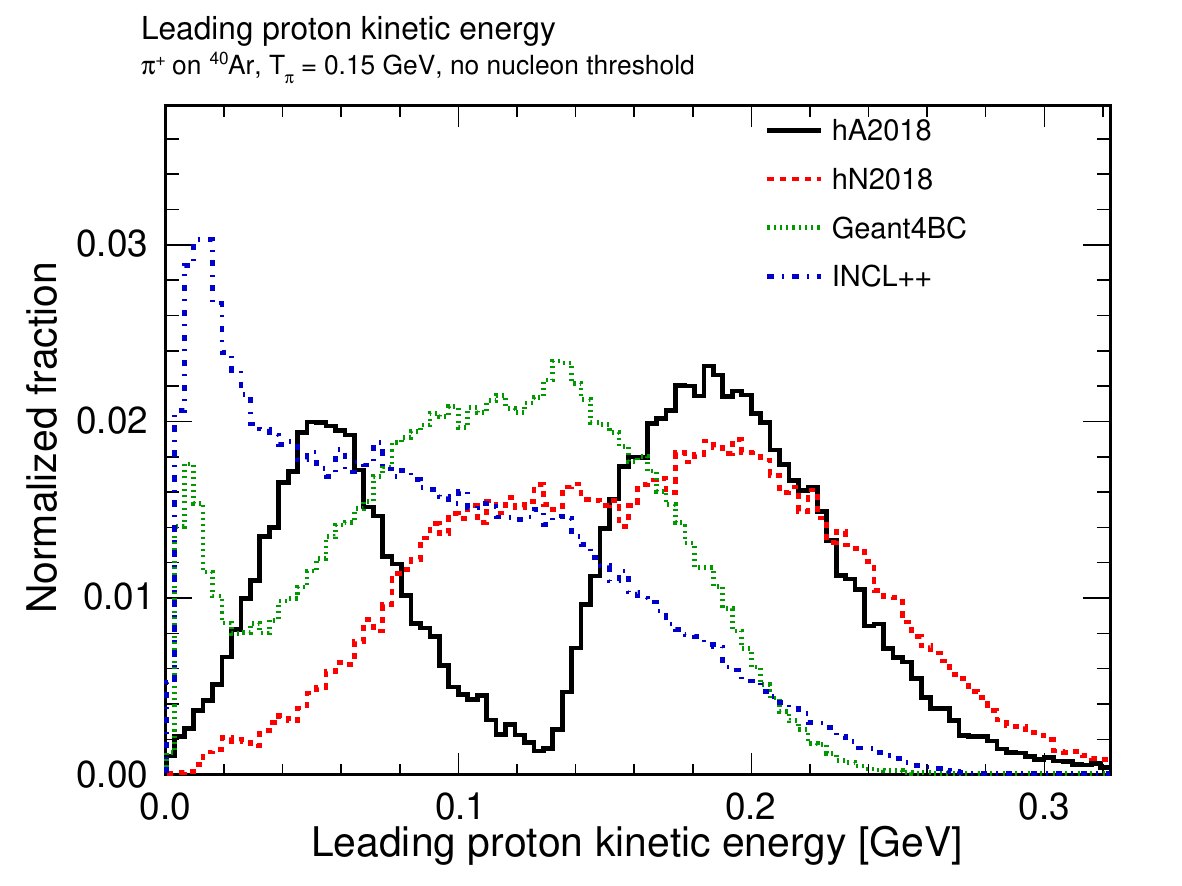}
    \includegraphics[width=0.45\linewidth]{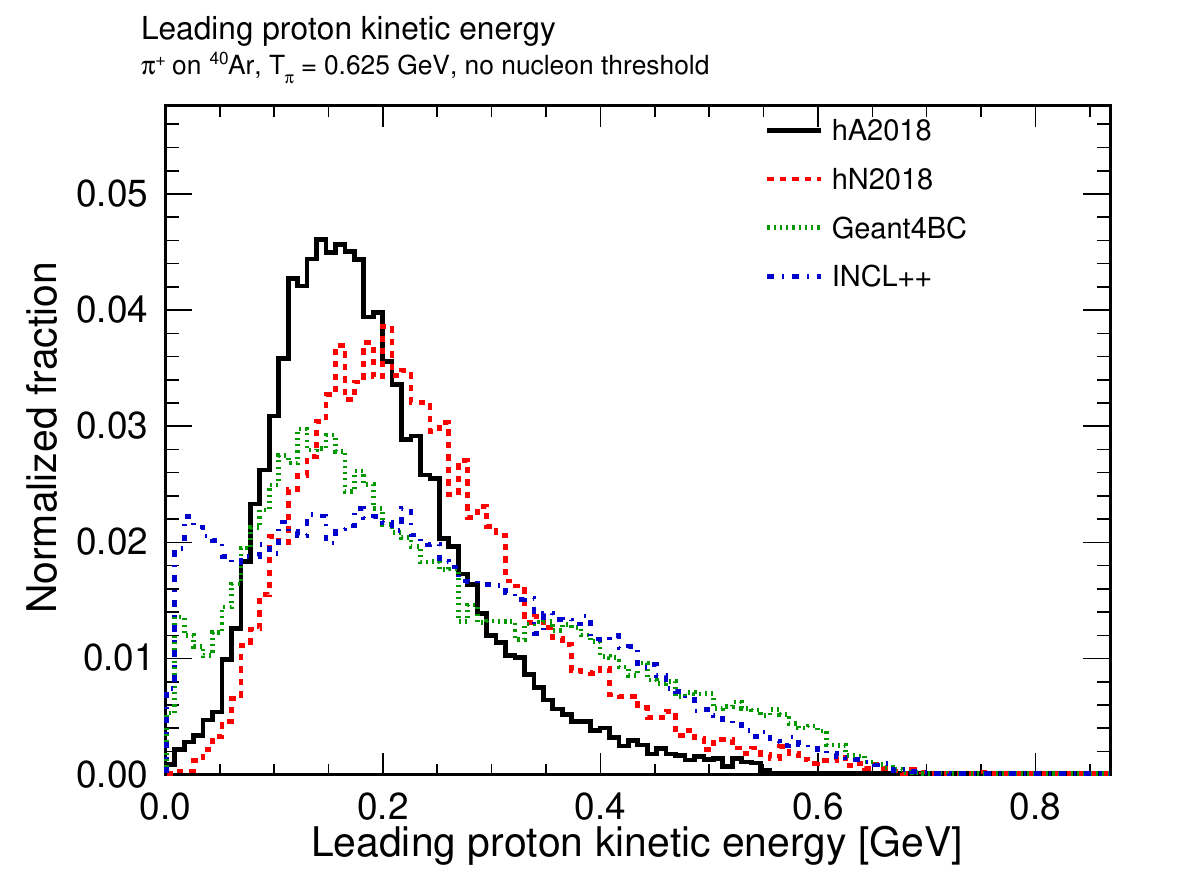}
    \caption{Plots of the leading proton kinetic energy for incident $\pi^+$ with $T_\pi=0.15\text{ GeV or }0.625\text{ GeV}$. The models peak at different energies, though this discrepancy is less pronounced at higher $T_\pi$. Note that the bimodal structure in hA2018 for $T_\pi=0.15\text{ GeV}$ is due to the fact that the pion rest mass energy is not added to the kinetic energy budget of final state particles in a pion absorption in the MN absorption simulation, but it is included in the QD absorption simulation. Hence, the peaks are separated by $m_\pi\approx 0.14\text{ GeV}$.}
    \label{fig:piAbsArLeadP}
\end{figure*}

\begin{figure*}[h]
    \centering
    \includegraphics[width=0.45\linewidth]{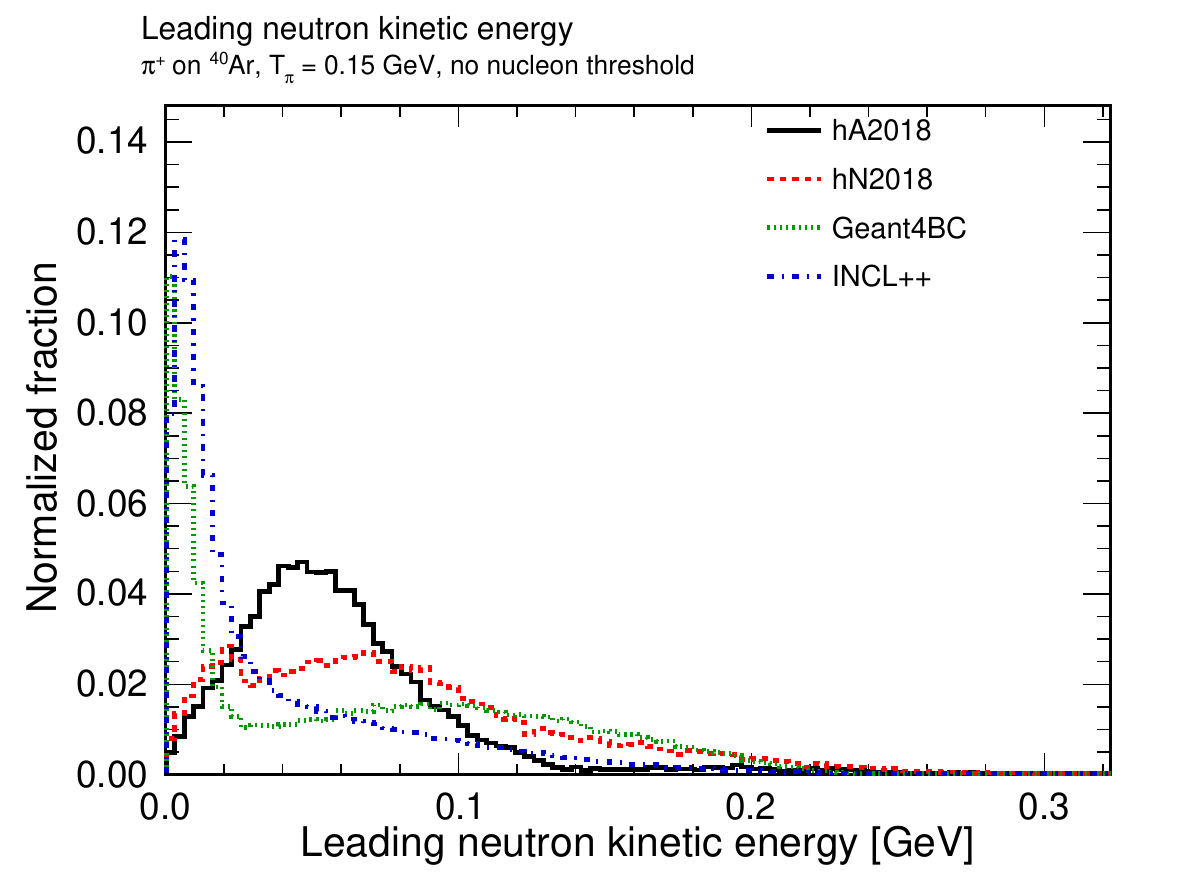}
    \includegraphics[width=0.45\linewidth]{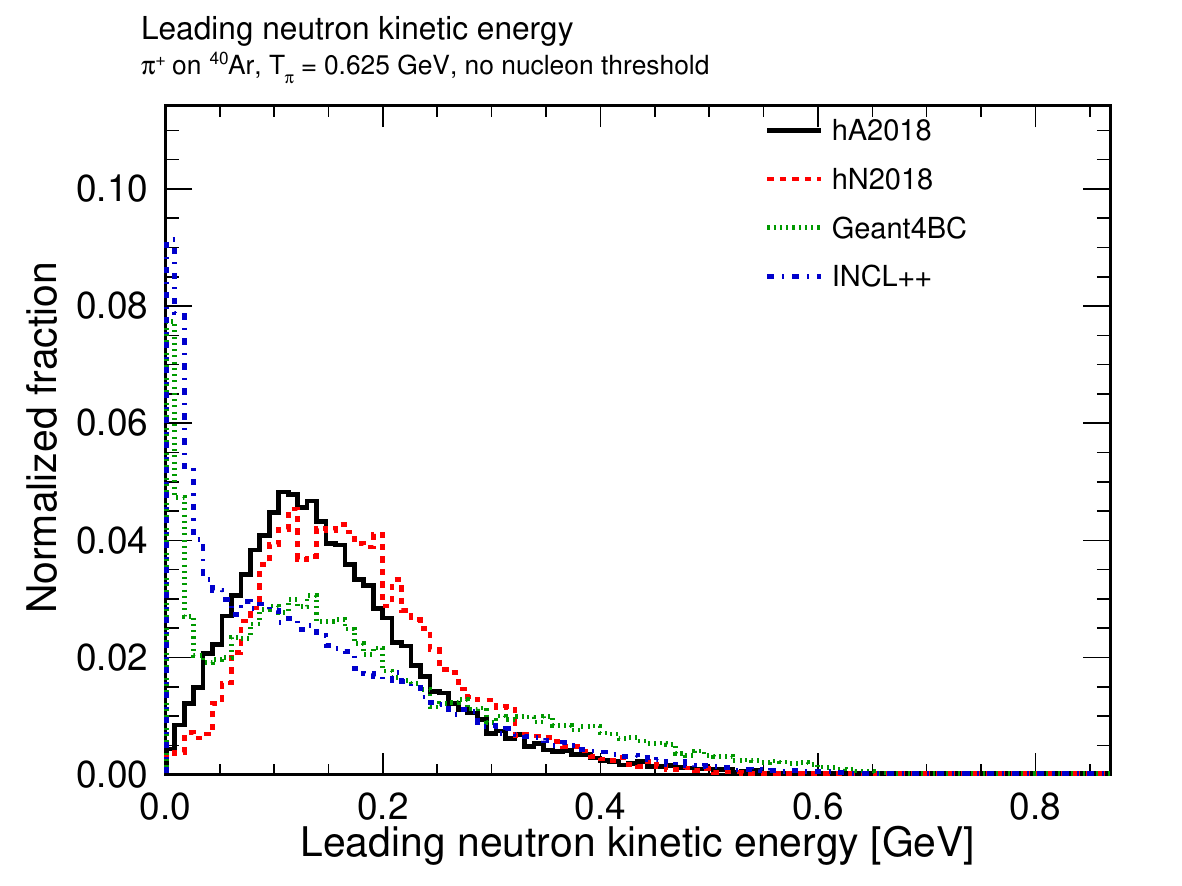}
    \caption{Plots of the leading neutron kinetic energy for incident $\pi^+$ with $T_\pi=0.15\text{ GeV or }0.625\text{ GeV}$. The energies tend to be lower for leading neutrons than leading protons, though the peaks are at similar energies for hA2018, \textsc{Geant4} BC, and \textsc{INCL++} between the leading proton and the leading neutron. The greatest shape difference is the sharp peak at very low leading neutron kinetic energies present in \textsc{Geant4} BC and \textsc{INCL++}, which isn't present for \textsc{GENIE} hA2018 or hN2018, and it is far more prominent for neutrons than protons. This indicates that in $\pi^+$ absorption events, final state multinucleon clusters and additional nuclear effects notwithstanding, \textsc{Geant4} BC and \textsc{INCL++} preferentially put more energy into protons than neutrons compared to \textsc{GENIE} hA2018 and hN2018.}
    \label{fig:piAbsArLeadN}
\end{figure*}

\begin{figure*}[h]
    \centering
    \includegraphics[width=0.45\linewidth]{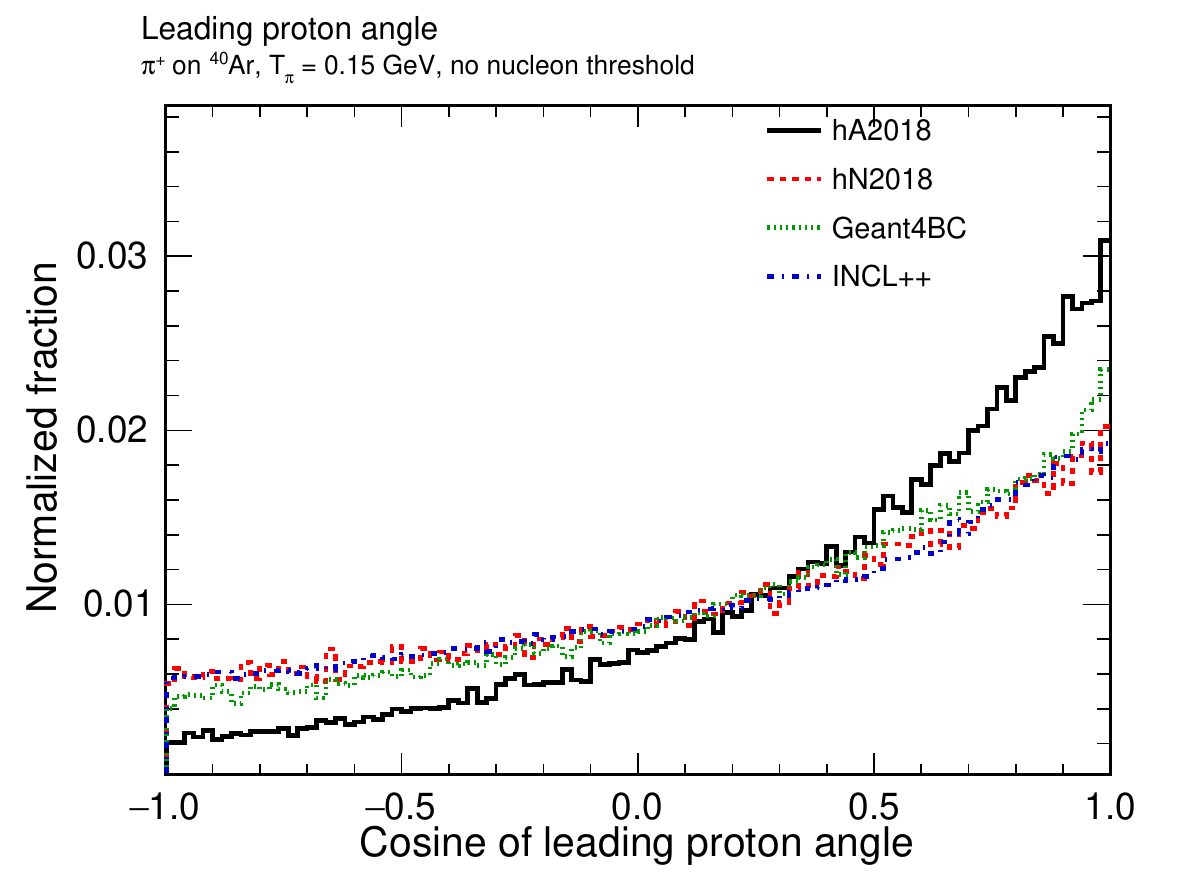}
    \includegraphics[width=0.45\linewidth]{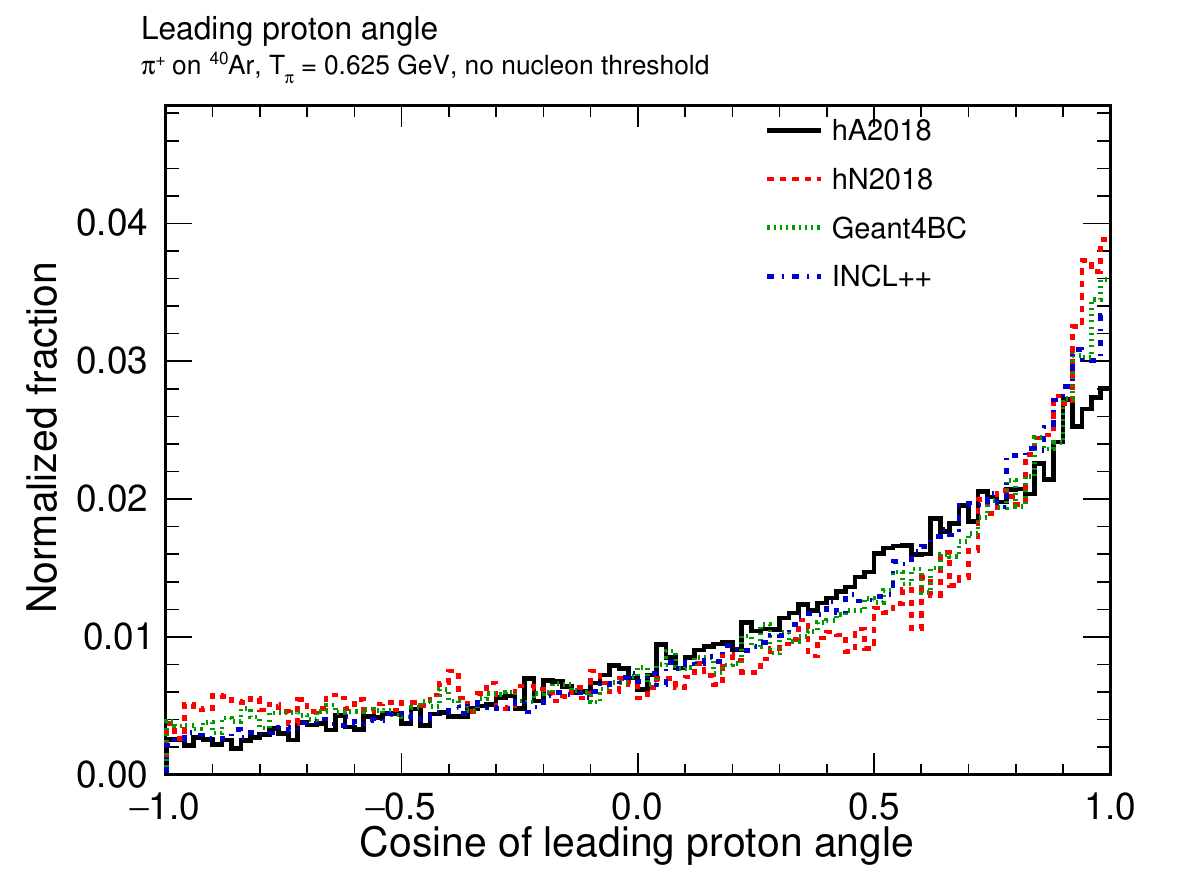}
    \caption{Plots of the leading proton angle relative to the incident pion beam for pions with 0.15 GeV or 0.625 GeV of kinetic energy. \textsc{GENIE} hA2018 predicts more forward leading protons for $T_\pi=0.15\text{ GeV}$ than hN2018, \textsc{Geant4} BC, and INCL++, which are all in rough agreement. This difference largely evens out at $0.625\text{ GeV}$, though the cascade models end up predicting a higher proportion of forward protons at the most forward angles compared to hA2018.}
    \label{fig:piAbsArLeadPCosTheta}
\end{figure*}

\begin{figure*}[h]
    \centering
    \includegraphics[width=0.45\linewidth]{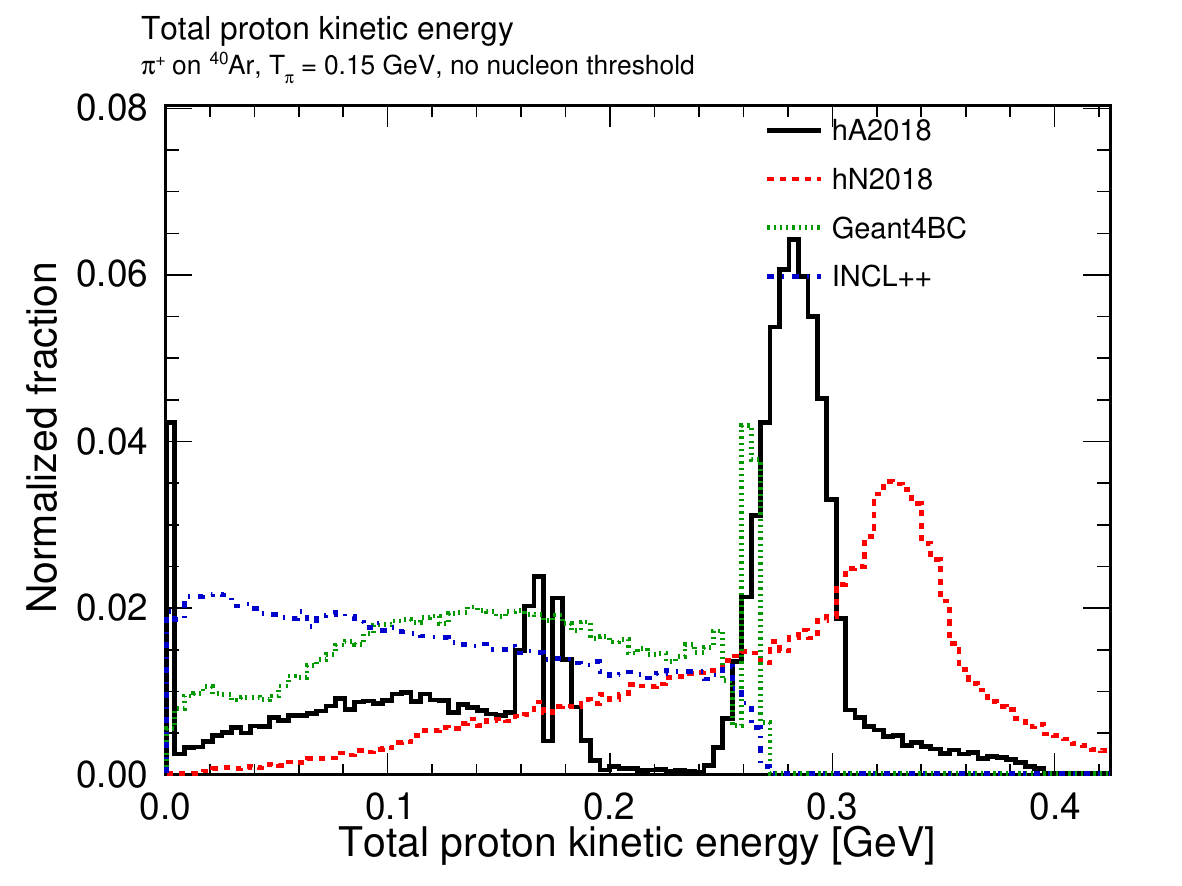}
    \includegraphics[width=0.45\linewidth]{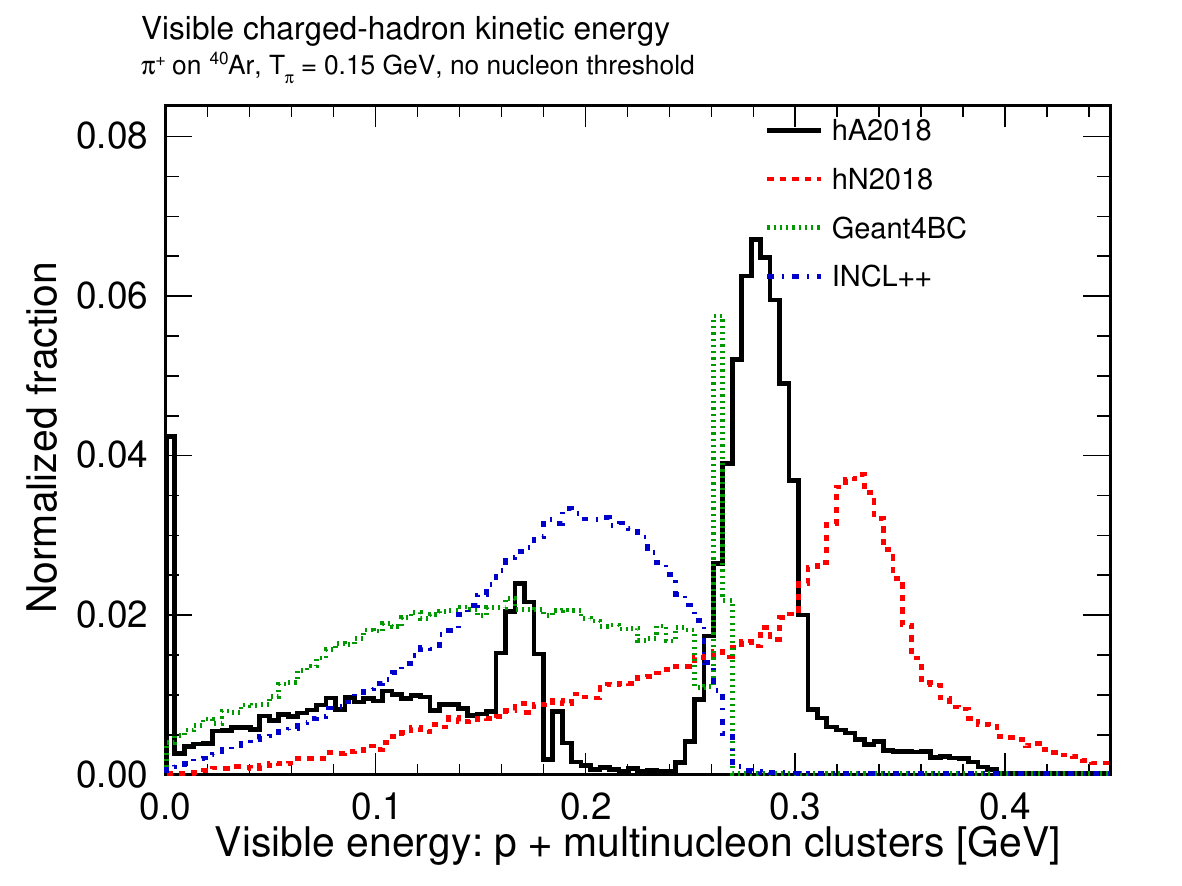}
    \includegraphics[width=0.45\linewidth]{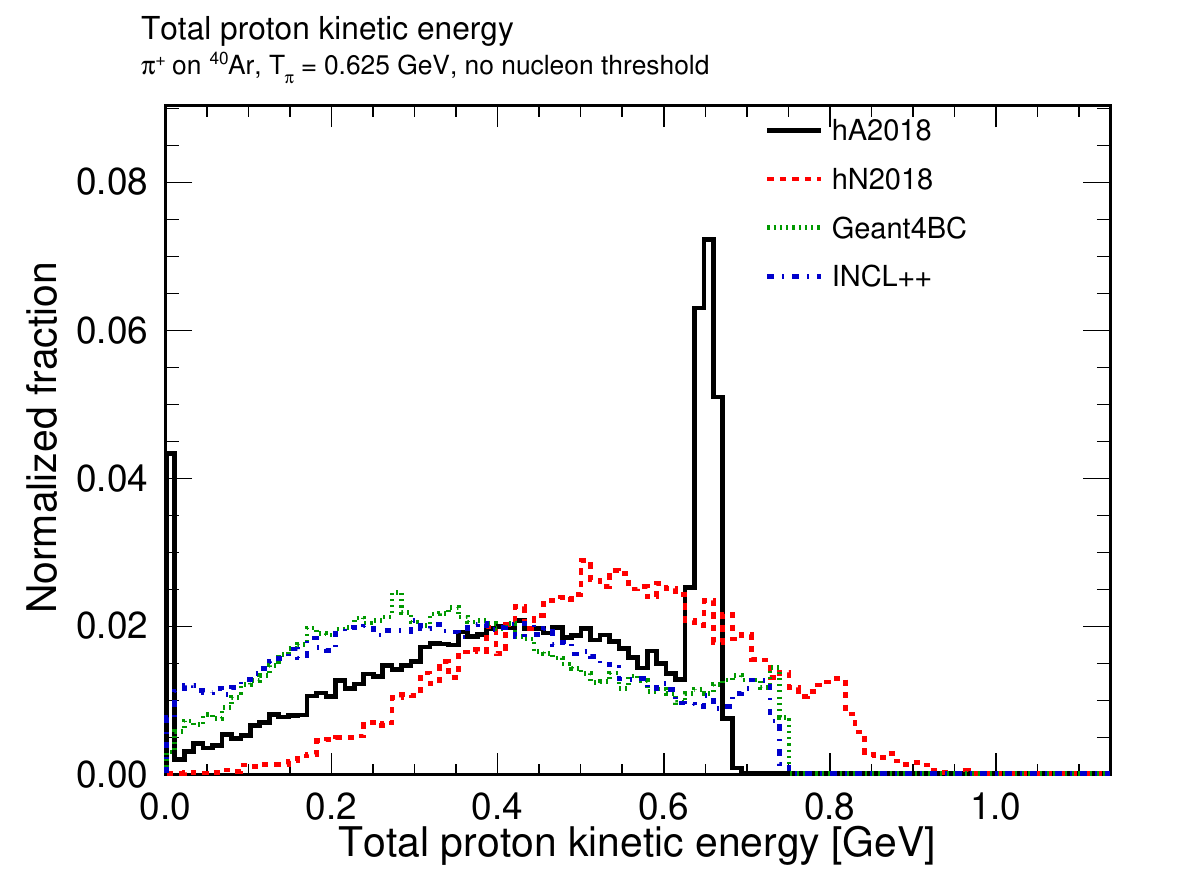}
    \includegraphics[width=0.45\linewidth]{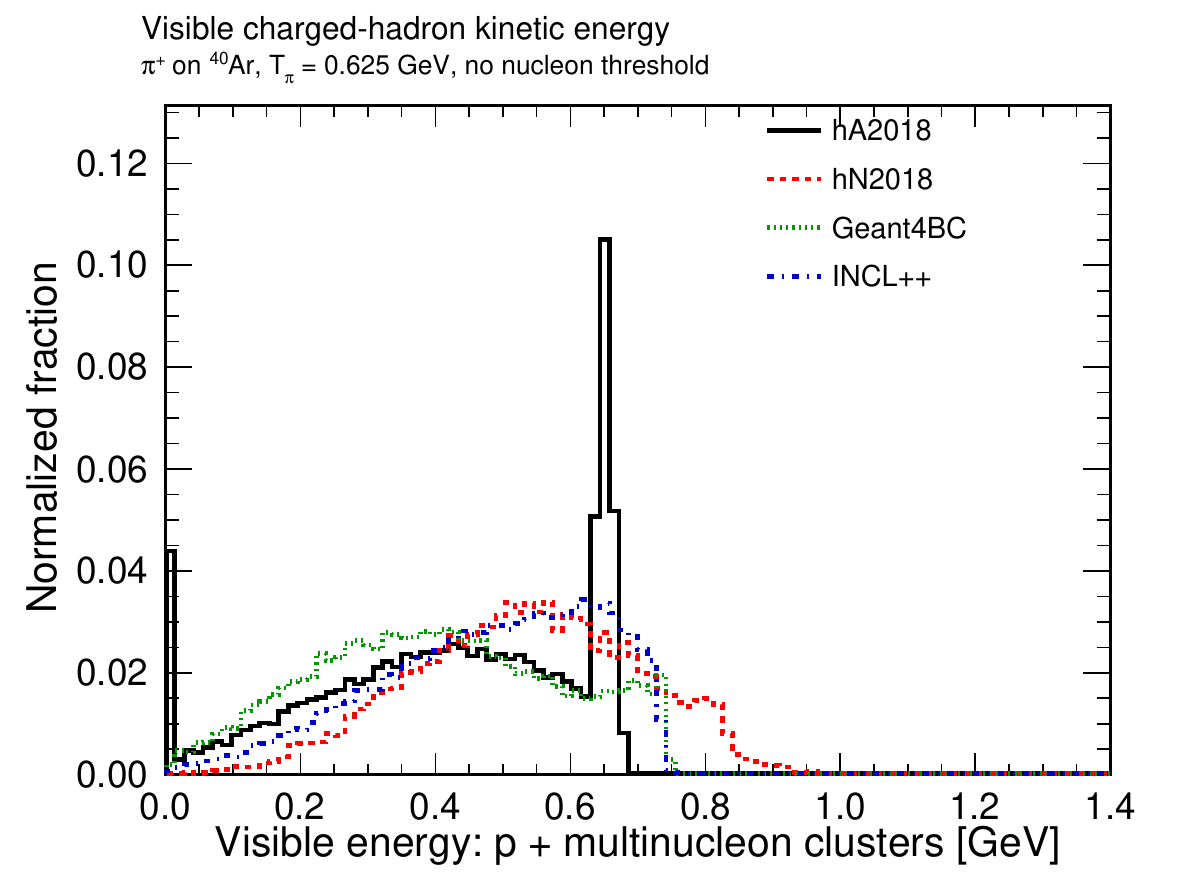}
    \caption{Plots of the total final state proton kinetic energy, for incident $\pi^+$ with $T_\pi=0.15\text{ GeV or }0.625\text{ GeV}$, with (right) or without (left) multinucleon clusters included in the final state. The visible energy (right) is a useful metric for LArTPC detectors, which are only able to reliably detect charged particles. We see hA2018 is heavily peaked near the incident pion kinetic energy. It is not at the pion total energy due to the MN absorption simulation property discussed in Fig. \ref{fig:piAbsArLeadP}. At lower $T_\pi$, \textsc{GENIE} hN2018 has a high-energy tail which exceeds the total pion energy; this indicates some unphysical addition of energy. At higher incident $T_\pi$, \textsc{Geant4} BC and \textsc{INCL++} agree remarkably well in total proton kinetic energy, while \textsc{GENIE} hN2018 and \textsc{INCL++} agree in visible energy. In the visible energy plots at $T_\pi=0.625\text{ GeV}$, hA2018 terminates near $T_\pi$, \textsc{Geant4} BC and \textsc{INCL++} terminate near $E_\pi$, and hN2018 extends beyond $E_\pi$.}
    \label{fig:piAbsArTotP}
\end{figure*}

\begin{figure*}[h]
    \centering
    \includegraphics[width=0.45\linewidth]{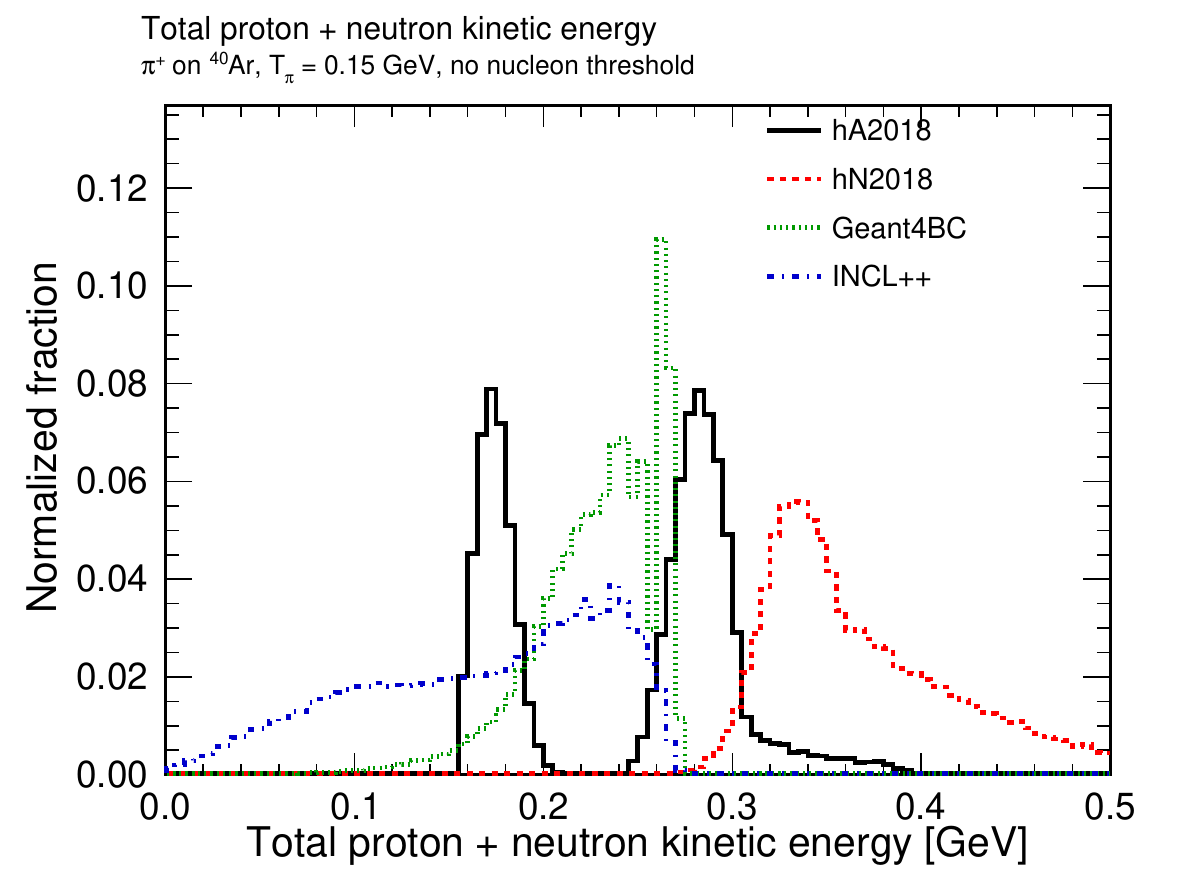}
    \includegraphics[width=0.45\linewidth]{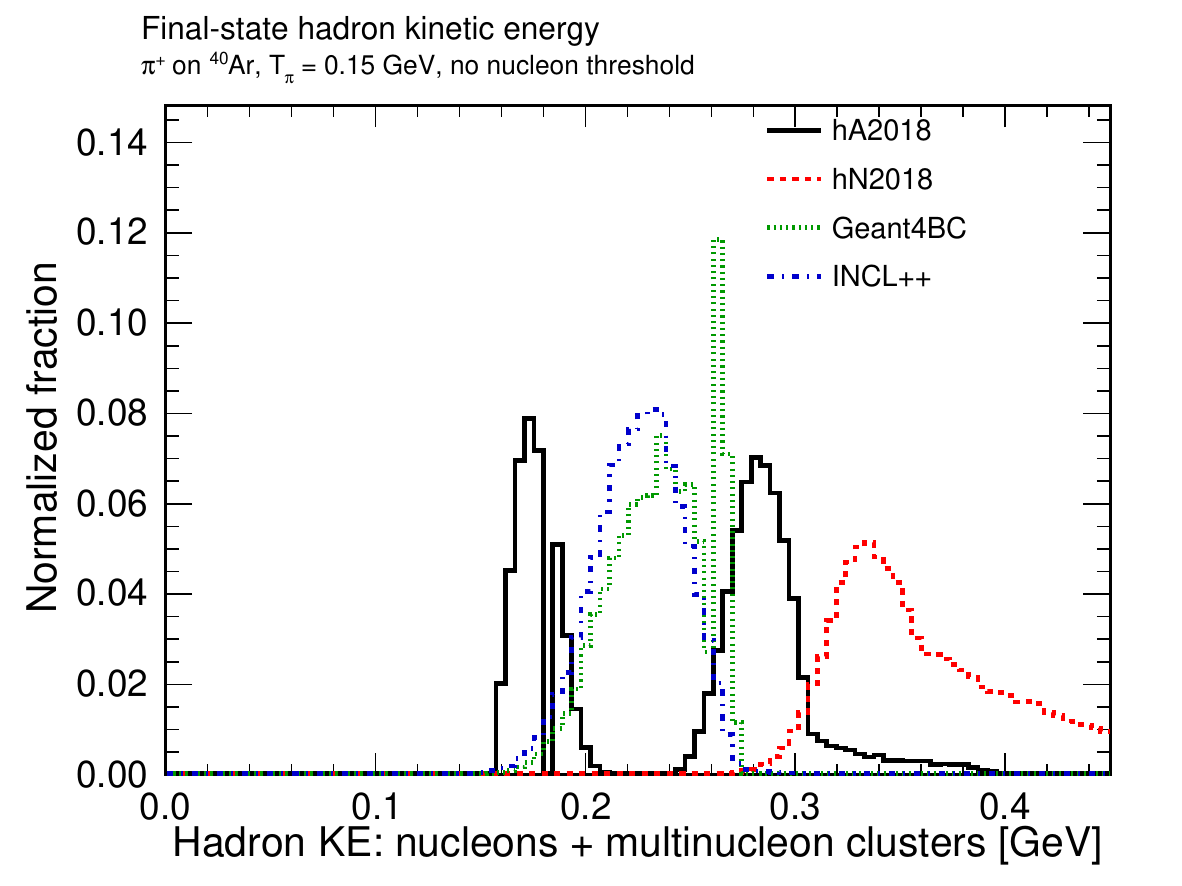}
    \includegraphics[width=0.45\linewidth]{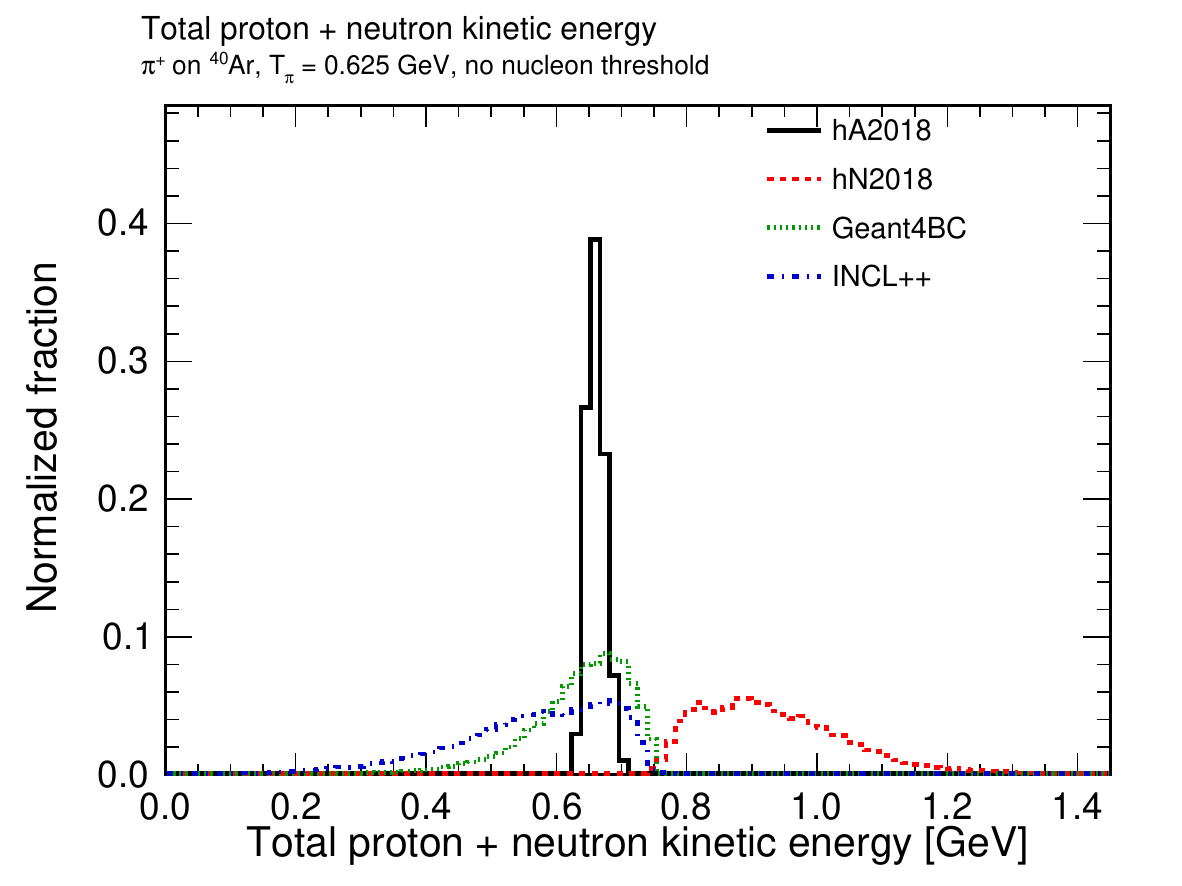}
    \includegraphics[width=0.45\linewidth]{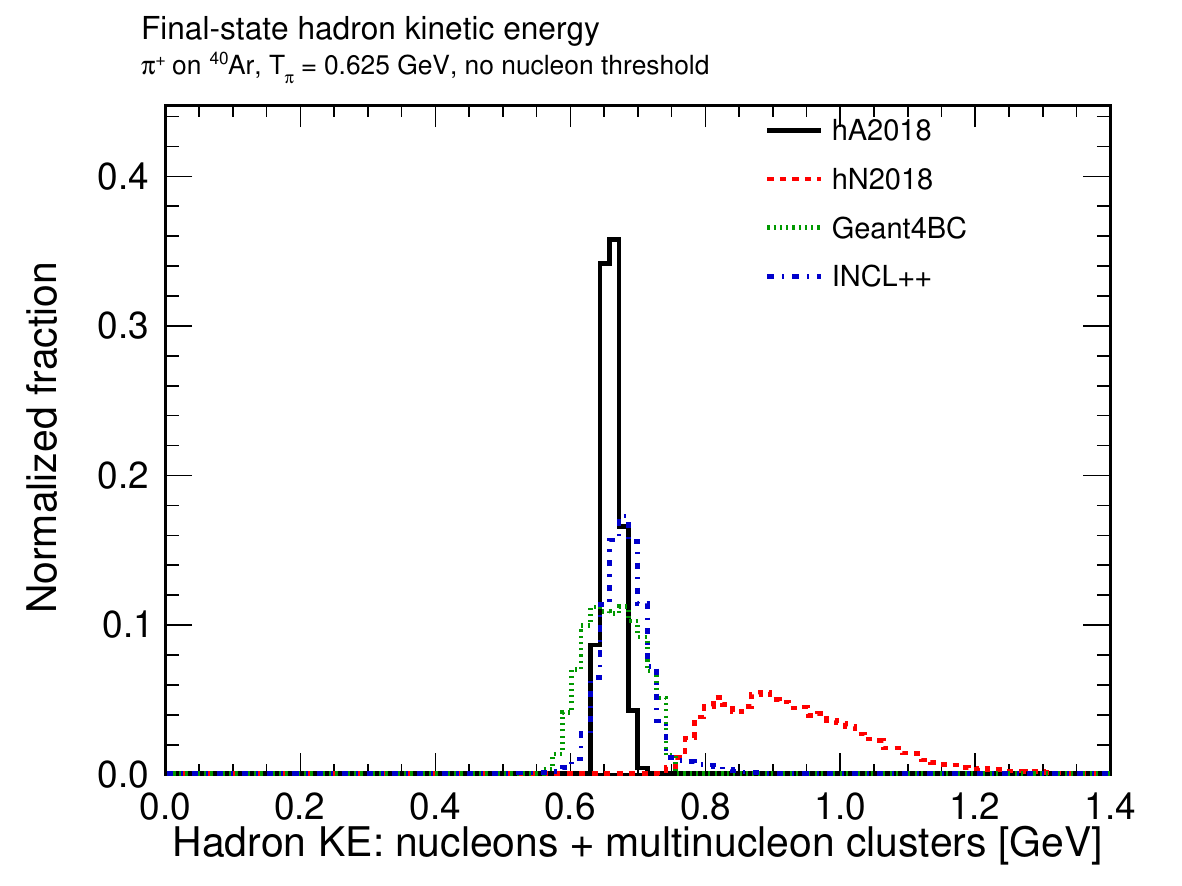}
    \caption{Plots of the total final state proton and neutron kinetic energy, for incident $\pi^+$ with $T_\pi=0.15\text{ GeV or }0.625\text{ GeV}$, with (right) or without (left) multinucleon clusters included in the final state. \textsc{Geant4} BC and \textsc{INCL++} model the total energy going into protons and neutrons differently, with \text{INCL++} having a more pronounced negative skew, while they agree (and are more symmetric) when including final state multinucleon clusters. We note that \textsc{GENIE} hA2018 and hN2018 do not model multinucleon clusters in the final state, so their distributions are unchanged when including multinucleon clusters in the final state kinetic energy accounting. \textsc{GENIE} hN2018 has the final state kinetic energy almost entirely in an unphysical region, much greater than the total incident pion energy. While the models give different predictions at $T_\pi=0.15\text{ GeV}$, at $T_\pi=0.625\text{ GeV}$, hA2018, \textsc{INCL++}, and \textsc{Geant4} BC all are symmetric around the same mean but with different widths in  the total hadronic kinetic energy.}
    \label{fig:piAbsArTotPN}
\end{figure*}

This section includes various final state kinetic energy distributions of interest for $\pi^+$ absorption on argon at incident pion energies relevant for SBN and DUNE simulated by the hA2018, hN2018, \textsc{INCL++}, and \textsc{Geant4} BC hadron scattering models. In particular, we look at the leading proton kinetic energy (Fig. \ref{fig:piAbsArLeadP}); leading neutron kinetic energy (Fig. \ref{fig:piAbsArLeadN}); leading proton angle (Fig. \ref{fig:piAbsArLeadPCosTheta}); total proton kinetic energy (Fig. \ref{fig:piAbsArTotP}); total visible kinetic energy, including protons and multinucleon clusters such as deuterons and helium nuclei (Fig. \ref{fig:piAbsArTotP}); total proton and neutron kinetic energy (Fig. \ref{fig:piAbsArTotPN}); and total hadronic kinetic energy, including nucleons and multinucleon clusters (Fig. \ref{fig:piAbsArTotPN}). For all of these distributions, there is a negligible effect from including the kinetic energy threshold of 5 MeV per nucleon on final state nucleons and clusters as introduced in Section \ref{sec:results}, so we show these distributions without the threshold applied.

Each figure shows the given distribution with the incident $T_{\pi^{+}}$ at 0.15 GeV or 0.625 GeV. In each plot, these distributions are shown for \textsc{GENIE} hA2018, \textsc{GENIE} hN2018, \textsc{INCL++}, and \textsc{Geant4} BC. All of these distributions are filled once per pion absorption interaction and normalized to 1 in order to factor out normalization differences due to different pion absorption cross sections between the models.

Through these figures, we observe features due to several effects. For example, \textsc{Geant4} BC and \textsc{INCL++} include multinucleon clusters, such as deuterons and helium nuclei, in their final states, while hA2018 and hN2018 only include protons and neutrons in the final states. This feature is seen in the differences in \textsc{INCL++} and \textsc{Geant4} BC between the proton kinetic energy and visible energy plots in Fig. \ref{fig:piAbsArTotP} and the nucleon kinetic energy and total hadronic kinetic energy plots in Fig. \ref{fig:piAbsArTotPN}. In these comparisons, we see that including the multinucleon clusters in the final state energy accounting causes the distributions for \textsc{INCL++} and \textsc{Geant4} BC to lose their low-energy tails in favor of a more symmetric shape.

In Fig. \ref{fig:piAbsArLeadN}, we note a pronounced shape difference between the models. There is a sharp peak at very low leading neutron kinetic energies present in \textsc{Geant4} BC and \textsc{INCL++} that isn't present in \textsc{GENIE} hA2018 or hN2018, and it is far more prominent for leading neutrons than leading protons as seen in Fig. \ref{fig:piAbsArLeadP}. This indicates that in $\pi^+$ absorption events, final state multinucleon clusters and additional nuclear effects notwithstanding, \textsc{Geant4} BC and \textsc{INCL++} preferentially put more energy into protons than neutrons compared to \textsc{GENIE} hA2018 and hN2018.

In Fig. \ref{fig:piAbsArLeadPCosTheta}, \textsc{GENIE} hA2018 predicts more forward leading protons for $T_\pi=0.15\text{ GeV}$ than hN2018, \textsc{Geant4} BC, and INCL++, which are all in rough agreement. This difference largely evens out at $0.625\text{ GeV}$, though the cascade models end up predicting a higher proportion of forward protons at the most forward angles compared to hA2018.

At $T_\pi=0.15$ GeV, \textsc{GENIE} hA2018 in Figs. \ref{fig:piAbsArLeadP}, \ref{fig:piAbsArTotP}, and \ref{fig:piAbsArTotPN} exhibits a bimodal shape because the pion rest mass energy is not added to the kinetic energy of final state particles in a pion absorption interaction when simulating MN absorption. This is not the case for the QD absorption simulation in hA2018. This property is addressed in future versions of \textsc{GENIE}, where the pion mass is added to the final state kinetic energy for both QD and MN absorption.

In Fig. \ref{fig:piAbsArTotPN}, \textsc{GENIE} hN2018 has much higher final state kinetic energy than $E_\pi$, indicating unphysical behavior in the simulation. Given that the models agree in final state visible energy in Fig. \ref{fig:piAbsArTotP} at $T_\pi=0.625\text{ GeV}$, we infer that the excess kinetic energy in hN2018 is given to neutrons, at least at higher $T_\pi$.

Overall, the four models produce more similar results at higher $T_\pi$ than lower $T_\pi$, indicating a convergence in the performance of the cascade models and single-step empirical model when there are more steps in the cascade and more nucleons and energy in the final state.

\end{document}